\documentclass[aps,preprint,groupedaddress,nofootinbib,showpacs,floatfix]{revtex4}
\usepackage{amsmath}
\usepackage{graphicx}
\usepackage{amssymb}
\usepackage{hyperref}
\usepackage{color}
\usepackage{soul,xcolor} 
\usepackage{pdfcomment} 

\newcommand{\beq}{\begin{equation}}
\newcommand{\eeq}{\end{equation}}
\newcommand{\bea}{\begin{eqnarray}}
\newcommand{\eea}{\end{eqnarray}}

\newcommand{\RNum}[1]{\uppercase\expandafter{\romannumeral #1\relax}}

\usepackage{epsfig,amssymb,amsmath,psfrag,epstopdf,color}
\usepackage{booktabs}
\usepackage{multirow}
\usepackage{ulem}
\usepackage{amsthm}

\def\3g{{\gamma\gamma\gamma}}
\begin{document}

\setstcolor{red} 

\newcommand{\CTHERAII}{CT14$_{\textrm{HERAII}}$}

\preprint{MSUHEP-19-024}
\preprint{TTK-19-49}
\preprint{Cavendish-HEP-19/19}

\title{An exploratory study of the impact of CMS double-differential top distributions on the gluon parton distribution function}

\author{Micha\l{} Czakon$^{(a,1)}$, Sayipjamal Dulat$^{(b,2)}$, Tie-Jiun Hou$^{(c,3)}$, Joey Huston$^{(d,4)}$, Alexander Mitov$^{(e,5)}$, 
Andrew S.~Papanastasiou$^{(e,f,6)}$, Ibrahim Sitiwaldi$^{(a,7)}$, Zhite Yu$^{(d,8)}$, and C.--P. Yuan$^{(d,9)}$}

\email{
$^{1}$mczakon@physik.rwth-aachen.de,
$^{2}$sdulat@hotmail.com, 
$^{3}$houtiejiun@mail.neu.edu.cn, 
$^{4}$Huston@msu.edu, 
$^{5}$adm74@cam.ac.uk, 
$^{6}$andrewp@hep.phy.cam.ac.uk, 
$^{7}$ibrahim010@sina.com, 
$^{8}$yuzhite@msu.edu, 
$^{9}$yuan@pa.msu.edu
}  
	
\affiliation{
$^{(a)}$ Institut f\"ur Theoretische Teilchenphysik und Kosmologie, RWTH Aachen University, D-52056 Aachen, Germany\\ 		
$^{(b)}$ School of Physics Science and Technology Xinjiang University, Urumqi, Xinjiang 830046 China. \\
$^{(c)}$ Department of Physics, College of Sciences, Northeastern University, Shenyang , Liaoning 110819 China \\
$^{(d)}$ Department of Physics and Astronomy, Michigan State University, East Lansing, MI 48824 U.S.A.\\ 
$^{(e)}$ Cavendish Laboratory, University of Cambridge, Cambridge CB3 0HE, UK\\ 
$^{(f)}$ MRC Institute of Genetics and Molecular Medicine, University of Edinburgh, Crewe Road, Edinburgh EH4 2XU, UK\\
}%

\date{\today}

\begin{abstract}

LHC $t\bar{t}$ data have the potential to provide constraints on the gluon distribution, especially at high $x$, with both ATLAS and CMS performing differential measurements. Recently, CMS has measured double-differential $t\bar{t}$ distributions at 8 TeV. In this paper we examine the impact of this data set on the gluon distribution. To that end we develop novel, double-differential NNLO predictions for that data. No significant impact is found when the CMS data is added to the CT14HERA2 global PDF fit, due to the larger impact of the inclusive jet data from both the Tevatron and  the LHC. If the jet data are removed from the fit, then an impact is observed. If the CMS data is scaled by a larger weight, representing the greater statistical power of the jet data, a roughly equal impact on the gluon distribution is observed for the $t\bar{t}$ as for the inclusive jet data. For data samples with higher integrated luminosity at 13 TeV, a more significant  impact of the double-differential $t\bar{t}$ data may be observed. 

\end{abstract}


\maketitle

\tableofcontents

\section{\label{sec:Introduction}Introduction} 

One of the limitations in searches for potential new physics at the LHC is the theoretical uncertainty in predictions for the standard model backgrounds to the new physics. In general, new physics is expected to occur at high masses, and thus requires the colliding partons to have relatively large fractions ($x$) of the parent protons' momenta.  These theoretical uncertainties include those related to the parton distribution functions (PDFs) at high $x$, especially those of the gluon distribution, the most poorly known PDF in this kinematic region. Until recently, the only data included in global PDF fits sensitive to the value of the high $x$ gluon  were those from inclusive jet production. Older parton distribution functions have used only jet data from the Tevatron; with newer generations of PDFs, jet data from the LHC has been added and generally has a significance equal to, or greater than, the Tevatron jet data, due to the wider kinematic range and the smaller systematic errors. 

For high transverse momentum jet production, however, the gluon distribution is sub-dominant, with $qq$ scattering being the dominant sub-process, followed by $gq$ scattering. Top pair production, on the other hand, is dominated by the $gg$ initial state, and thus provides a direct handle on the gluon distribution. For top pair production at high mass, rapidity or transverse momentum, the sensitivity continues to high momentum fraction $x$ values. 

Both ATLAS and CMS have measured 
top pair production with variables such as the $t\bar{t}$ mass, rapidity ($y$), either of the individual top quark (anti-quark), or of the pair, and the transverse momentum ($p_T$), again either of the individual top quark (anti-quark), or of the combination~\cite{Aad:2015mbv,Khachatryan:2015oqa,Khachatryan:2016mnb,Aad:2019ntk}. Some of this data has been included in PDF fits (see, for example, Refs.~\cite{Hou:2019gfw, Czakon:2016olj, Ball:2017nwa, Bailey:2019yze}). Each distribution, or combination, has a sensitivity to the gluon distribution. Recently, CMS has measured double-differential top pair distributions, using combinations of the above variables~\cite{Sirunyan:2017azo}, which have the potential to provide a greater sensitivity to the initial state gluon distribution, when combined with the recent NNLO calculation of such double-differential distributions. The NNLO calculation of top pair production of course also depends greatly on the value of $\alpha_s(m_Z)$, which itself is anti-correlated with the high $x$ gluon distribution. 
In this paper, we explore the relative sensitivity and importance of the double-differential top-pair distributions, to the high $x$ gluon distribution, with current data, and with extrapolations to what might be expected from future data. 

This paper is organized as follows:
Sec.\ref{sec:overview} describes  the double-differential measurements of CMS. 
Sec.\ref{sec:NNLO} then discusses the theoretical framework for the calculation of the double-differential theoretical predictions, and their inclusion in fastNNLO.    
Sec.\ref{sec:correlation} then explores the correlation between the measured distributions and the parton $x$ values of the gluon distribution. The correlations indicate the kinematic range over which the data may have some influence on the gluon distribution in the global PDF fits. 
Correlation, however, is not sufficient by itself to describe the impact. 
In Sec.\ref{sec:PDF}, ePump~\cite{Schmidt:2018hvu}, is used to update the PDFs in the CTEQ-TEA fitting framework. We discuss the impact of adding the CMS double-differential top data to the CT14HERA2 global fit~\footnote{CT14HERA2 was the latest published CTEQ-TEA set at the time of writing of this paper, with the CT18 paper \cite{Hou:2019efy} in progress. The gluon distribution for CT14HERA2 is similar to that obtained in CT14, except at very high $x$ where CT14 has a harder gluon.}, with and without jet data, from the Tevatron and the LHC, included in the original CT14HERA2 data set.  

Finally, Sec.\ref{sec:conclusion} concludes, and offers a projection of the impact of additional data at the LHC.

\section{\label{sec:overview} Experimental overview}

In this work, we consider the normalized double-differential top-quark data from 
CMS~\cite{Sirunyan:2017azo}, which consists of the following normalized $t\bar t$ distributions: the transverse momentum of the top quark ($p^t_T$) as a function of the top rapidity ($y_t$), the top-antitop system transverse momentum ($p_T^{t\bar{t}}$) as a function of the $t\bar{t}$ mass ($m_{t\bar{t}}$), the pseudo-rapidity separation of the top pair ($\Delta \eta_{t\bar{t}}$) as a function of the $t\bar t$ mass ($m_{t\bar t}$), the rapidity of the top quark ($y_t$) as a function of the $t\bar t$ mass ($m_{t\bar t}$), the $t\bar t$ rapidity ($y_{t\bar t}$) as a function of the $t\bar t$ mass ($m_{t\bar t}$), the azimuthal separation ($\Delta \phi_{t\bar{t}}$) of the top and anti-top  as a function of the $t\bar{t}$ mass ($m_{t\bar{t}}$). The last distribution is particularly sensitive to  the effects of soft gluon radiation, and not to the parton $x$ values,  
for a given $m_{t\bar{t}}$ value,
so will not be used in the following comparison to fixed-order predictions. The data sets, the number of data points in each set, and the internal CTEQ-TEA reference number are given in Table 1 below.  The data were taken at 8 TeV with an integrated luminosity of $19.7 fb^{-1}$. The statistical and systematic errors are typically of similar size, with the largest systematic error being due to the jet energy scale. In the original CMS paper~\cite{Sirunyan:2017azo}, the data were compared to NLO fixed-order predictions, to NLO+parton shower predictions and to fixed-order approximate NNLO predictions (for several observables). 
For this paper, comparisons are made to full NNLO predictions. 

\begin{table}[htbp]
	\begin{center}
		\caption{The double-differential $t\bar{t}$ data sets used in this study.  The ID number refers
to the internal references inside the CTEQ-TEA fitting code. }
		\begin{tabular}{| c | c | c |}
			\hline
			ID & data &  no. of data points \\
			\hline
			573 & $\sigma^{-1}\, d^2\sigma/dy_t dp^t_T$  & 16  \\
			\hline
			574 & $\sigma^{-1}\, d^2\sigma/dm_{t\bar t} dp^{t\bar t}_T$  & 16  \\
			\hline
			575 & $\sigma^{-1}\, d^2\sigma/dm_{t\bar t} d\Delta \eta_{t\bar t}$  & 12  \\
			\hline
			576 & $\sigma^{-1}\, d^2\sigma/dm_{t\bar t} dy_t$  & 16  \\
			\hline
			577 & $\sigma^{-1}\, d^2\sigma/dm_{t\bar t} dy_{t\bar t}$  & 16 \\
			\hline
		\end{tabular}
		\label{table:table1}
	\end{center}
\end{table}

\section{\label{sec:NNLO} NNLO calculation of differential top-pair distributions}

In this work we calculate the NNLO QCD corrections to one- and two-dimensional  top quark-pair differential distributions at the LHC. The distributions are defined in terms of the following top quark kinematic variables: $p_T^t,\, m_{tt},\, p_T^{t\bar t},\, y_{t},\, y_{t\bar t}$ and $\Delta\eta_{t\bar t}$. The $p_T^t$ and $y_{t}$ distributions are averaged over the corresponding top and antitop distributions. Our binning follows the CMS collaboration's 8 TeV measurement \cite{Sirunyan:2017azo}. 

We use $m_t=173.3$ GeV and utilize the dynamic scales derived in Ref.~\cite{Czakon:2016dgf} (see also Ref.~\cite{Czakon:2018nun}):
\begin{eqnarray}
\mu_0 &=& {m_T\over 2} \equiv {1\over 2}\sqrt{p_{T,t}^2+m_t^2}\label{eq:scale-mT}\,,\\
\mu_0 &=& {H_T\over 4} \equiv {1\over 4}\left(\sqrt{p_{T,t}^2+m_t^2} + \sqrt{p_{T,{\bar t}}^2+m_t^2}\right)\label{eq:scale-HT}\,.
\end{eqnarray}
Specifically, we compute the one-dimensional average top $p_T^t$ distribution with the scale Eq.~(\ref{eq:scale-mT}) while all other one-dimensional distributions and all two-dimensional ones are computed with the help of the scale Eq.~(\ref{eq:scale-HT}). Scale variation is derived with the help of the usual 7-point variation of the factorization sand renormalization scales around the central scale $\mu_0$.

The calculations performed in this work are used to produce tables in the {\tt fastNLO} format \cite{Kluge:2006xs,Britzger:2012bs}. We note that this is the first time two-dimensional $t\bar t$ distributions are implemented in this format. More details about our {\tt fastNLO} tables can be found in Ref.~\cite{Czakon:2017dip}. These tables have the advantage that predictions can be recalculated very fast with any PDF set or for any value of $\alpha_S$. As a cross-check, we have also provided in electronic format two binned predictions based on the {\tt NNPDF30\_nnlo\_as\_0118} \cite{Ball:2014uwa} and {\tt CT14nnlo\_as\_0111} \cite{Dulat:2015mca} PDF sets. All predictions can be downloaded from the following webpage (http://www.precision.hep.phy.cam.ac.uk/results/ttbar-fastnlo/). 

In this work we follow the {\tt STRIPPER} approach \cite{Czakon:2010td,Czakon:2011ve,Czakon:2014oma} previously applied to top-pair production in Refs.~\cite{Czakon:2014xsa,Czakon:2015owf,Czakon:2016ckf,Czakon:2016dgf,Behring:2019iiv}. We have implemented it in a flexible, fully-differential partonic Monte Carlo program which, in principle, is able to calculate any infrared safe partonic observable. Further technical details can be found in Ref.~\cite{Czakon:2019tmo}.
Two-dimensional distributions in NNLO QCD have recently also been calculated in Ref.~\cite{Catani:2019hip} for a different LHC setup.

\section{\label{sec:correlation}Double-differential top data and the correlation to the gluon distribution}

\begin{figure}[h] 
	\includegraphics[width=0.45\textwidth]{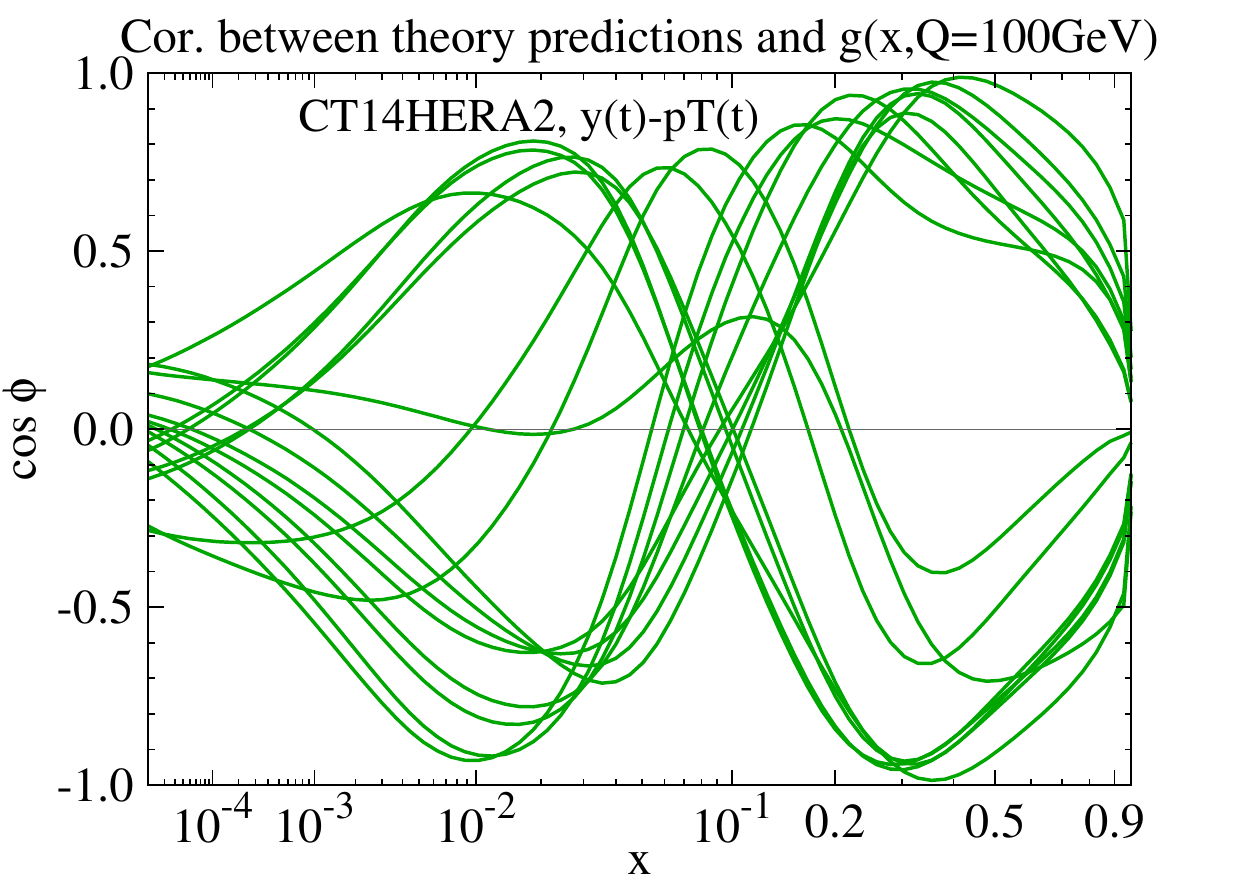}
	\includegraphics[width=0.45\textwidth]{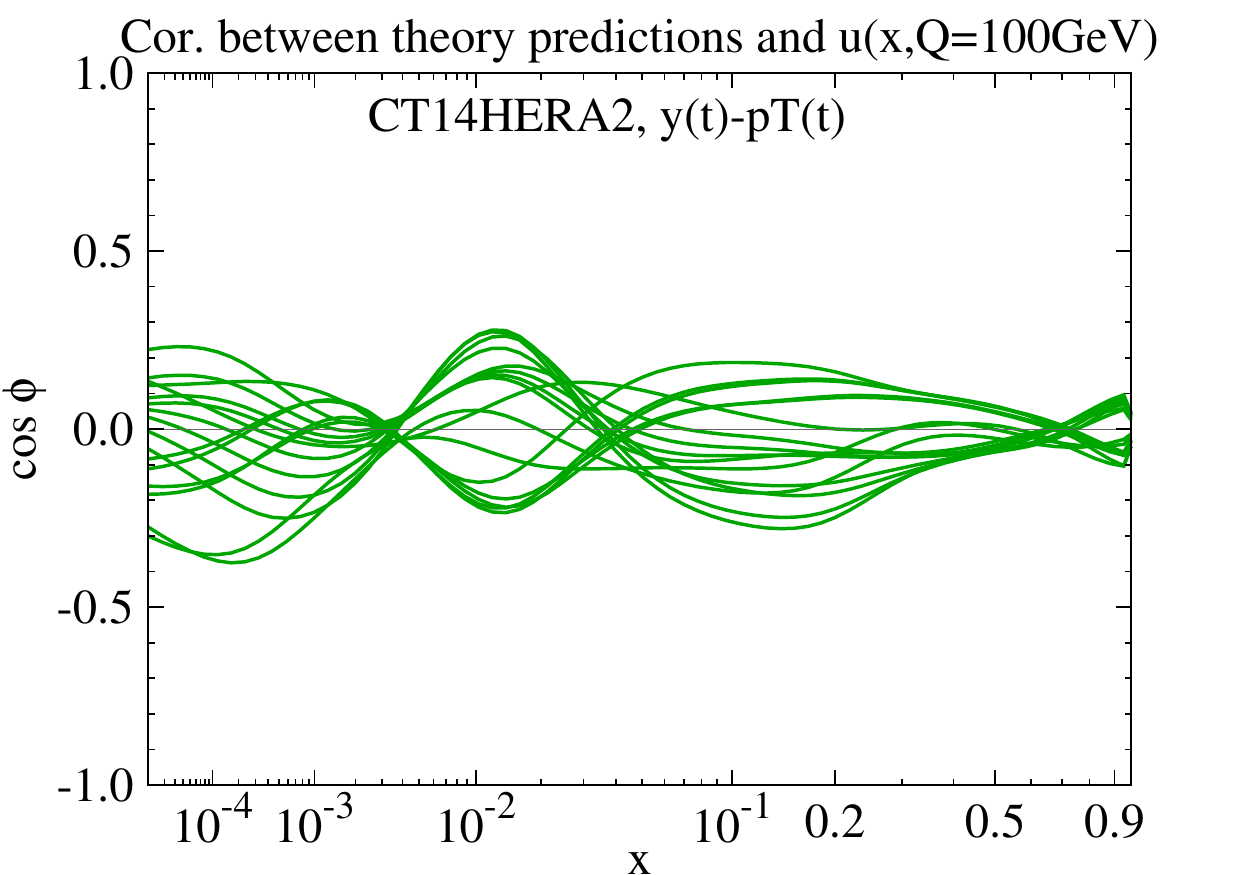}
	\caption{The correlation cosines between  the $y_t$ and $p_T^t$ double differential distributions (573) and the CT14HERA2 gluon (left) and up quark (right) PDF distributions as a function of $x$. Each curve corresponds to a separate data point. }
	\label{Fig:correlation573}
\end{figure}

\begin{figure}[h] 
	\includegraphics[width=0.45\textwidth]{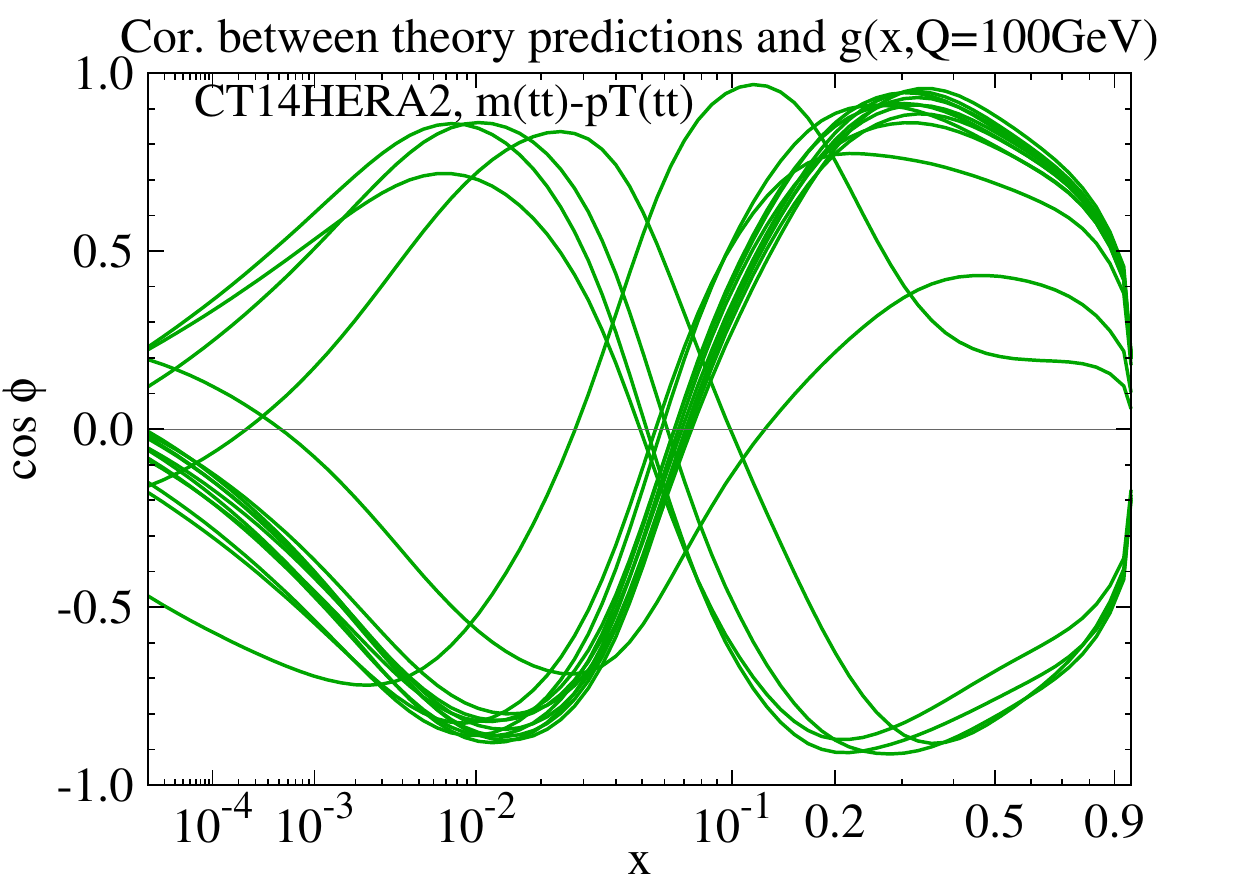}
	\includegraphics[width=0.45\textwidth]{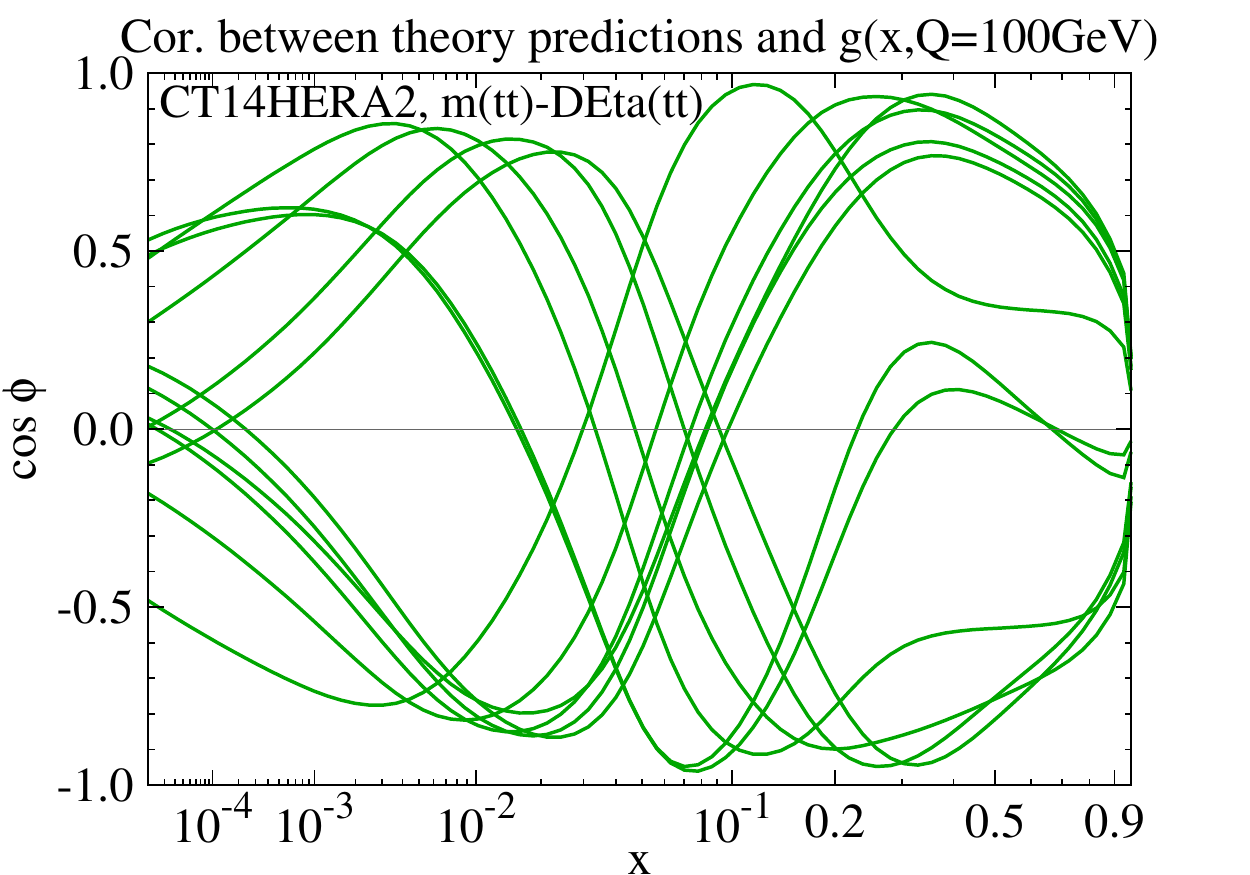}
	\includegraphics[width=0.45\textwidth]{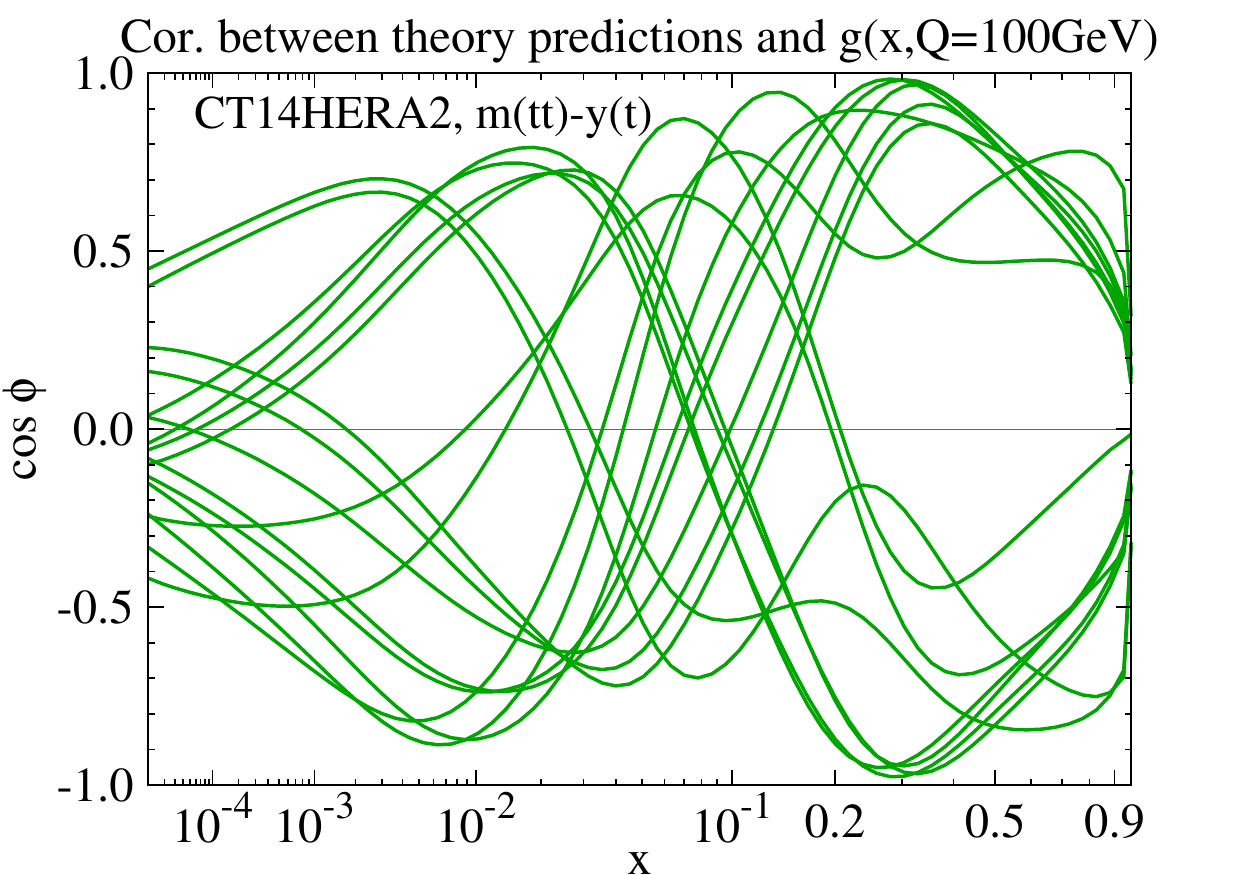}
	\includegraphics[width=0.45\textwidth]{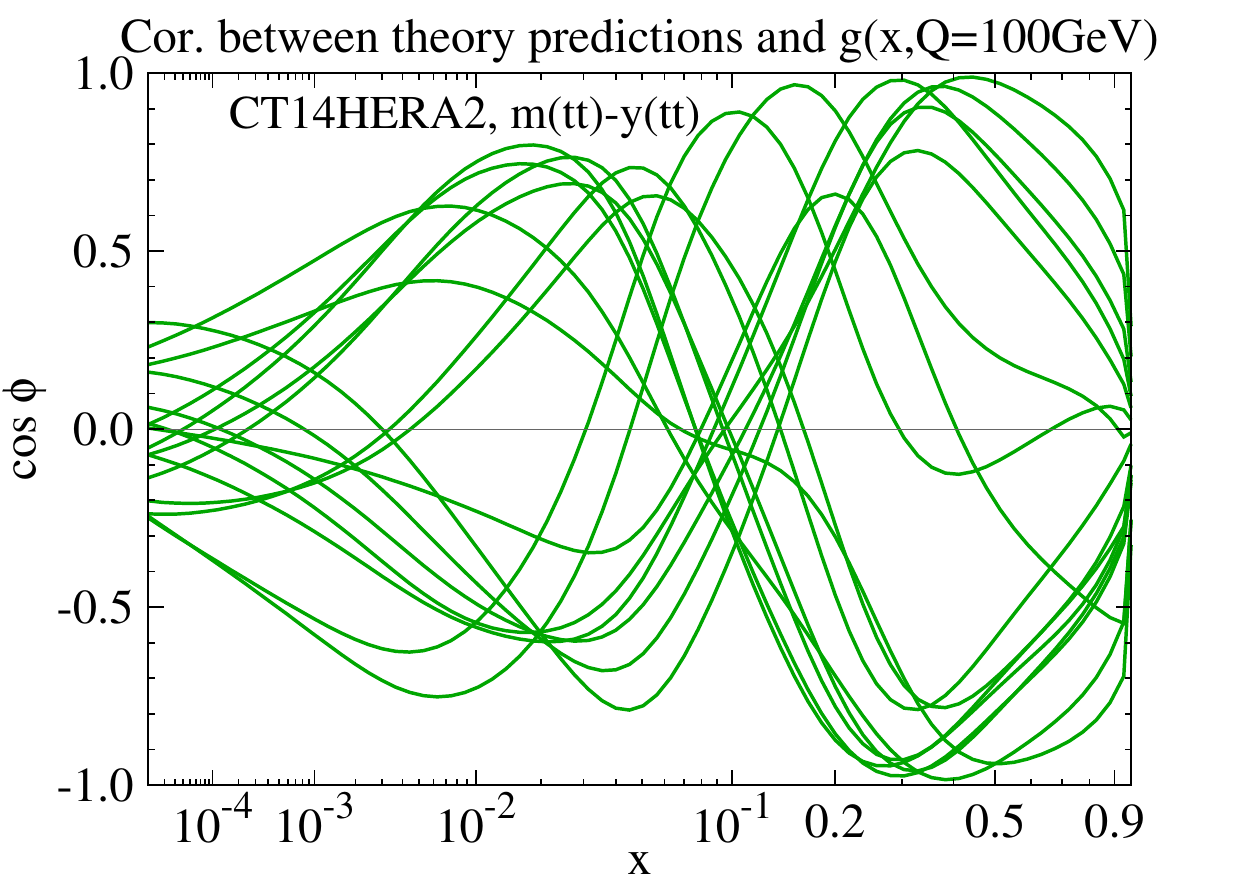}
	\caption{The correlations between the $t \bar t$ double differential $m_{t\bar t}$-$p_T^{t\bar t}$, $m_{t\bar t}$-$\Delta\eta_{t\bar t}$, $m_{t\bar t}$-$y_t$ and $m_{t\bar t}$-$y_{t\bar t}$  distributions (IDs 574 to 577) 
		 and the CT14HERA2 gluon PDFs, as a function of $x$. Each curve corresponds to a separate data point.}
	\label{Fig:correlation574-577g}
\end{figure}

The double-differential $t \bar t$ data 
are expected to have the strongest correlation with the gluon PDF,  as the dominant  $t \bar t$ production mechanism at the LHC is through $gg$ fusion. This argument can be demonstrated quantitatively by examining the correlation cosines for the $t \bar t$ data as a function of the gluon momentum fraction $x$. The quantity $\cos\varphi$, the correlation cosine,  characterizes whether the PDF degrees
of freedom of quantities $X$ and $Y$ are correlated ($\cos\varphi\approx1$),
anti-correlated ($\cos\varphi\approx-1$), or uncorrelated ($\cos\varphi\approx0$)\cite{Campbell:2017hsr}. In this case, $X$ and $Y$ are the gluon distribution and various double-differential $t \bar t$ distributions, respectively. 

As an example, the correlation cosines between $t \bar t$ data and the gluon and up quark distributions are  shown in Fig.~\ref{Fig:correlation573} for the CMS data set labeled 573, ($\sigma^{-1}\, d ^2\sigma/dy_t dp_T^t$)~\footnote{The correlations with the other PDF flavors are found to be weak, similar to that observed for the up quark.}. There are 16 data points in this data set, so the plot contains 16 curves.  It is apparent that gluon PDF has a stronger correlation with the $t \bar t$ data, and in particular that it is mainly the gluon PDF for $x>0.01$ range that has strong correlations. Note that approximately half of the data points have a strong correlation with the gluon distribution at an $x$ value above 0.1 and a strong anti-correlation with the gluon near $x$ of 0.01, while (approximately) the other half of the data points have the opposite behavior. There is not much correlation in the $x$ region around 0.1, and as will be seen later, the constraints on the gluon distribution by the top data sets tend to be weaker here. 

The high $x$ gluon in particular still has a great deal of uncertainty in global PDF fits. The range around 0.01 is also of interest as it plays a role in Higgs boson production through $gg$ fusion.   The correlations between the other CMS $t \bar t$ observables and the gluon PDF are shown in Fig.~\ref{Fig:correlation574-577g}, where the same conclusion holds~\footnote{It is interesting that data sets 576 and 577 have strong correlations more uniformly spread in $x$ than do the other distributions.}. Although this demonstrates that the CMS double differential (2D) $t \bar t$ observables depend highly on gluon PDFs in these $x$ ranges, it does not necessarily mean that the $t \bar t$ data are going to have a strong impact on the determination of the gluon PDF in a global fit.

The impact of a data set on a global PDF fit has been discussed in Ref.~\cite{Wang:2018heo}, as involving not only a correlation between the data and specific PDFs, in a given $x$ range, but also a sensitivity of the data to those PDFs. The sensitivity is determined by the number of data points,  the kinematic range they cover, and the  magnitudes of the statistical and systematic errors (and the correlations of the latter).
It can be shown that~\cite{Wang:2018heo,Hou:2019gfw}
one of the strongest sensitivities (per data point) for the gluon distribution is given by the CMS double-differential top data (not included in CT14 or CT14HERA2, but will be in CT18~\cite{Hou:2019efy}). 
Although, the CMS double-differential top data
has one of the highest sensitivities per data point, the data sets that have the largest sensitivities are the HERA I+II data set and the CMS 7 TeV inclusive jet data set. The former has a relatively low average sensitivity per data point but has 1120 data points.  The latter has a moderate sensitivity to the gluon distribution per data point, but has 185 data points, most with reasonably small statistical and systematic errors. In the next section, we will examine the actual impact of the top and jet data sets on a global PDF fit using the program ePump. As shown in Ref.~\cite{Schmidt:2018hvu,Hou:2019gfw}, ePump can quickly provide quantitative information on the impact of a given data set to updating PDFs and their error bands, including the information on the relevant parton flavor and $x$ range. We will show below that although $t\bar{t}$ data has a high sensitivity to gluon PDF per data point, the sensitivity of the whole data set is quite small due to the small number of data points.

\section{\label{sec:PDF}Results of PDF fitting} 

As introduced in Ref.~\cite{Schmidt:2018hvu}, ePump is a convenient software tool that allows an  examination of the impact of a new data set, without the need to perform a complete global PDF fit.
The $\chi^2$ and dof for each of the CMS double-differential top data sets, compared to NNLO predictions using CT14HERA2, are shown in Table \ref{table:ttb2Dchi2}, first without including the data set in the fit, and then including the data set via ePump. The data provided by the CMS experiment~\cite{Sirunyan:2017azo} are normalized distributions, with correlated systematic uncertainties and correlation matrices of statistical uncertainties. Due to the loss of one degree of freedom when constructing normalized distributions, the correlation matrices are singular, so when we use the data to update PDFs (or to calculate the $\chi^2$), we discard the bin with the largest values of kinematic variables and the corresponding correlation coefficients for each data set, as instructed by the experimental paper~\cite{Sirunyan:2017azo}.

\begin{table}[htbp]
	\begin{center}
		\caption{The $\chi^2$ for each 2D $t\bar t$ data set, calculated with the original global-fit CT14HERA2 PDFs and ePump  updated CT14HERA2 PDFs.}
		\begin{tabular}{| c | c | c | c | c |}
			\hline
			ID & data &  dof  & $\chi^2$ before updating & $\chi^2$ after updating \\
			\hline
			573 & $\sigma^{-1}\, d^2\sigma/dy_t dp^t_T$  & 15 & 35.5 & 34.9 \\
			\hline
			574 & $\sigma^{-1}\, d^2\sigma/dm_{t\bar t} dp_T^{t\bar t}$  & 15 & 82.3 & 80.6 \\
			\hline
			575 & $\sigma^{-1}\, d^2\sigma/dm_{t\bar t} d\Delta \eta_{t\bar t}$  & 11 & 22.1 & 22.0 \\
			\hline
			576 & $\sigma^{-1}\, d^2\sigma/dm_{t\bar t} dy_t$  & 15 & 20.2 & 20.1 \\
			\hline
			577 & $\sigma^{-1}\, d^2\sigma/dm_{t\bar t} dy_{t\bar t}$  & 15 & 23.8 & 23.5 \\
			\hline
		\end{tabular}
		\label{table:ttb2Dchi2}
	\end{center}
\end{table}

Several aspects can be immediately noticed. First, the $\chi^2$/dof for each of the data sets is on the order of 1.3-2, except for the  ($m_{t\bar t}, p_T^{t\bar t}$) data set (ID 574) which has a $\chi^2$/dof of over 5. Second, there is minimal improvement in the $\chi^2$/dof when each data set is included in the global fit, which indicates that the global fit PDFs are not changed greatly by including the $t\bar{t}$ data.

\begin{figure}[h] 
\includegraphics[width=0.45\textwidth]{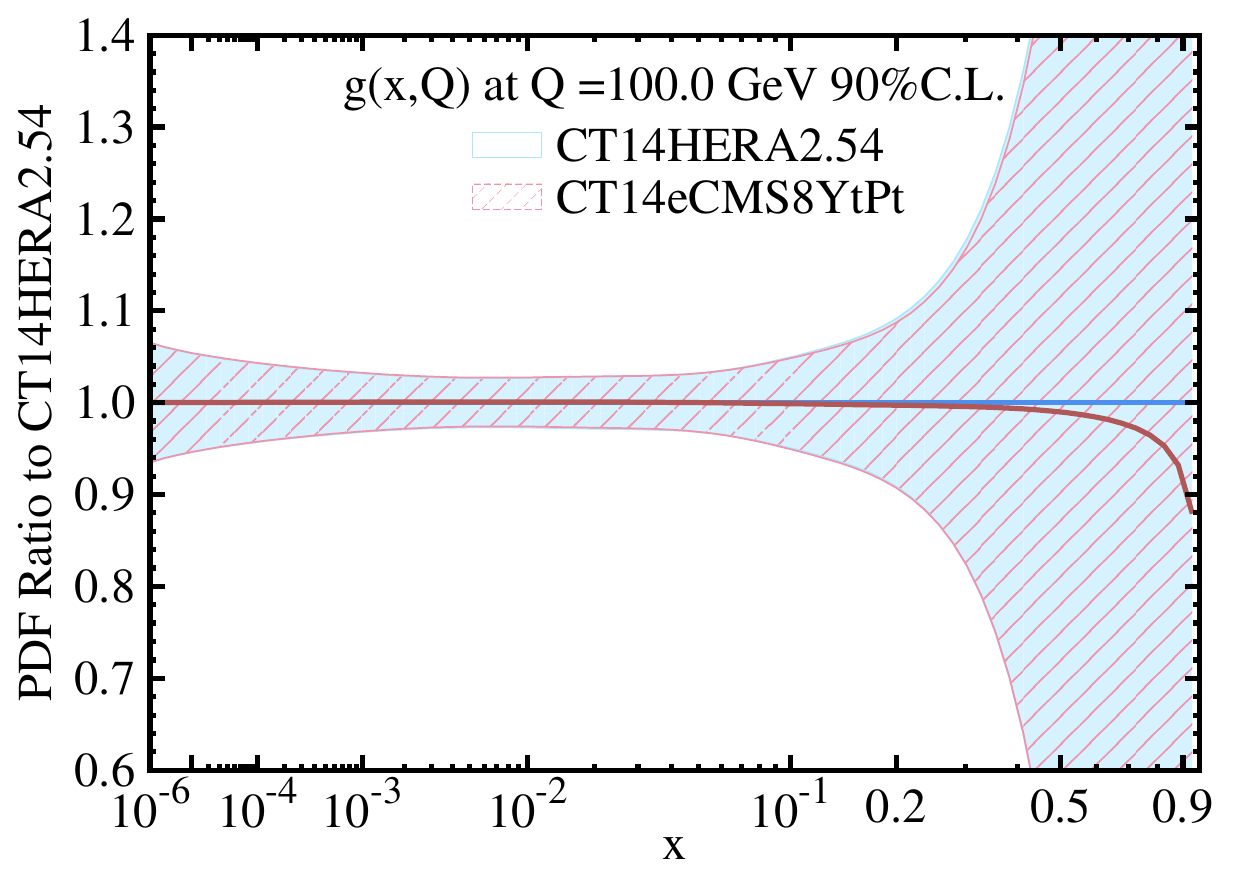}
\includegraphics[width=0.45\textwidth]{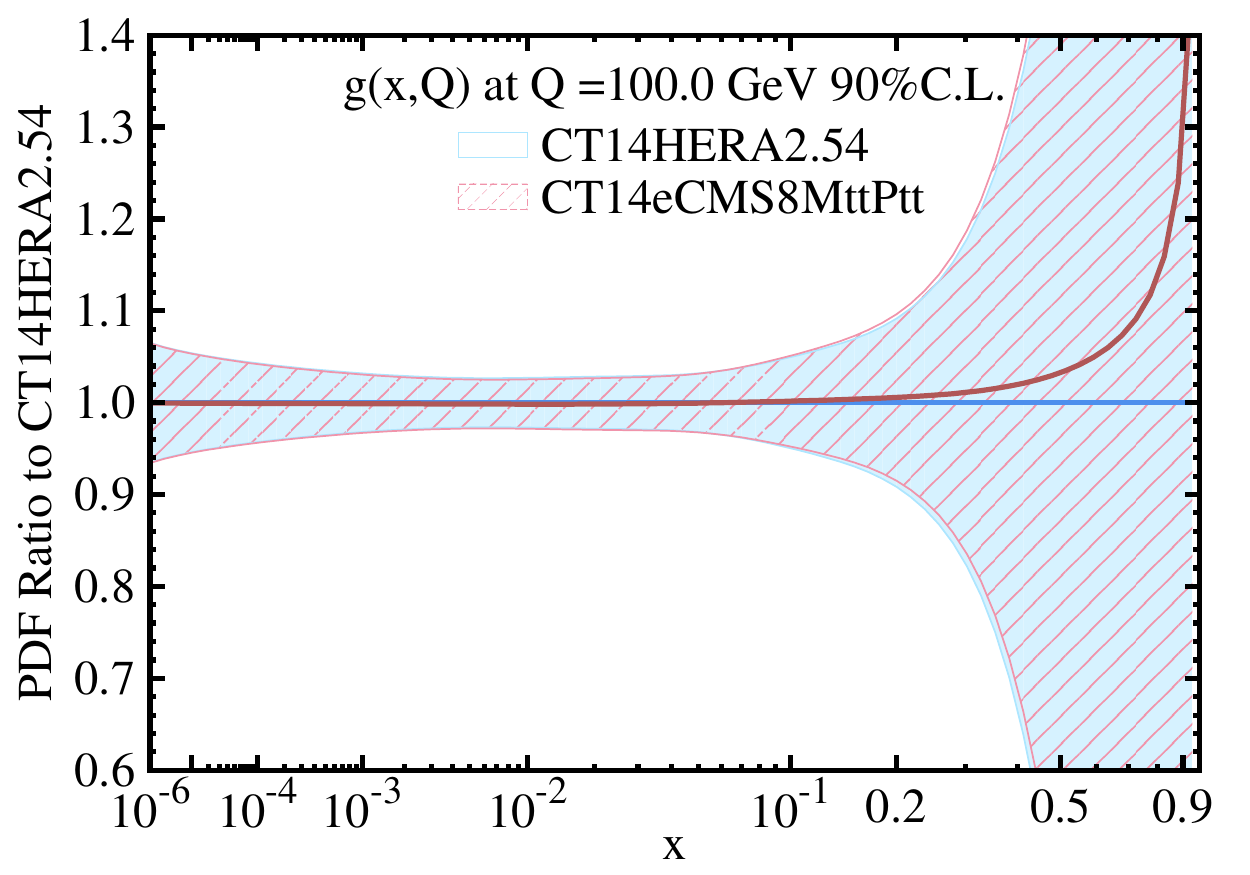}
\includegraphics[width=0.45\textwidth]{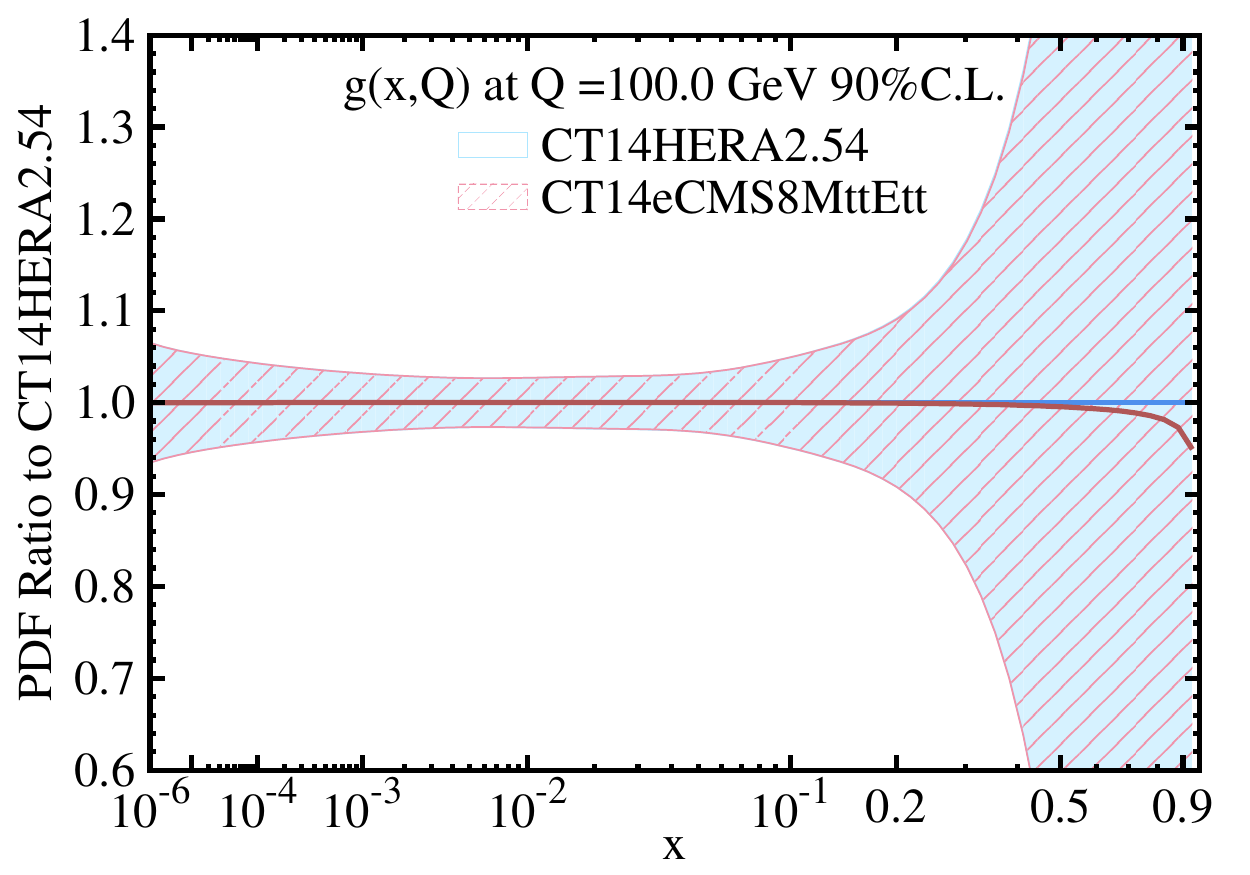}
\includegraphics[width=0.45\textwidth]{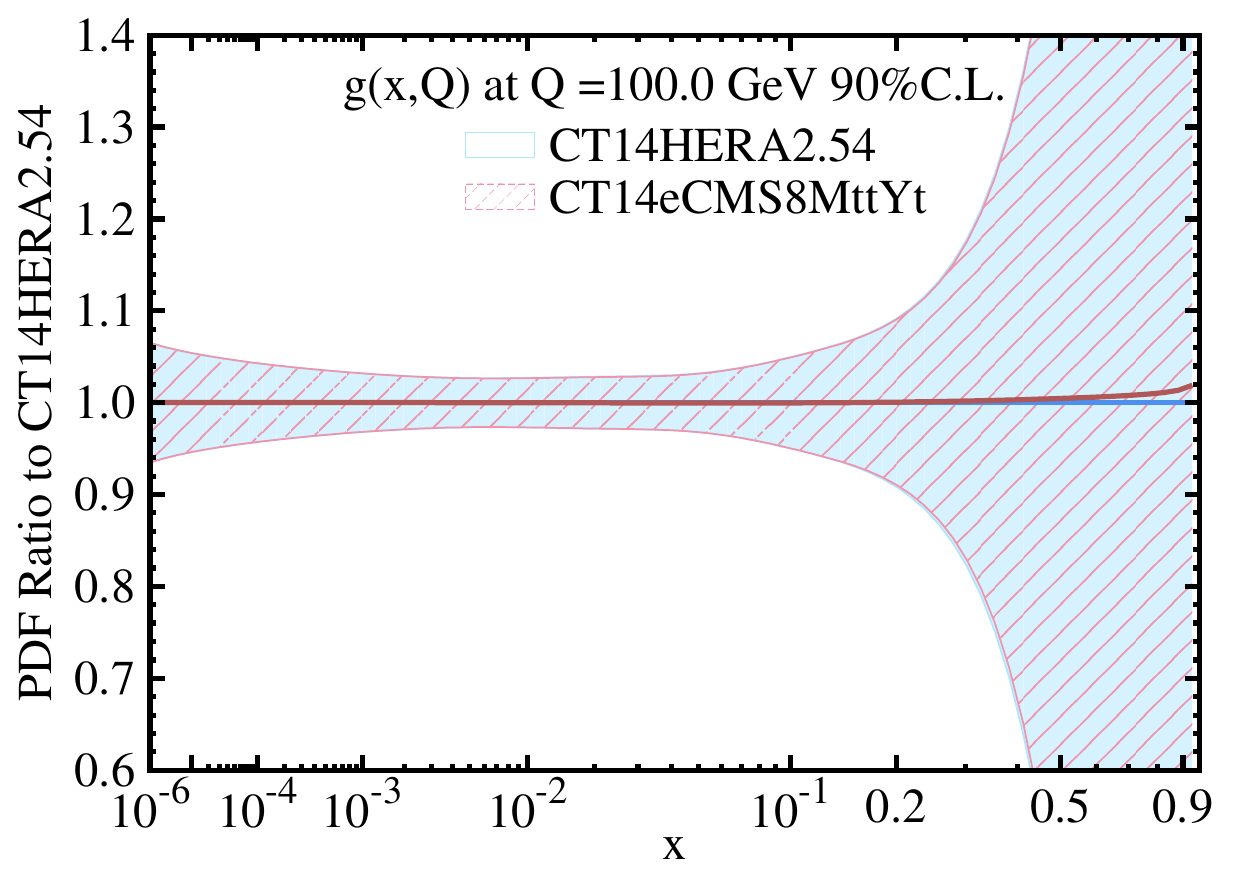}
\includegraphics[width=0.45\textwidth]{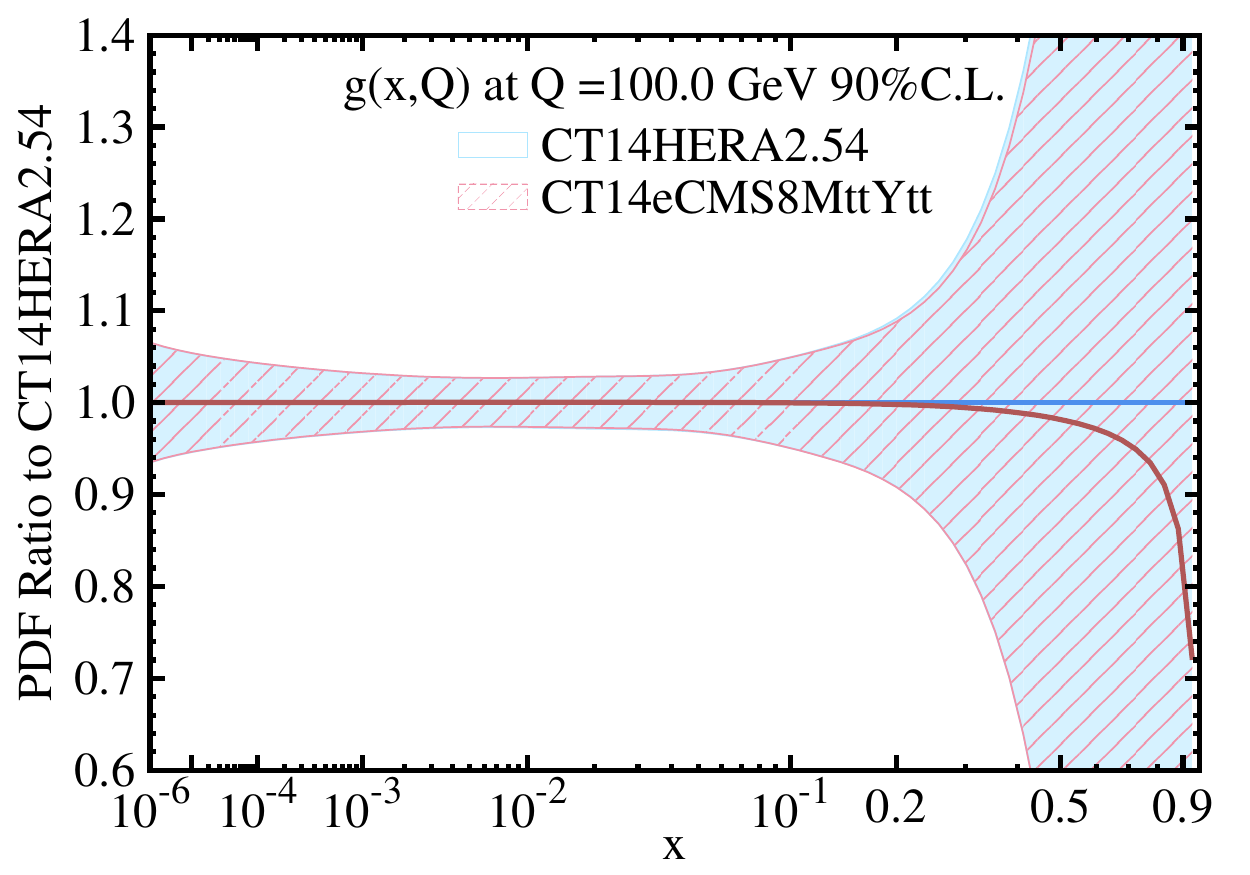}
\caption{ 
Ratios of the central value and uncertainty of the $g$ PDF to the CT14HERA2 central value, before and after updating the CT14HERA2 PDFs 
by adding each of the CMS 8 TeV  double-differential $t\bar t$ data set one at a time. The suffix ``.54" is to stress that there are 54 eigen-sets used in CT14HERA2 global fit, without the two gluon extreme sets. The letter ``e" is to note that the PDF was obtained by ePump updating. For example, CT14eCMS8YtPt denotes the updated PDFs with the inclusion of the ($y_{t}, p^t_T$) data set (ID 573). 
}
	\label{Fig:ct14p577g}
\end{figure}

\begin{figure}[h] 
	\includegraphics[width=0.45\textwidth]{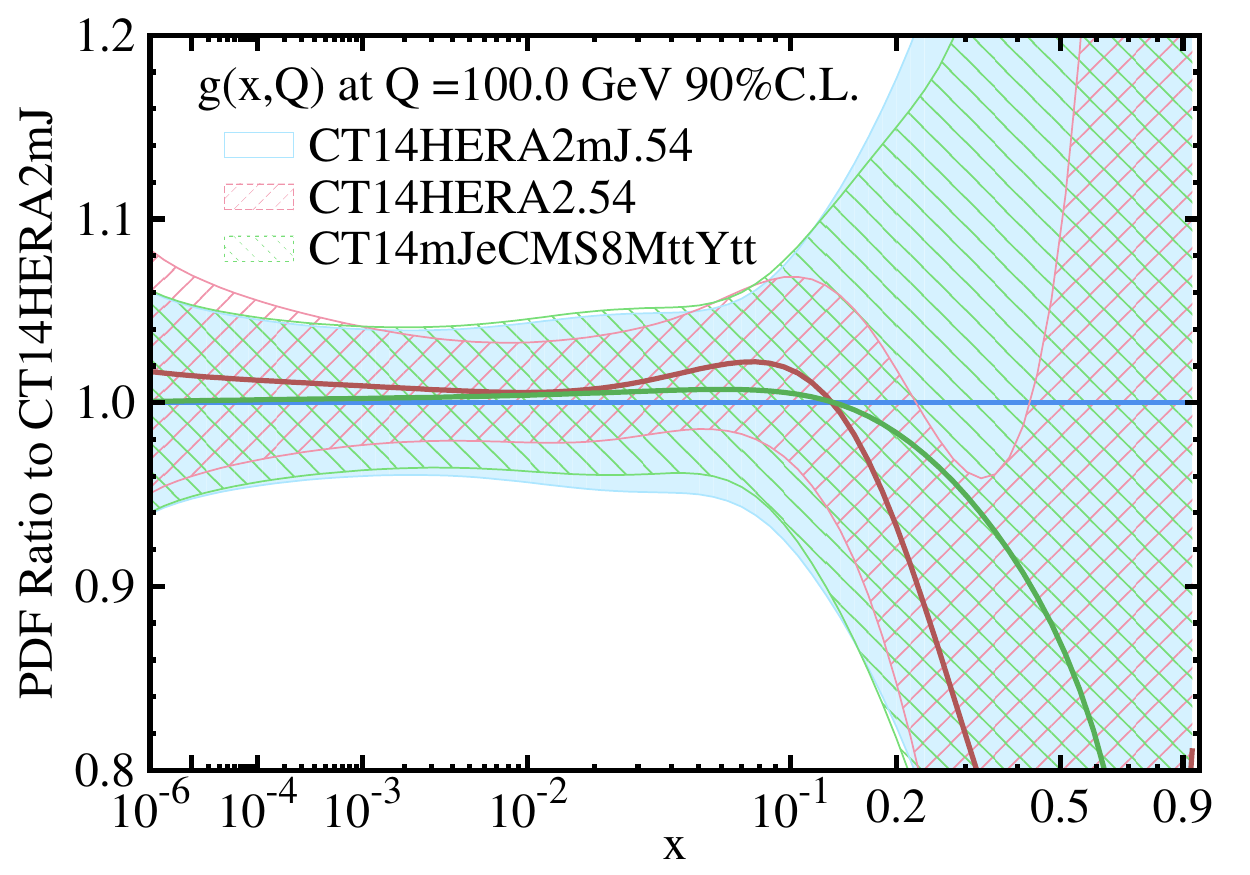}
	\includegraphics[width=0.45\textwidth]{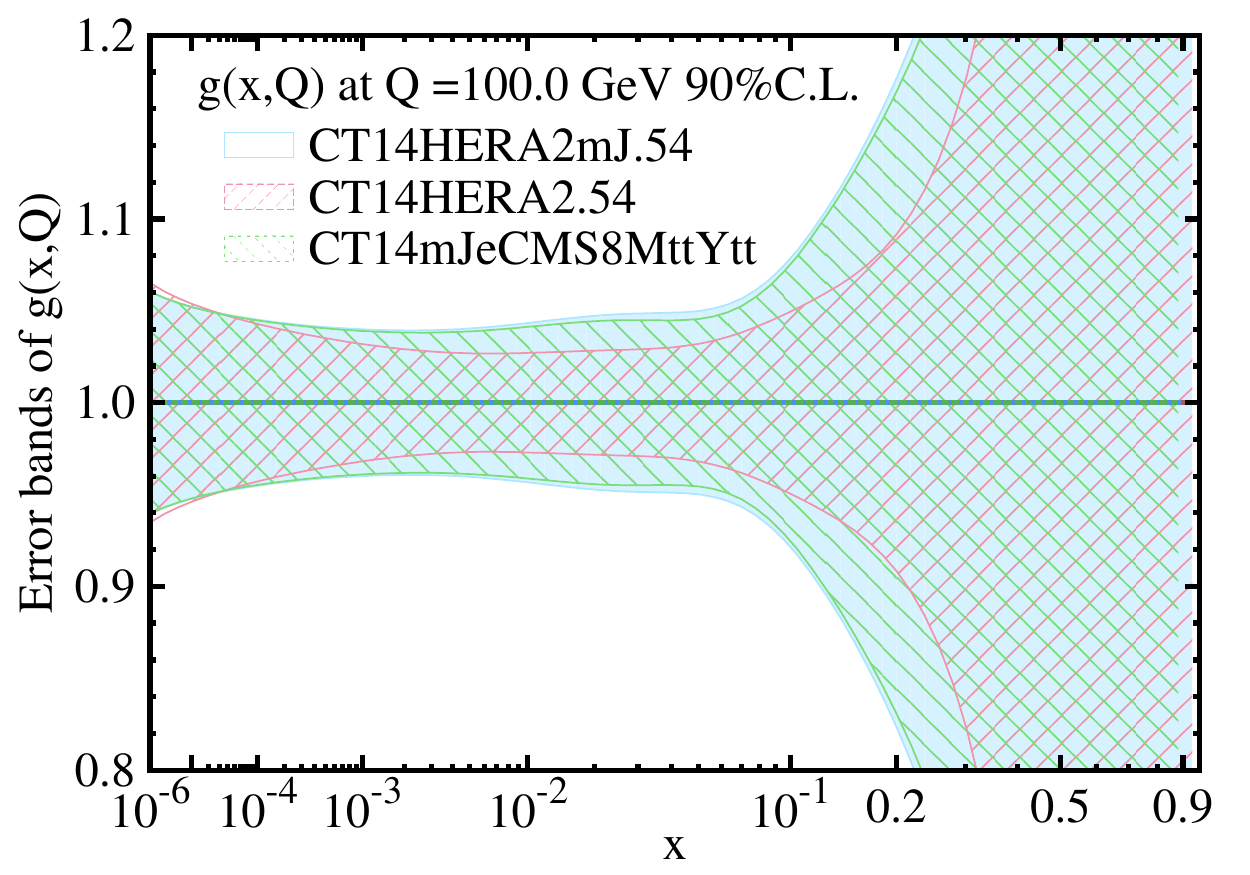}
	\caption{The updated gluon PDF and its error band when the CMS 8 TeV ($m_{t\bar t}$,$y_{t\bar t}$) double differential data (ID 577) is added to CT14HERA2mJ using ePump with weight=1. The suffix ``.54" is to stress that there are 54 eigen-sets used in CT14HERA2 global fit, without the two gluon extreme sets. The letter ``e" is to note that the PDF was obtained by ePump. Left: PDF central values. Right: Error bands.}
	\label{Fig:mj577w1g}
\end{figure}

\begin{figure}[h] 
	\includegraphics[width=1.0\textwidth]{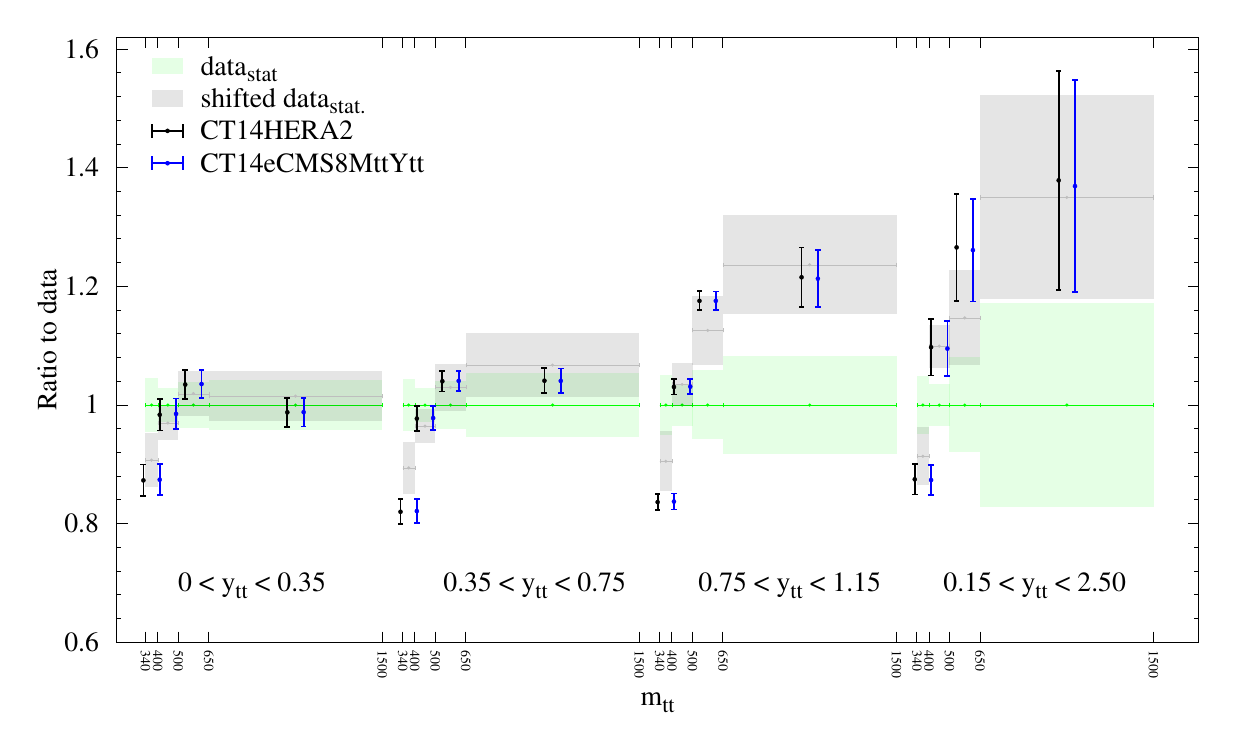}
	\caption{
	The comparison between the NNLO theory prediction of CT14HERA2 and experiment data for each data point for the CMS 8 TeV double differential ($m_{t\bar t}$,$y_{t\bar t}$)  data set. 
The shifted data shown are with respect to the CT14HERA2 global fit. The shifted data for the updated PDFs, labeled by CT14eCMS8MttYtt,  do not differ much and thus are not shown in the figure. 
}
	\label{Fig:CT14comparison_5}
\end{figure}

\begin{figure}[h] 
	\includegraphics[width=1.0\textwidth]{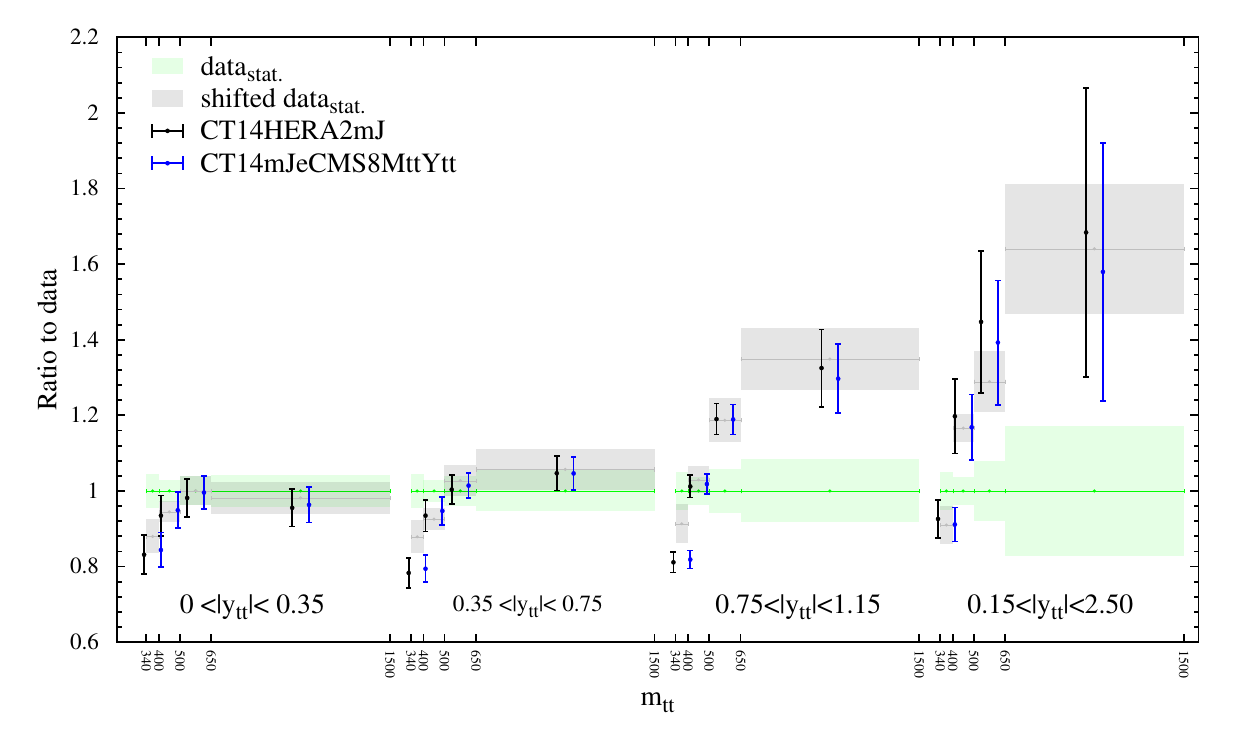}
	\caption{
	The comparison between the NNLO theory prediction of CT14HERA2mJ and experiment data for each data point for the CMS 8 TeV double differential ($m_{t\bar t}$,$y_{t\bar t}$)  data set. 
The shifted data shown are with respect to the CT14HERA2 global fit. The shifted data for the updated PDFs, labeled by CT14mJeCMS8MttYtt,  do not differ much and thus are not shown in the figure. 
}
	\label{Fig:mjcomparison_5}
\end{figure}

\begin{figure}[htbp] 
\includegraphics[width=0.45\textwidth]{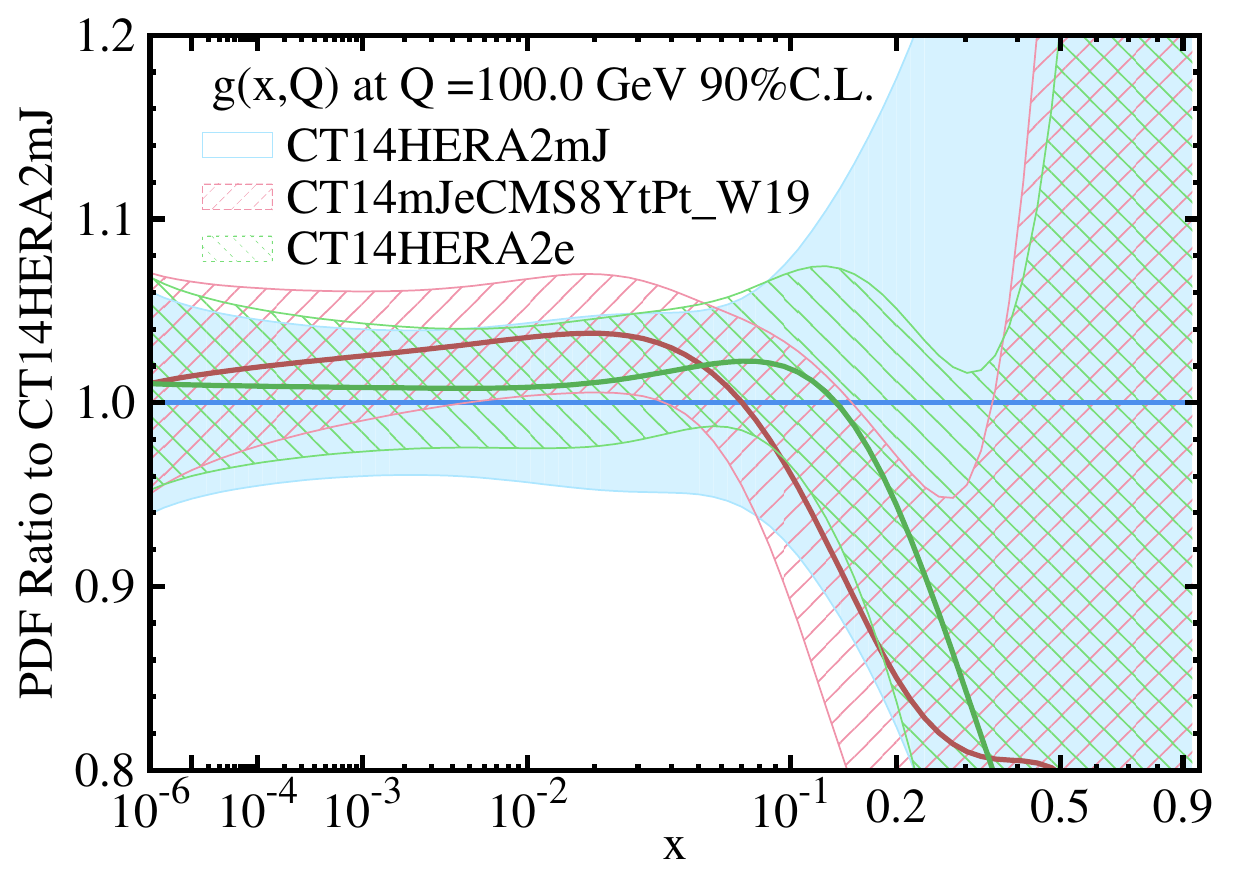}
\includegraphics[width=0.45\textwidth]{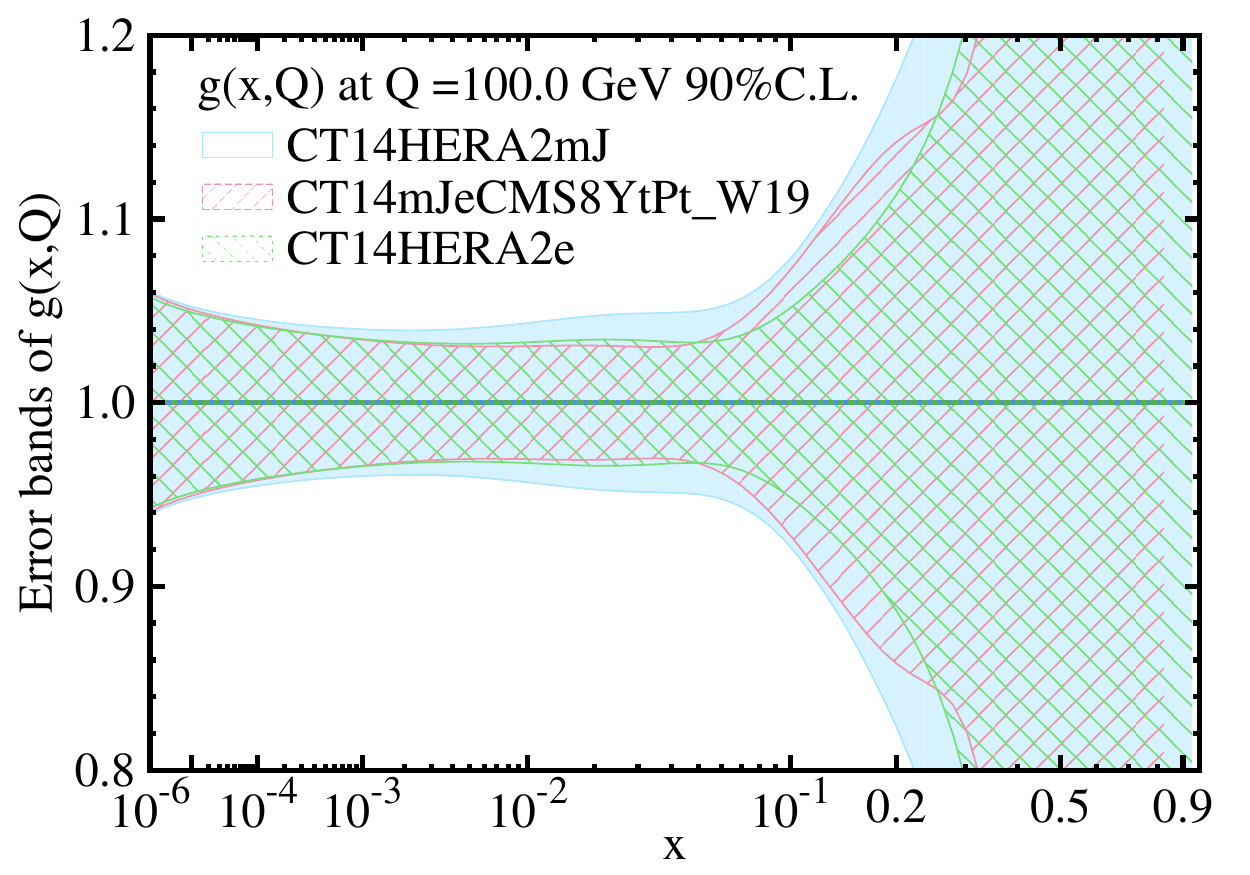}
\caption{The updated gluon PDF and its error band when adding the ($y_t$,$p_T^t$) data set (ID 573), using ePump with weight=19, compared to CT14HERA2mJ and CT14HERA2e. Hereafter, CT14HERA2e is obtained by adding jet data back to CT14HERA2mJ using ePump, which is very similar to the CT14HERA2 PDF set \cite{Hou:2019gfw}. Left: PDF central values. Right: Error bands.}
\label{Fig:mj573w19g}
\end{figure}

\begin{figure}[htbp] 
\includegraphics[width=0.45\textwidth]{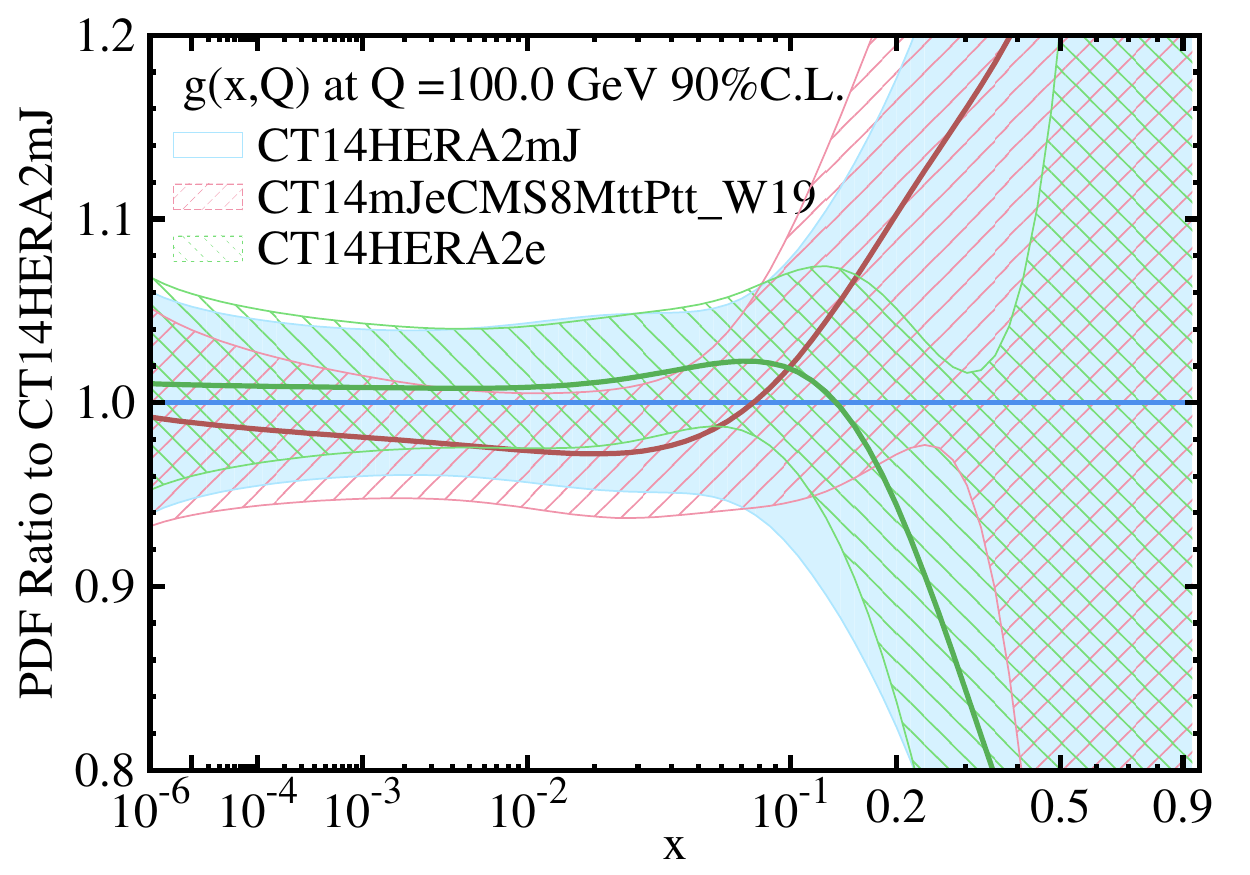}
\includegraphics[width=0.45\textwidth]{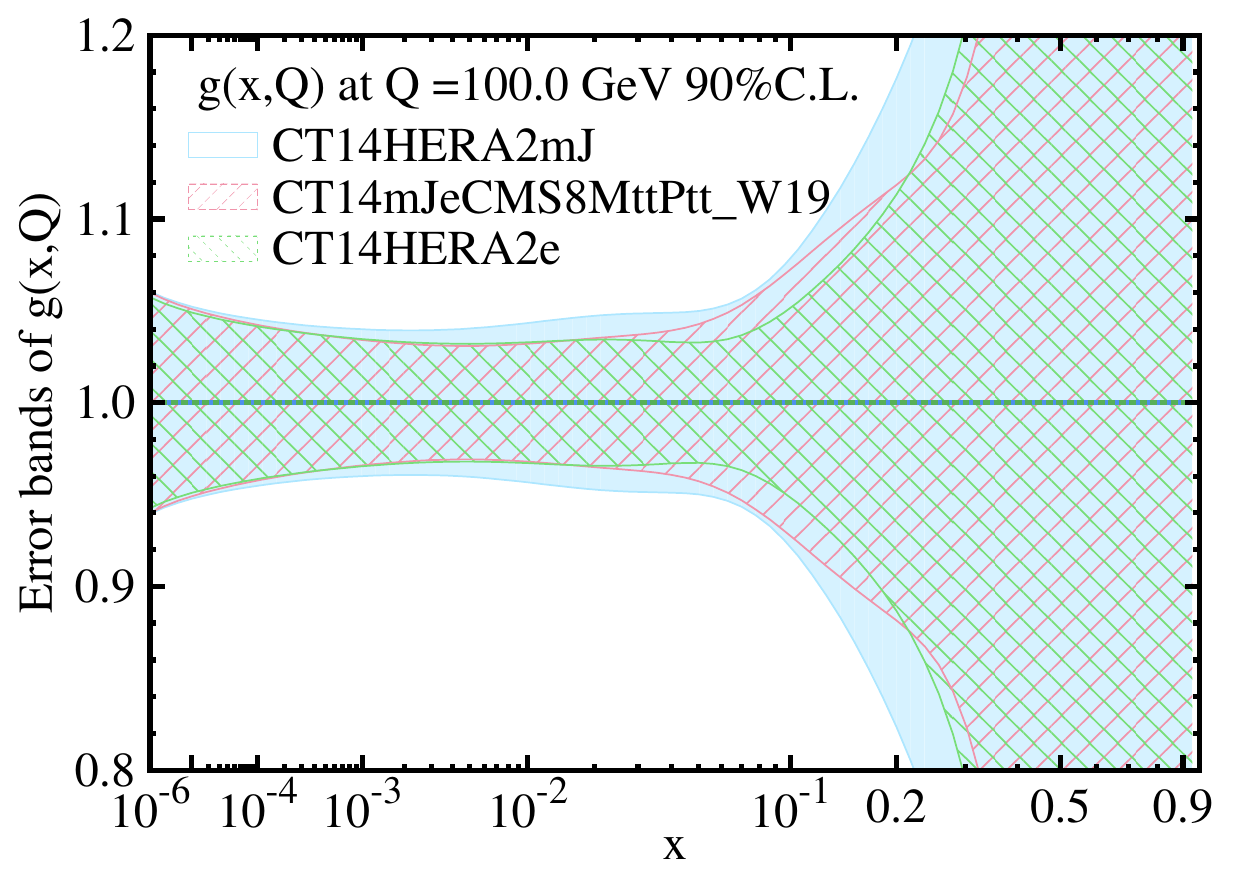}
\caption{The updated gluon PDF and its error band when adding the ($m_{t\bar t}$ and $p_T^{t\bar t}$) data set (ID 574), using ePump with weight 19, compared to CT14HERA2mJ and CT14HERA2e. Left: PDF central values. Right: Error bands.}
\label{Fig:mj574w19g}
\end{figure}

\begin{figure}[htbp] 
\includegraphics[width=0.45\textwidth]{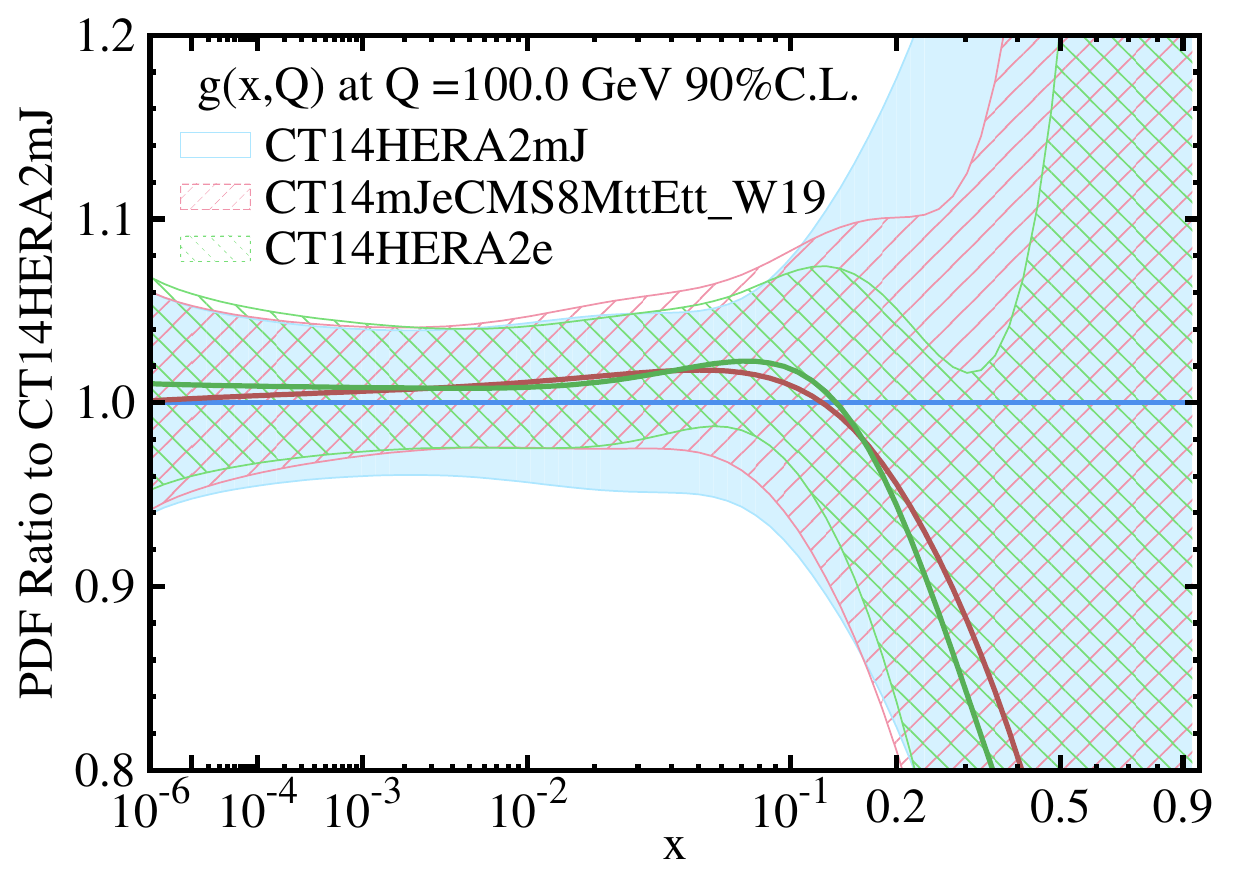}
\includegraphics[width=0.45\textwidth]{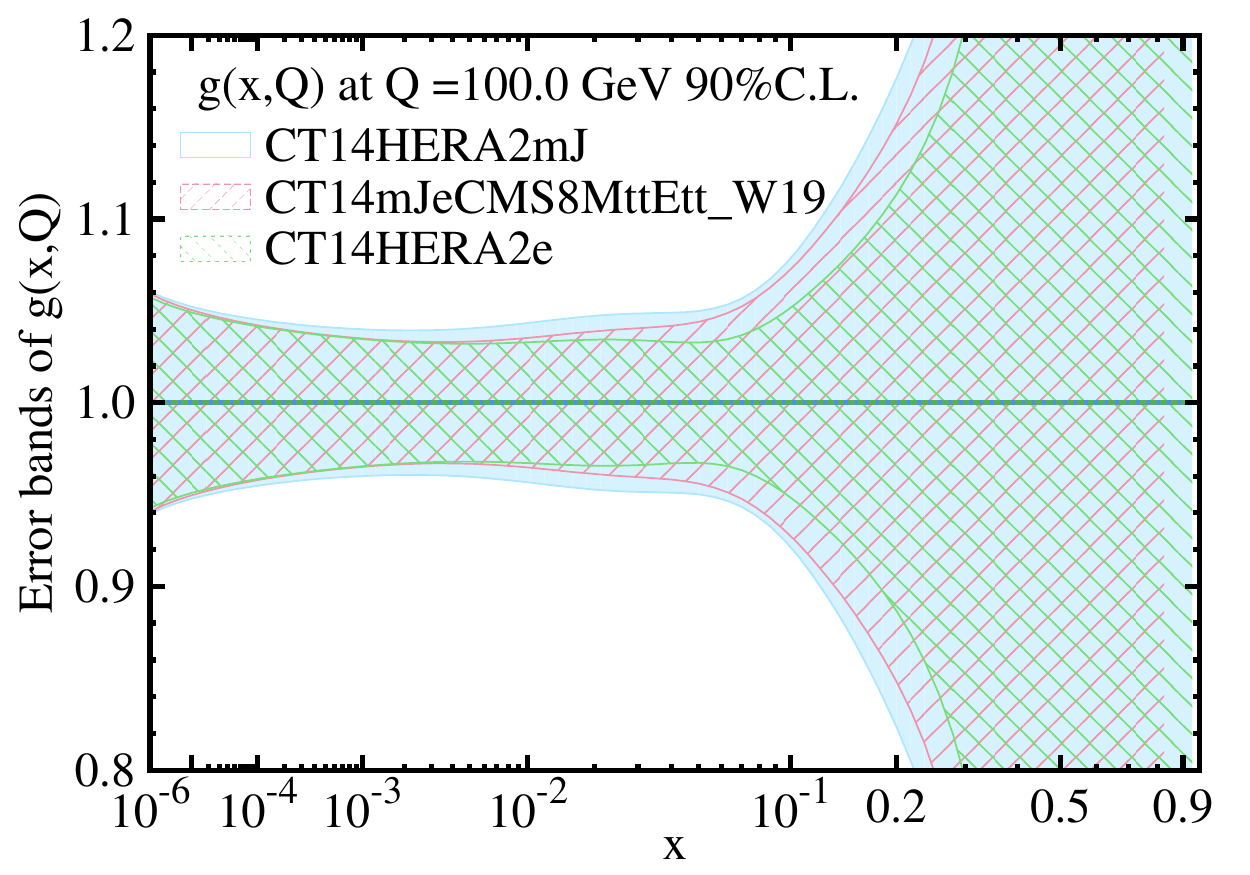}
\caption{The updated gluon PDF and its error band when adding the ($m_{t\bar t}$, $\Delta\eta_{t\bar t}$) data set (ID 575), using ePump with weight 19, to CT14HERA2mJ and CT14HERA2e. Left: PDF central values. Right: Error bands.}
\label{Fig:mj575w25g}
\end{figure}

\begin{figure}[htbp] 
\includegraphics[width=0.45\textwidth]{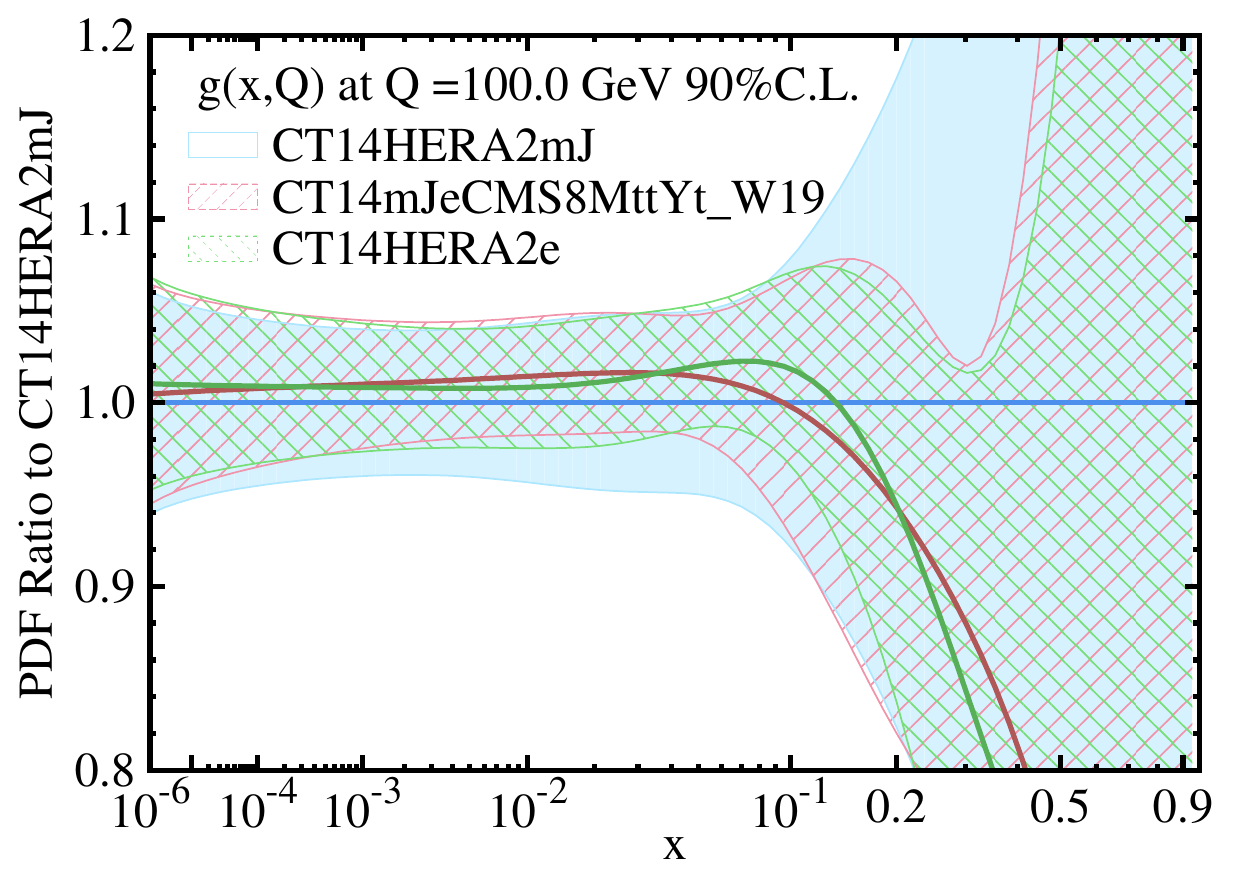}
\includegraphics[width=0.45\textwidth]{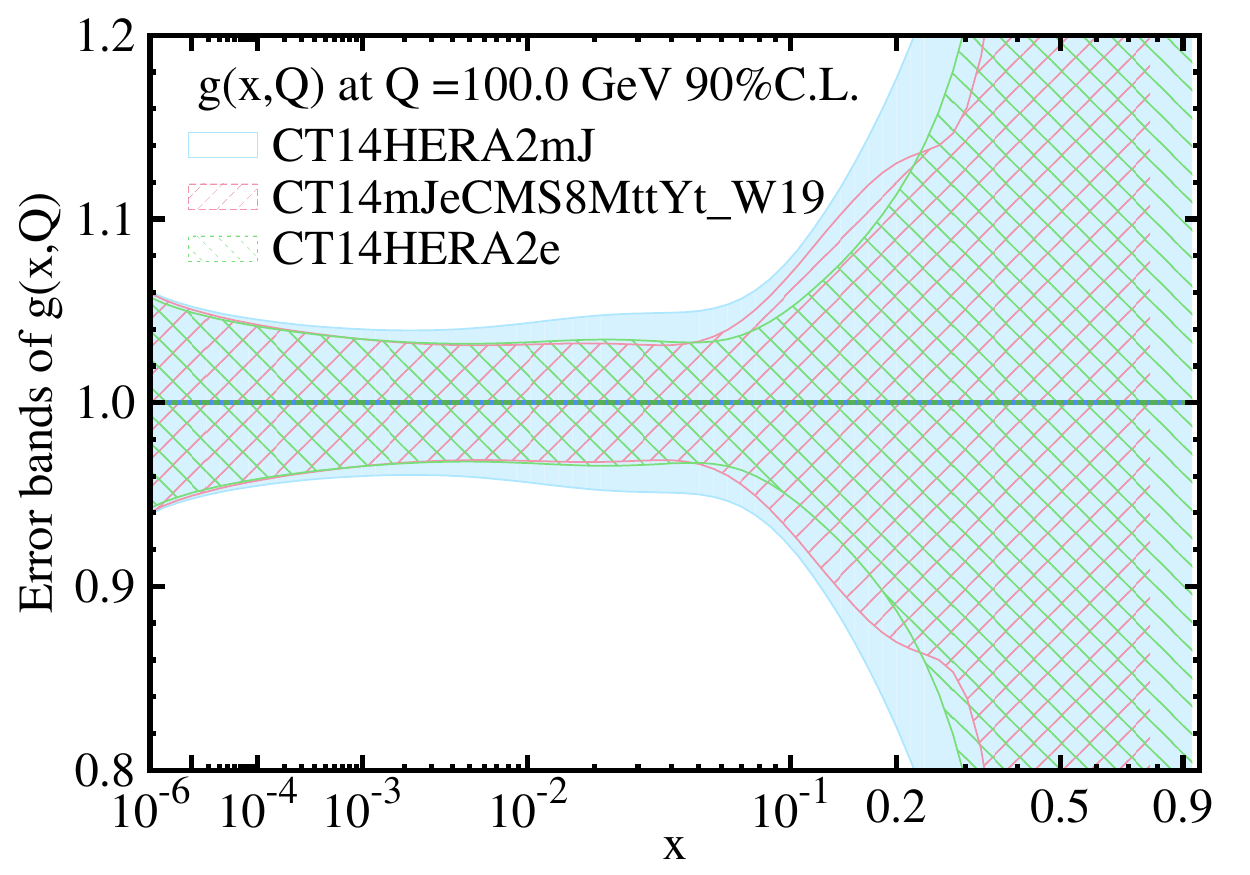}
\caption{The updated gluon PDF and its error band when adding the ($m_{t\bar t}$,$y_t$) data set (ID 576), using ePump with weight 19, compared to CT14HERA2mJ and CT14HERA2e. Left: PDF central values. Right: Error bands.}
\label{Fig:mj576w19g}
\end{figure}

\begin{figure}[h] 
\includegraphics[width=0.45\textwidth]{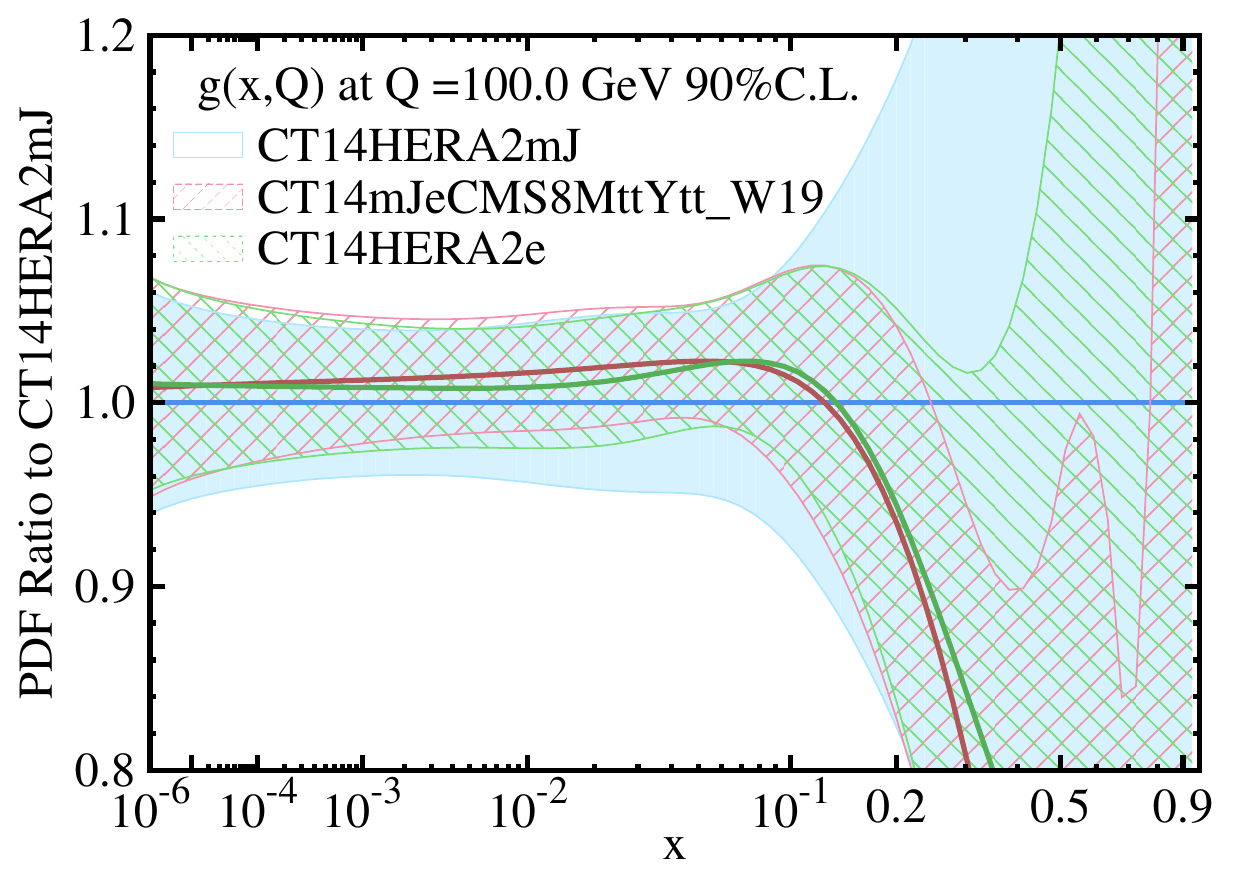}
\includegraphics[width=0.45\textwidth]{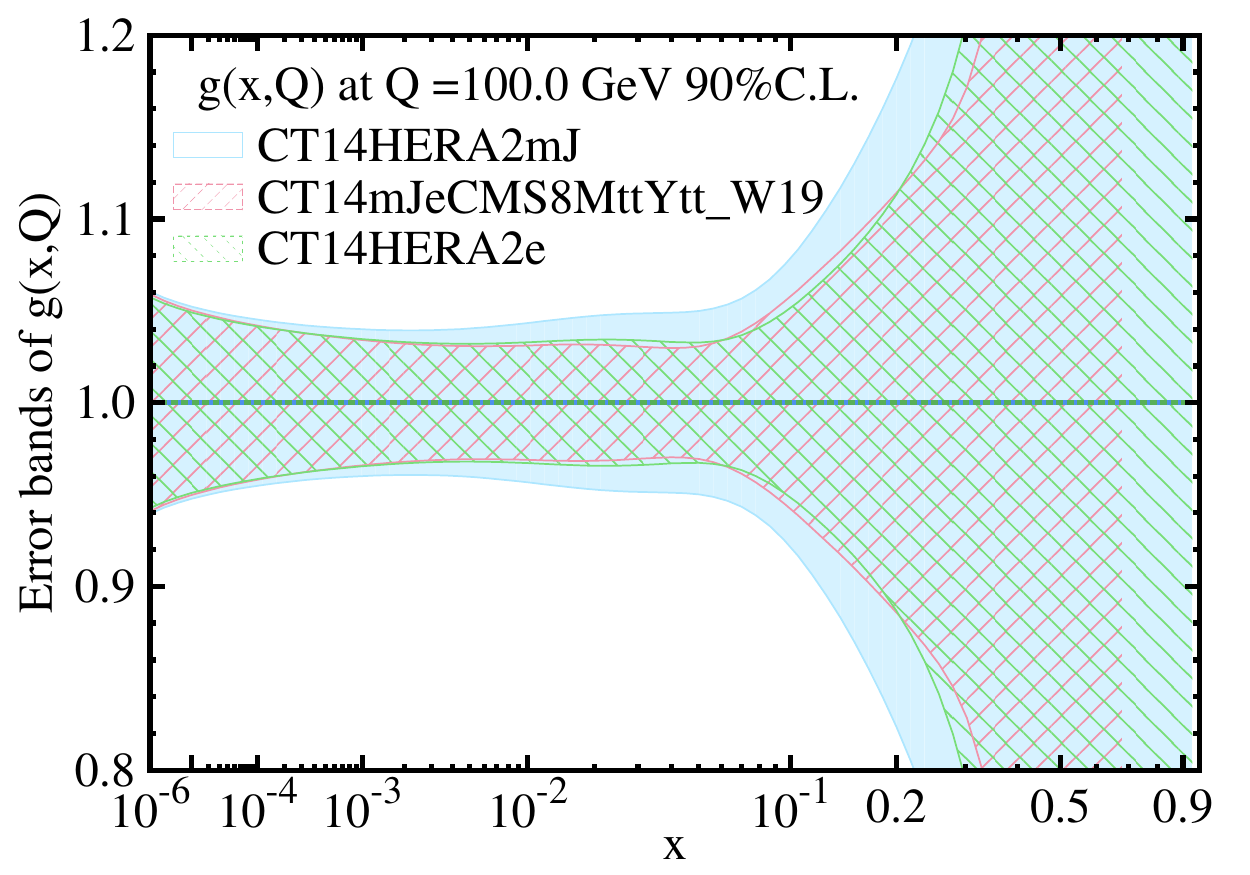}
\caption{The updated gluon PDF and its error band when adding the ($m_{t\bar t} \,$,$\, y_{t\bar t}$) data set (ID 577), using ePump with weight 19,  compared to CT14HERA2mJ and CT14HERA2e. Left: PDF central values. Right: Error bands.}
\label{Fig:mj577w19g}
\end{figure}

 When each double-differential $t\bar t$ data set is individually added to CT14HERA2, no strong impacts are observed on the gluon distribution. A similar result was noted for the influence of the single-differential top measurements~\cite{Hou:2019gfw}. Fig.~\ref{Fig:ct14p577g} shows the updated gluon PDF when each of the double-differential $t\bar t$ data sets is added to update the CT14HERA2 PDFs using the ePump code.
The central value of the gluon PDF changes only at large $x$ values, in a region basically unconstrained by present data, and the double differential ($m_{t\bar t}, p_T^{t\bar t}$) data set (ID 574) prefers a larger gluon PDF at large $x$, opposite to the other $t\bar{t}$ data sets. This feature will become more apparent later when the weight of the data set is increased in the PDF updating procedure. (See Figs.~\ref{Fig:mj573w19g}-\ref{Fig:mj577w19g}.) With that being said, the impact of the double-differential $t \bar t$ data set is marginal, and there is no noticeable change in the gluon PDF uncertainty. This is because the behavior of gluon PDF at large $x$ region is 
dominated by the effects of the other data sets,  among which are the jet data, included in the CT14HERA2 fit. If the jet data from the Tevatron and LHC are removed from this fit (which yields a new set of global fit PDFs, named  PDF CT14HERA2mJ), then the impact of the $t\bar t$
double-differential data is noticeably larger.
The ePump-updated gluon PDF of CT14HERA2mJ, using the 
($m_{t\bar t}, y^{t\bar t}$) data (ID 577)
as an example, is shown in Fig.~\ref{Fig:mj577w1g}. (Note that CT14HERA2mJ now serves as the reference set.)   
Some clear trends are observed. The gluon distribution at high $x$ (above 0.15) is larger than that preferred by CT14HERA2 (but still somewhat smaller than that preferred by CT14HERA2mJ). The PDF uncertainty is still larger than that of CT14HERA2 for $x \ge 10^{-4}$. A comparison of Fig.~\ref{Fig:ct14p577g} and ~\ref{Fig:mj577w1g}  shows that the double-differential $t\bar t$ data has an impact on the best-fit gluon-PDF in the large $x$ region, but in the presence of the jet data, the impact on gluon-PDF error band is diminished. Hence, we conclude that the overall sensitivity of the $t\bar t$ data set is less than the jet data due to the much smaller number of total data points in the $t \bar t$ data set, but as we shall see, the sensitivity of each $t \bar t$ data point is about the same as the jet data.

The level of agreement of the $t \bar t$ data with the NNLO predictions, using CT14HERA2 or CT14HERA2mJ, can be
observed by comparing the theory predictions and the data for each data point. Figs.~\ref{Fig:CT14comparison_5} and~\ref{Fig:mjcomparison_5} show the comparisons for the CMS 8TeV ($m_{t\bar t}$,$y_{t\bar t}$) data set, for CT14HERA2 and CT14HERA2mJ, respectively. In the comparison of the data to the NNLO prediction using CT14HERA2, the data points are shifted according to the optimal systematic error shifts leading to best agreement with the theoretical prediction.  The shifted data points are closer to the theory prediction, as expected. Similar results are obtained for the other double-differential observables. It can also be seen that the theory predictions of the $t\bar{t}$ double differential distributions do not  change much after the ePump updating, which is consistent with the minimal impact on PDFs observed for the CMS 8 TeV double-differential $t\bar{t}$ data.

In total, there are 305 jet data points included in the CT14HERA2 PDF fit, including data from CDF and D0 at the Tevatron, and ATLAS and CMS at the LHC~\footnote{Specifically, the data include 1.13 $fb^{-1}$ from CDF and 0.70 $fb^{-1}$ from D0, at a center-of-mass energy of 1.96 TeV and 4.5 $fb^{-1}$ from ATLAS and 5 $fb^{-1}$ from CMS, at a center-of-mass energy of 7 TeV.}. The statistical errors vary from less than 1\% at low transverse momentum to tens of percent at high $p_T$. In contrast, as shown in Table II, there are only 16 data points for all but one of the CMS double-differential top data with statistical errors that vary from 2\% to 17\%, and systematic errors on the order of 3-17\%. Thus, there is a factor of 19 times more jet data points than double-differential $t\bar t$ data points. 

An interesting exercise is to increase the weight for the CMS $t\bar t$ data in the PDF updating (using ePump), using either CT14HERA2mJ or CT14HERA2 global-fit PDFs as the base, to a level that corresponds either to the statistical power of the full jet data, or to that of the most important single jet data  in the PDF fit, the CMS 7 TeV data set~\cite{Hou:2019gfw}. A weight of 19 (=305/16)\footnote{We divide by 16 because that is the original number of data points. It is only due to the nature of dealing with normalized distributions that we subtract one degree of freedom when calculating the $\chi^2/{\rm dof}$ 
in Table~\ref{table:ttb2Dchi2}.
} would correspond to having a similar number of data points for the top data as for the entire jet data set, and can also be considered as corresponding to an effective decrease in the statistical and systematic errors of the top data. For completeness, we also compare  the impact of increasing the weight of the CMS $t\bar t$ data by a factor of 8 (=133/16) to that of the jet data set with the largest impact in the CT14HERA2 fit which was found to be the 7 TeV CMS jet data with 133 data points in total~\cite{Hou:2019gfw}. To provide intermediate results, weights of 3 and 5 are also considered. 
It should be stressed that increasing the weight is not exactly equivalent to an increased luminosity, since there is no change in the central values of the data, i.e. the jitter from the existing data due to limited statistics is preserved in the re-weighted data, reducing somewhat its impact on the PDF fit.
There is also a competing factor going in the opposite direction of the jitter.
As stated, the dominant subprocess for  $t\bar{t}$ production is through $gg$ fusion.  For the inclusive jet process, $gg$ fusion is sub-dominant at high $p_T$, overshadowed  by $gq$ and $qq$ scattering.  Thus, at high $x$, $t\bar t$ events should inherently have a larger impact on the gluon distribution (per event) than inclusive jet production. Thus, due to these two sources, an exact scaling with number of  data points is not expected, but still may be informative.
The results of $\chi^2$ are shown in  Tables~\ref{table:ttb2Dchi2HERA2w} and \ref{table:ttb2Dchi2mJw} for the cases of the CT14HERA2 and CT14HERA2mJ PDFs, respectively.

\begin{table}[htbp]
	\begin{center}
		\caption{The $\chi^2$ for each 2D $t\bar{t}$ data set, calculated with the original global-fit CT14HERA2 PDFs (i.e., $w=0$) and  ePump updated CT14HERA2 PDFs with different weights. For simplicity, the same weights are also applied to the ($m_{t\bar t}, \Delta \eta_{t\bar t}$)
			data set (ID 575), although it contains 12 (before removing one bin), not 16, data points. 
		} 
		\begin{tabular}{| c | c | c | c | c | c | c | c | c |}
			\hline
			ID & data &  dof   & \multicolumn{6}{c|}{$\chi^2$} \\
			  & & & $w=0$ & $w=1$ & $w=3$ & $w=5$ & $w=8$  & $w=19$ \\
			\hline
			573 & $\sigma^{-1}\, d^2\sigma/dy_t dp^t_T$  & 15	& 35.5 	& 34.9 & 33.9 & 33.1 & 32.2 & 30.0\\
			\hline
			574 & $\sigma^{-1}\, d^2\sigma/dm_{t\bar t} dp^{t\bar t}_T$       & 15 	& 82.3 & 80.6 & 77.8 & 75.6	& 73.1 & 68.1 \\
			\hline
			575 & $\sigma^{-1}\, d^2\sigma/dm_{t\bar t} d\Delta \eta_{t\bar t}$  & 11 & 22.1 & 22.0 & 21.8 & 21.6 & 21.3 & 20.3\\
			\hline
			576 & $\sigma^{-1}\, d^2\sigma/dm_{t\bar t} dy_t$  & 15 & 20.2 & 20.1 & 20.0 & 19.9 & 19.8 & 19.5 \\
			\hline
			577 & $\sigma^{-1}\, d^2\sigma/dm_{t\bar t} dy_{t\bar t}$ & 15 & 23.8 & 23.5 & 23.1 & 22.9 & 22.6 & 22.1  \\
			\hline
		\end{tabular}
		\label{table:ttb2Dchi2HERA2w}
	\end{center}
\end{table}

\begin{table}[htbp]
	\begin{center}
		\caption{
The $\chi^2$ for each 2D $t\bar{t}$ data set, calculated with the original global-fit CT14HERA2mJ PDFs (i.e., $w=0$) and  ePump updated CT14HERA2mJ PDFs with different weights. For simplicity, the same weights are also applied to  the ($m_{t\bar t}, \Delta \eta_{t\bar t}$)
data set (ID 575), although it contains 12 (before removing one bin), not 16, data points. 	
}
		\begin{tabular}{| c | c | c | c | c | c | c | c | c |}
			\hline
		ID & data &  dof   & \multicolumn{6}{c|}{$\chi^2$} \\
			  & & & $w=0$ & $w=1$ & $w=3$ & $w=5$ & $w=8$  & $w=19$ \\
			\hline
			573 & $\sigma^{-1}\, d^2\sigma/dy_t dp^t_T$    & 15 	& 50.9 	& 41.1 & 33.5 & 30.1 & 27.3 & 22.3\\
			\hline
			574 & $\sigma^{-1}\, d^2\sigma/dm_{t\bar t} dp^{t\bar t}_T$  & 15	& 70.9 & 69.0 & 66.7 & 65.5	& 64.4 & 63.1 \\
			\hline
			575 & $\sigma^{-1}\, d^2\sigma/dm_{t\bar t} d\Delta \eta_{t\bar t}$  & 11 & 26.4 & 25.9 & 25.0 & 24.2 & 23.2 & 20.4\\
			\hline
			576 & $\sigma^{-1}\, d^2\sigma/dm_{t\bar t} dy_t$  & 15 & 27.6 & 25.5 & 23.1 & 21.9 & 20.9 & 19.7 \\
			\hline
			577 & $\sigma^{-1}\, d^2\sigma/dm_{t\bar t} dy_{t\bar t}$ & 15 & 45.3 & 34.7 & 27.5 & 25.1 & 23.6 & 22.3  \\
			\hline
		\end{tabular}
		\label{table:ttb2Dchi2mJw}
	\end{center}
\end{table}

First, note that the starting $\chi^2$ values are larger for  CT14HERA2mJ than for CT14HERA2, especially for the $y_t,p^t_T$ (ID 573) and  $m_{t\bar t},y_{t\bar t}$ (ID 577) 
data sets. As will be shown later, this is because the gluon distribution the double-differential top data prefer is closer to that of CT14HERA2 than CT14HERA2mJ. 
As the weight increases, there is a slower decrease in $\chi^2$ for the CT14HERA2 fit than for the CT14HERA2mJ fit, due to the inclusion of jet data in the CT14HERA2 fit.
However,  the exact magnitude of the decrease in $\chi^2$ should not be taken too seriously, since, as previously stated,
increasing the weight does not change the central values of the data, i.e. the
jitter from the existing data due to limited statistics is preserved in the re-weighted data. 
We observe that the inclusion of the $y_t,p^t_T$ (ID 573) and  $m_{t\bar t},y_{t\bar t}$ (ID 577) 
data sets show a noticeable improvement in $\chi^2$ when included in the CT14HERA2mJ fit, but not in the CT14HERA2 fit. They show further improvement for CT14HERA2mJ on the use of higher weights.

We now consider the impact on the gluon distribution,  first considering the weight of 19, again weighting an individual double-differential $t\bar{t}$ data set to have the equivalence of the total jet data in CT14HERA2. 
The ePump updated gluon PDFs with this weight are shown in Figs.~\ref{Fig:mj573w19g}-\ref{Fig:mj577w19g}, where it can be seen that the $t\bar t$ data has a similar constraint on the gluon PDFs as does the jet data (included in the CT14HERA2 fit), both for the value of the central PDF and the size of the error band. The central gluon distribution that is thus obtained does not always agree with that obtained using the jet data (CT14HERA2), but the error bands are all of  similar size. 
A detailed look reveals that 
the ($m_{t\bar t}, \Delta \eta_{t\bar t}$), ($m_{t\bar t},y_t$) and 
($m_{t\bar t}, y^{t\bar t}$)
data sets (IDs 575-577) give a similar constraint on the central $g$ PDF as that of jet data; 
the ($y_t,p^{t}_T$) data set (ID 573) 
also has a similar trend but prefers somewhat harder gluon at moderate $x$; and the ($m_{t\bar t},p^{t\bar t}_T$) data set (ID 574) prefers softer gluon at $x<0.1$ and  harder gluon at high $x$ than jets and the other data sets.\footnote{In fact, the ($m_{t\bar t},p^{t\bar t}_T$) data set (ID 574) is the only CMS 8 TeV double differential $t\bar{t}$ data set that prefers a significantly harder gluon at high $x$, as compare to the CT14HERA2 PDFs.} 
As a result, the gluon PDF in the region sensitive to gluon-gluon Higgs boson production ($x \sim $0.01) is larger in the case of PDF-updating with the ($y_t,p^{t}_T$) data (ID 573) and smaller with the ($m_{t\bar t},p^{t\bar t}_T$) data (ID 574). 
Note that including each individual $t\bar{t}$ data set in the PDF-update does not modify gluon PDF at $x\sim 0.1$, which verifies our observation, based on  Figs.~\ref{Fig:correlation573} and~\ref{Fig:correlation574-577g}, that the correlation between the CMS 8 TeV double differential $t\bar{t}$ data and gluon PDF is very small at $x\sim 0.1$.
Furthermore, we would like to note that due to the different composition of hard scattering processes contributing to the production of $t \bar t$ and jet productions at the LHC, the weighted $t \bar t$ data provide a slightly narrower gluon-PDF error band for $x$ around 0.3, as compared to the jet data, cf. Figs.~\ref{Fig:mj573w19g} and~\ref{Fig:mj574w19g}, when using CT14HERA2mJ global fit as the base for PDF updating.

Based on the above results, we conclude that each data point of the CMS 8 TeV double-differential $t\bar{t}$ data has at least the same constraining power (or sensitivity) as that of jet data. 
The absence of any significant impact as a whole data set is due to the small number of total data points in the CMS 8 TeV $t \bar t$ data set as compared to that in the jet data. 
Hence, the total sensitivity of the  CMS 8 TeV $t \bar t$ data set is not great.
At higher integrated LHC luminosities, the $t\bar{t}$ data may provide some unique constraints on gluon PDF, especially for the large $x$ region (for $x > 0.3$).

\begin{figure}[htbp] 
\includegraphics[width=0.45\textwidth]{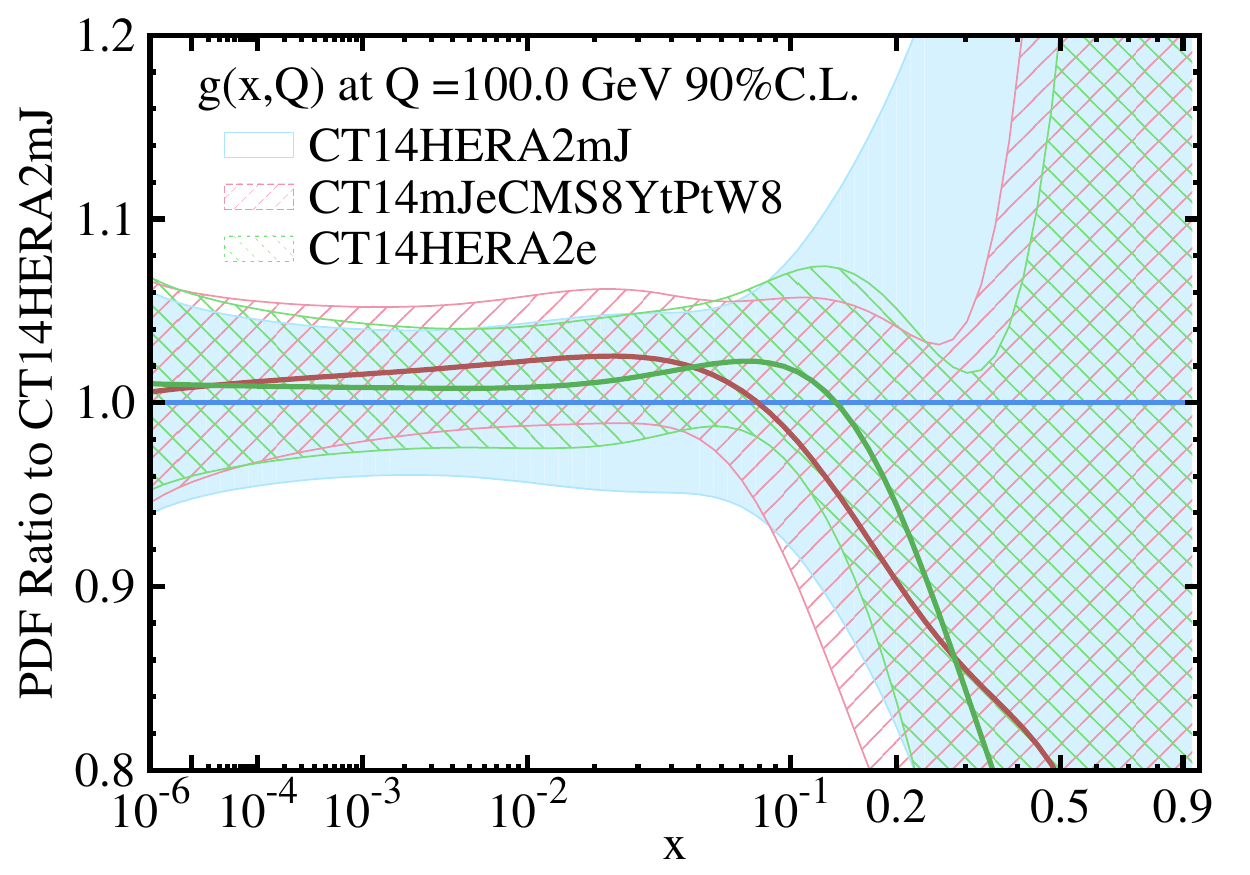}
\includegraphics[width=0.45\textwidth]{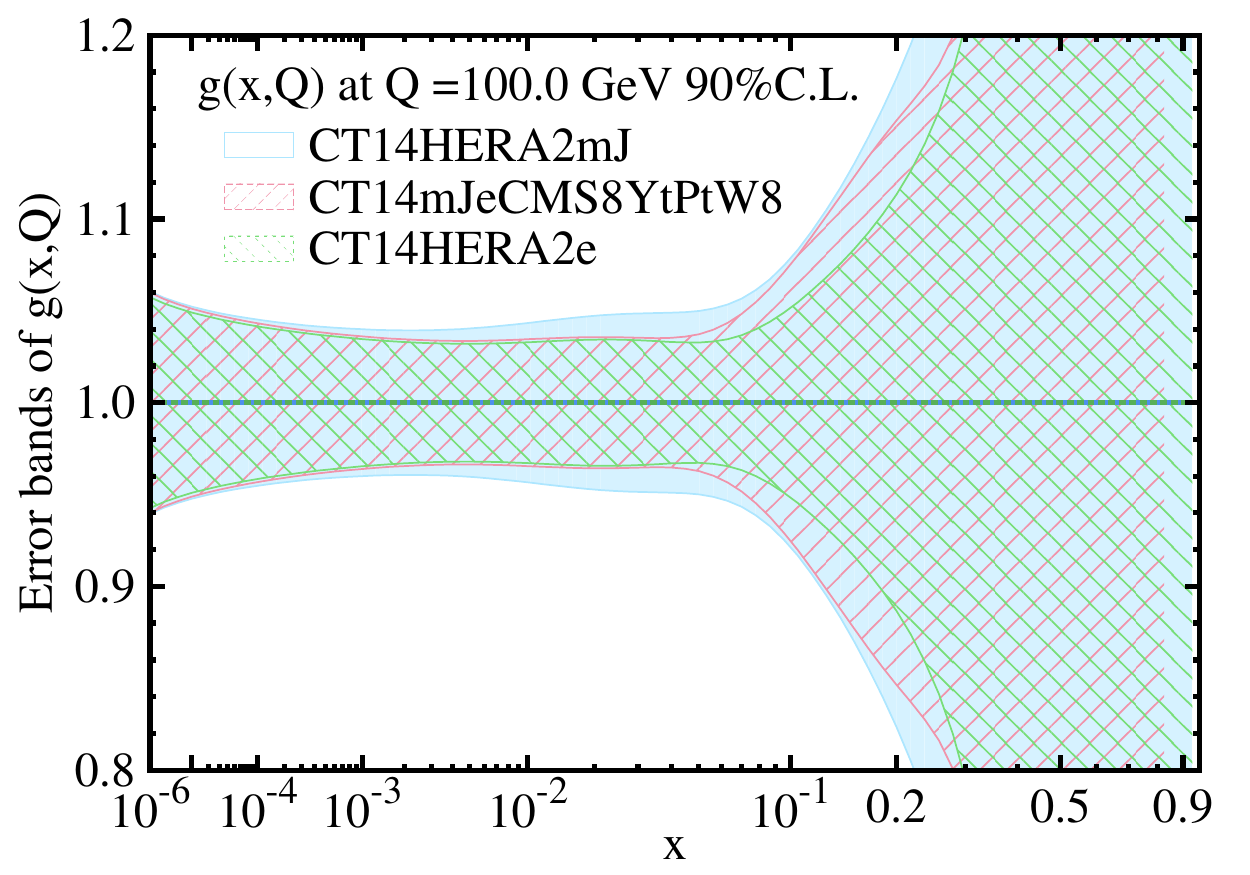}
\caption{The updated gluon PDF and its error band when adding 573 (which is differential in $y_t$ and $p_T^t$), using ePump with weight 8, compared to CT14HERA2mJ and CT14HERA2e. Left: PDF central values. Right: Error bands.}
\label{Fig:mj573w8g}
\end{figure}

\begin{figure}[htbp] 
\includegraphics[width=0.45\textwidth]{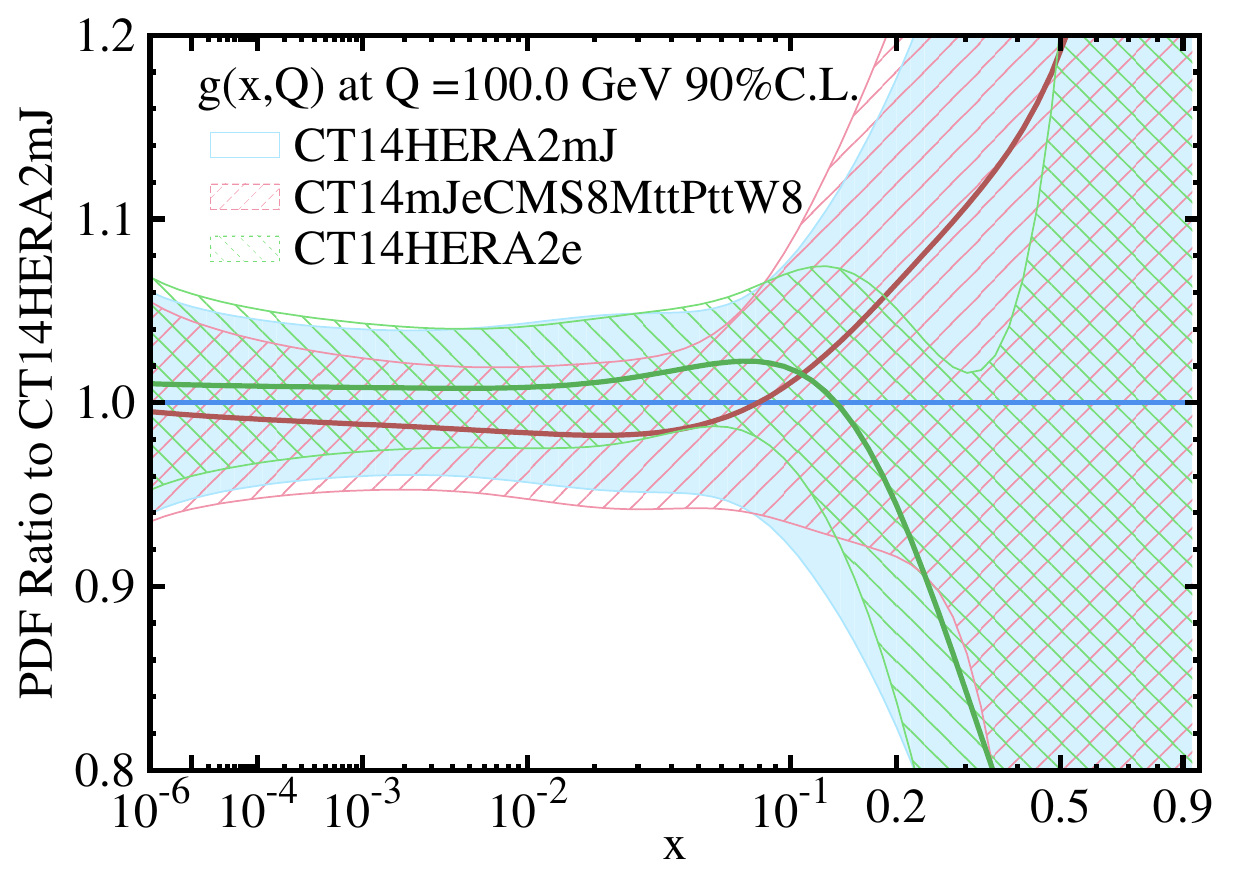}
\includegraphics[width=0.45\textwidth]{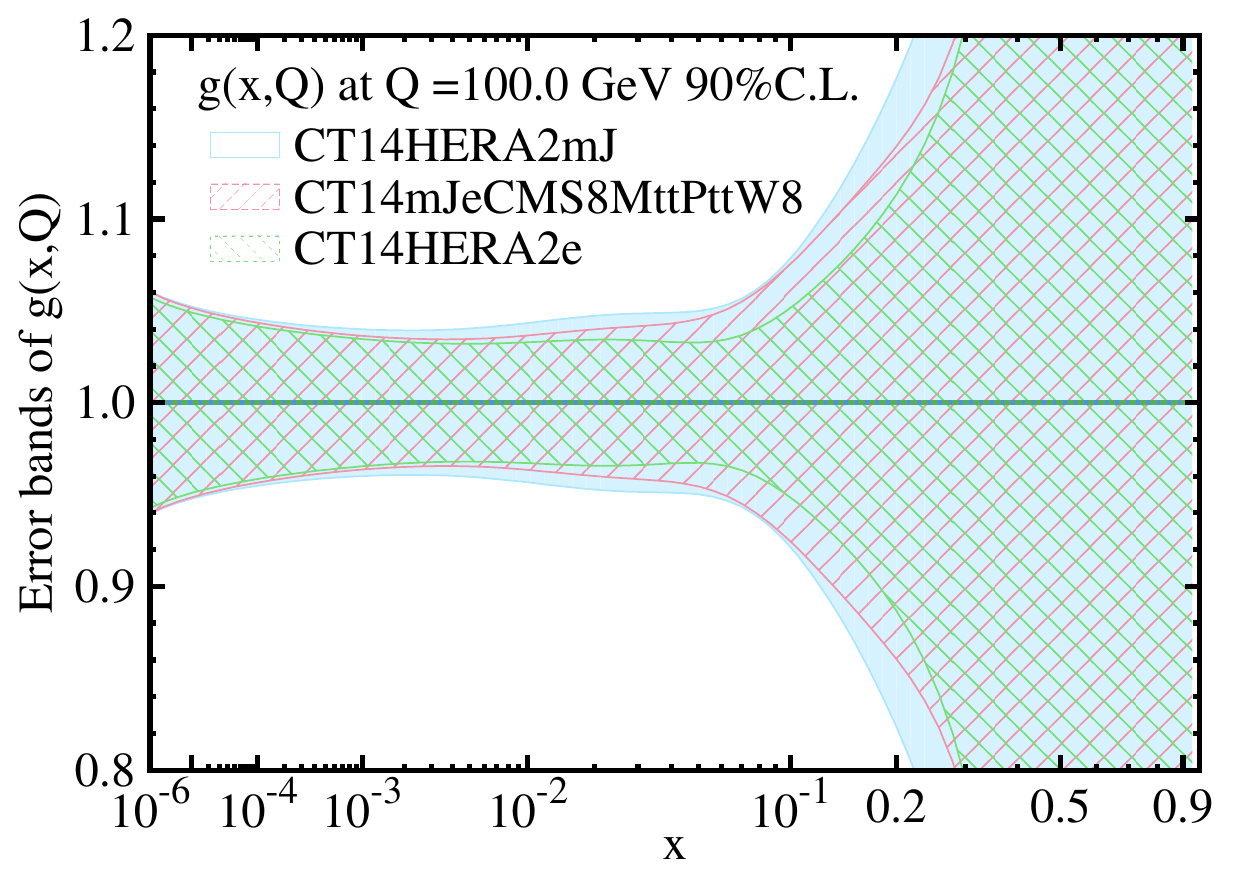}
\caption{The updated gluon PDF and its error band when adding 574 (which is differential in $m_{t\bar t}$ and $p_T^{t\bar t}$), using ePump with weight 8, compared to CT14HERA2mJ and CT14HERA2e. Left: PDF central values. Right: Error bands.}
\label{Fig:mj574w8g}
\end{figure}

\begin{figure}[htbp] 
\includegraphics[width=0.45\textwidth]{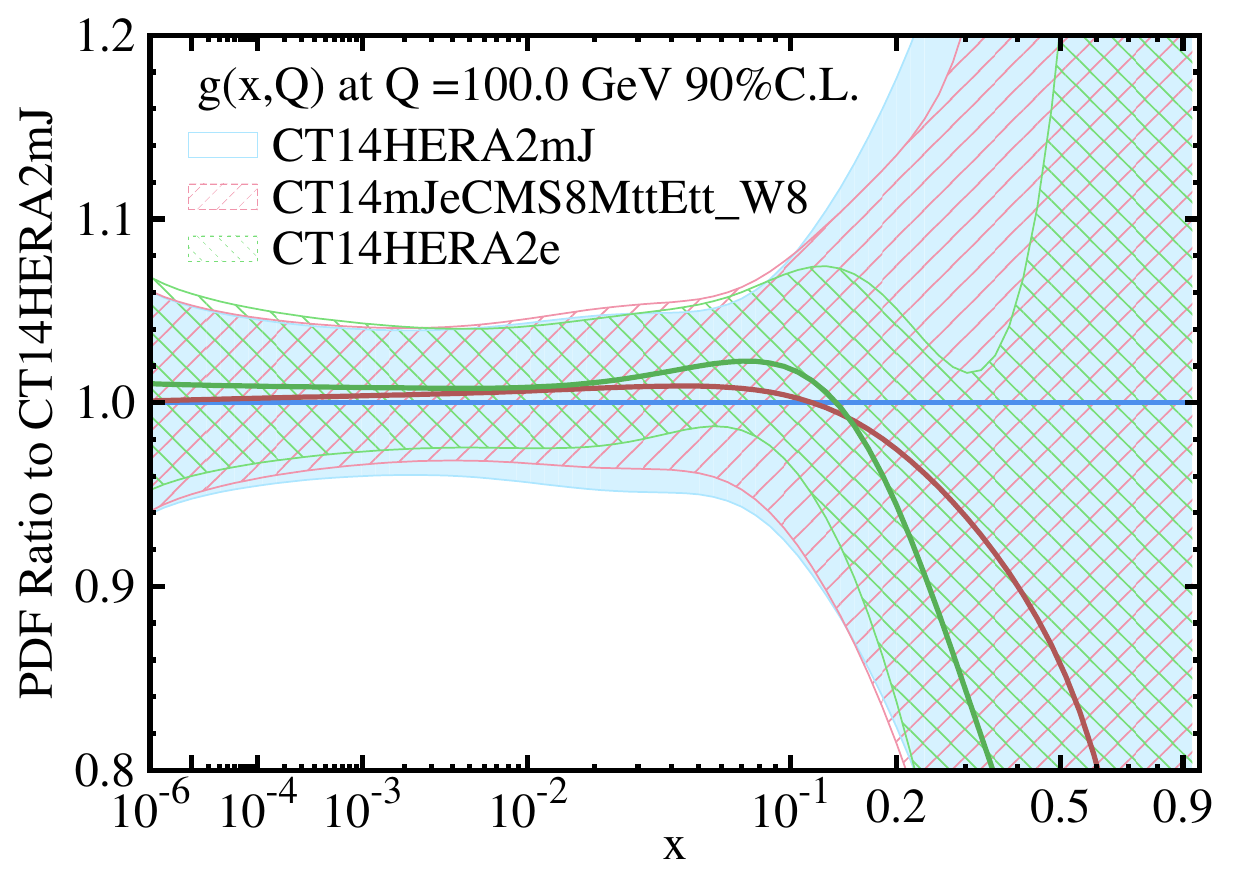}
\includegraphics[width=0.45\textwidth]{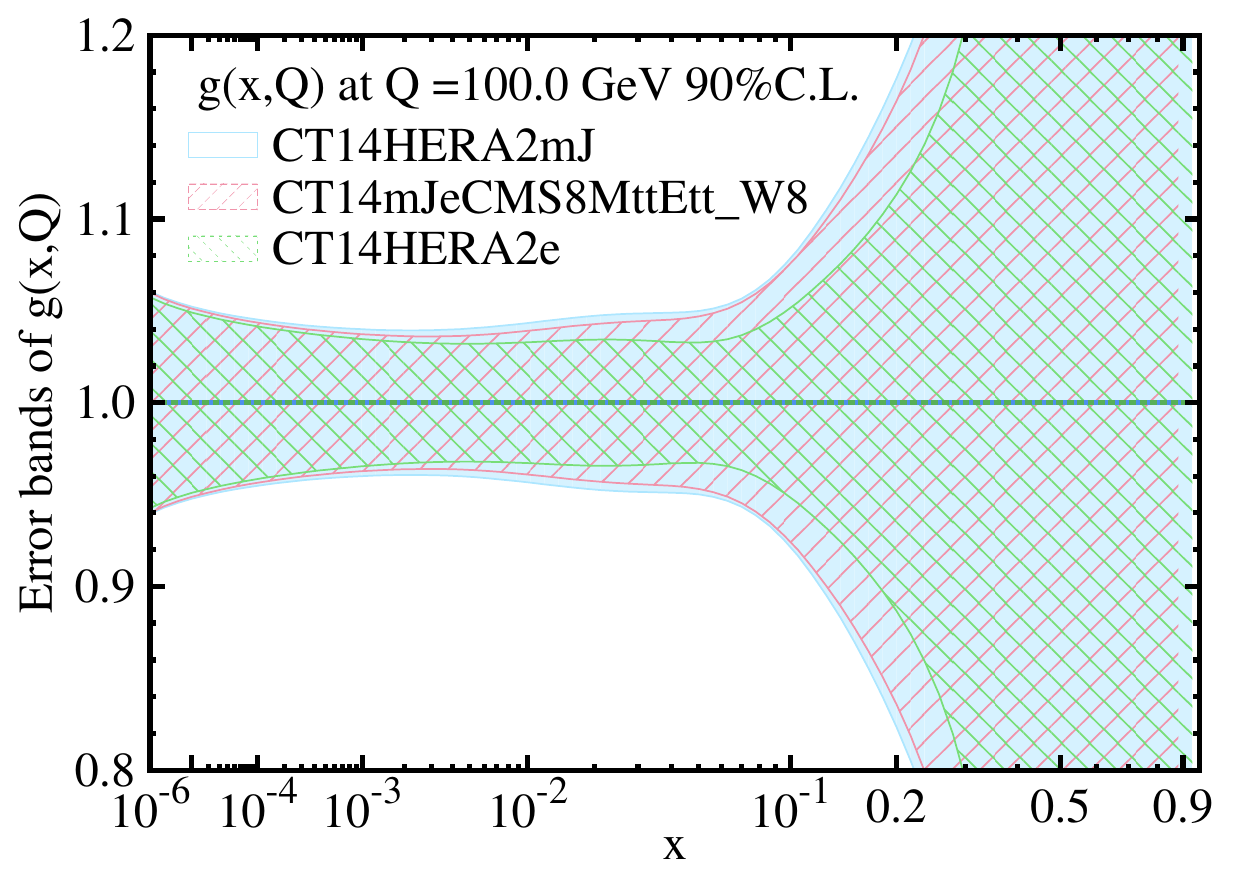}
\caption{The updated gluon PDF and its error band when adding 575 (which is differential in $m_{t\bar t}$ and $\Delta\eta_{t\bar t}$), using ePump with weight=8, compared to CT14HERA2mJ and CT14HERA2e. We still use weight 8 instead of 133/12=11 because these 12 data points were also constructed out of the same $t\bar t$ data. Left: PDF central values. Right: Error bands.}
\label{Fig:mj575w8g}
\end{figure}

\begin{figure}[htbp] 
\includegraphics[width=0.45\textwidth]{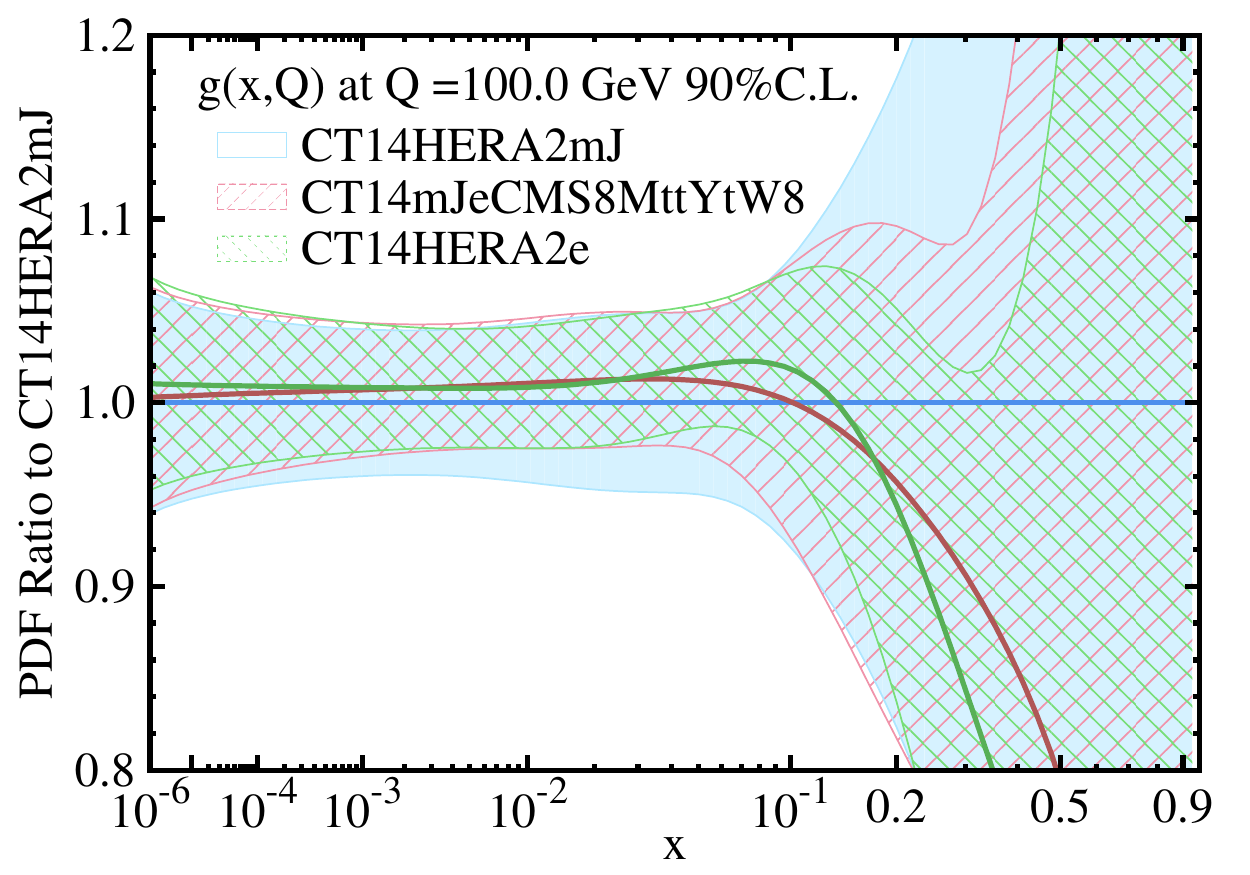}
\includegraphics[width=0.45\textwidth]{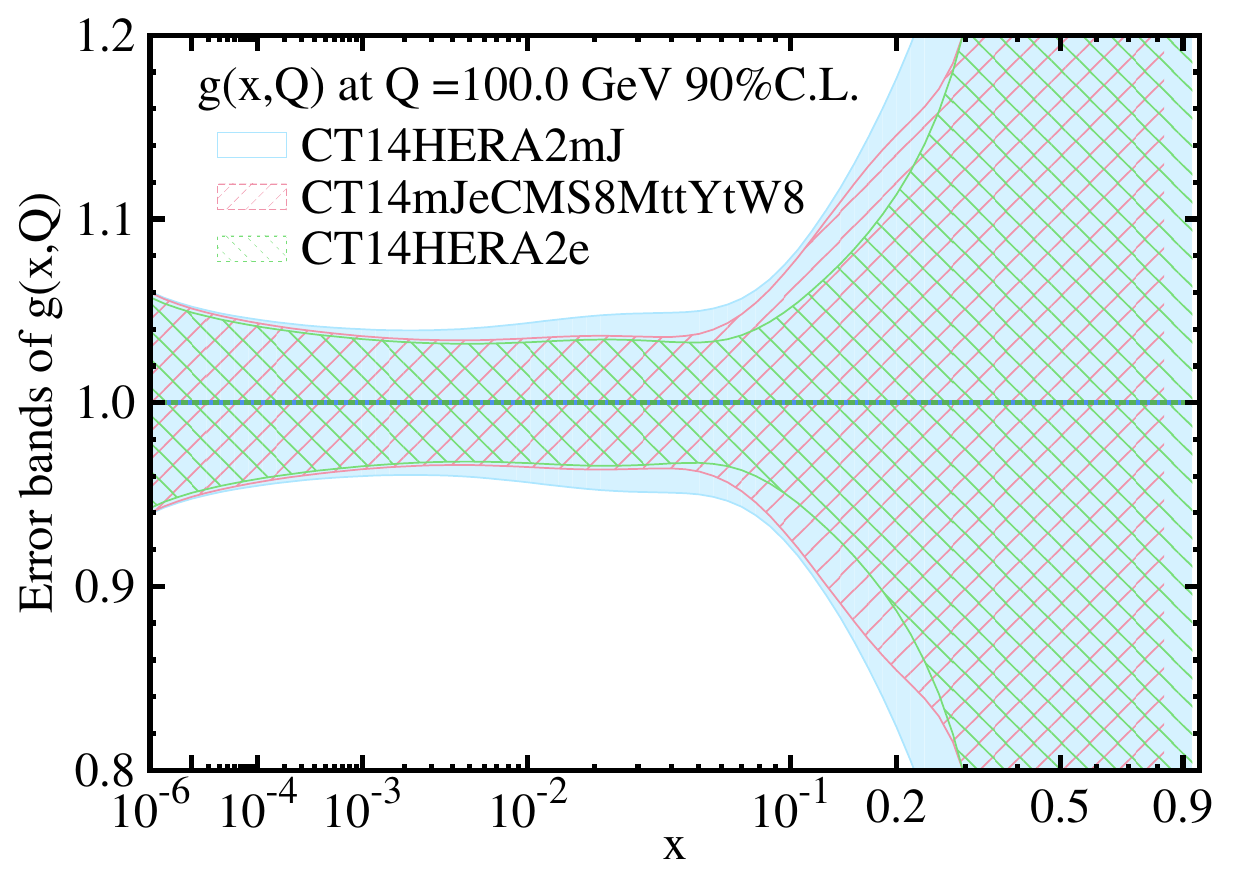}
\caption{The updated gluon PDF and its error band when adding 576 (which is differential in $m_{t\bar t}$ and $y_t$), using ePump with weight 8, compared to CT14HERA2mJ and CT14HERA2e. Left: PDF central values. Right: Error bands.}
\label{Fig:mj576w8g}
\end{figure}

\begin{figure}[h] 
\includegraphics[width=0.45\textwidth]{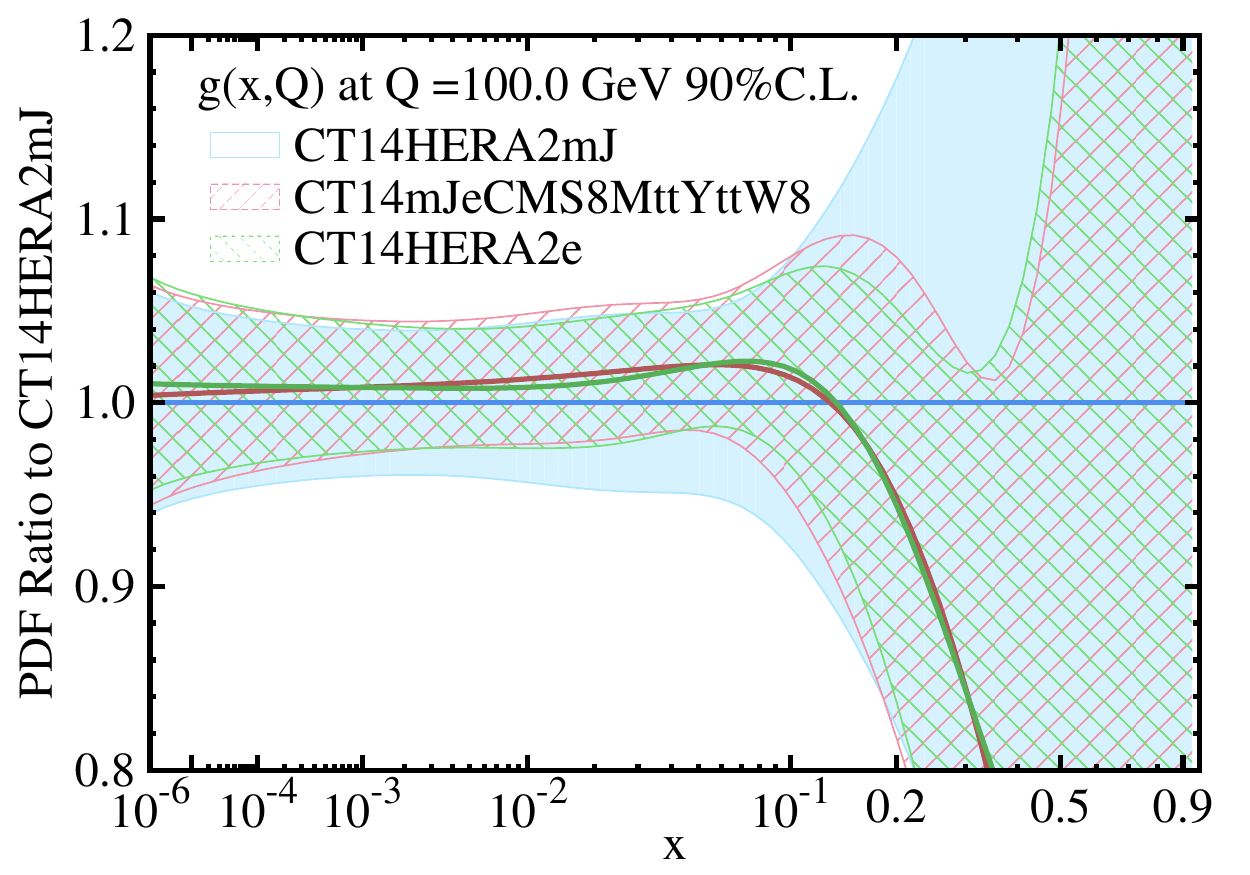}
\includegraphics[width=0.45\textwidth]{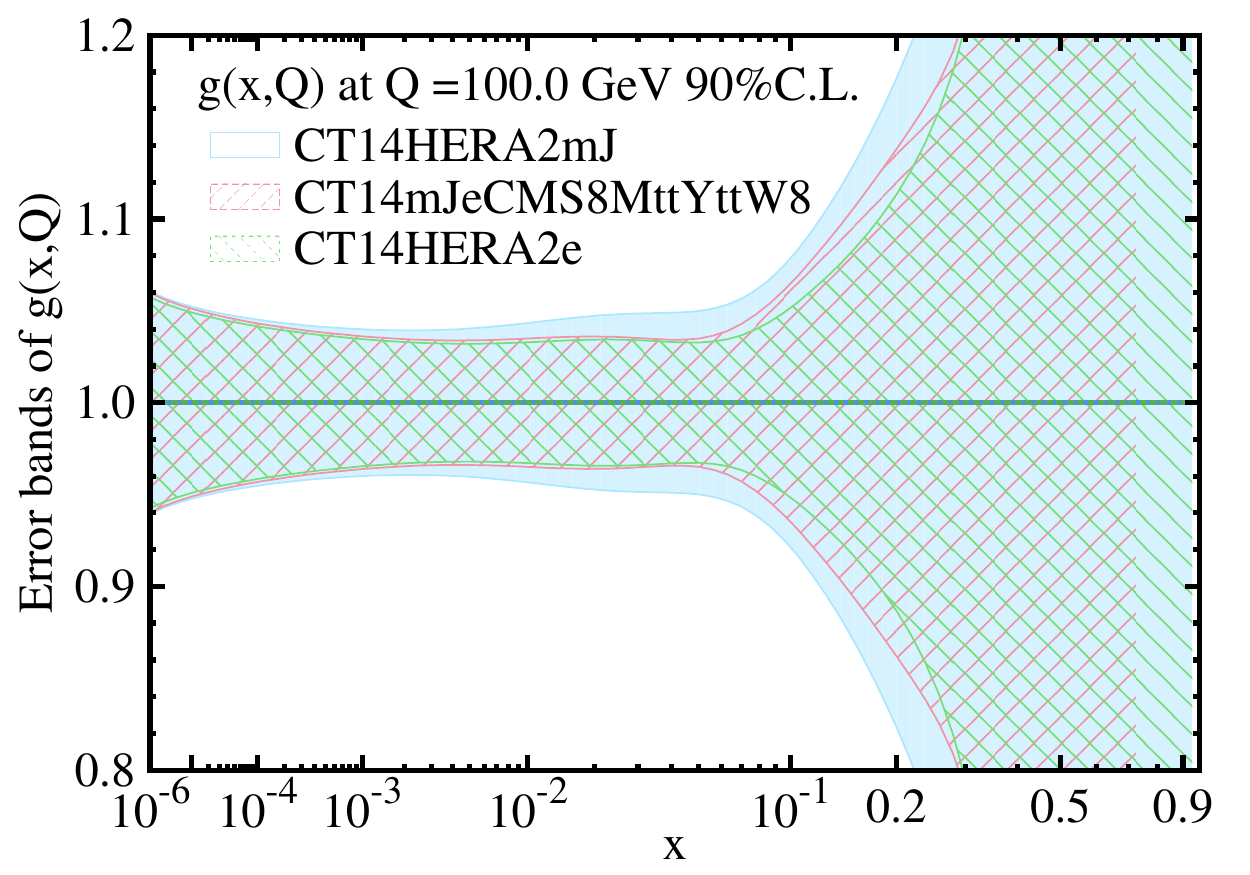}
\caption{The updated gluon PDF and its error band when adding data set 577 (which is differential in $m_{t\bar t}$ and $y_{t\bar t}$),  using ePump with weight 8, compared to CT14HERA2mJ and CT14HERA2e. Left: PDF central values. Right: Error bands.}
\label{Fig:mj577w8g}
\end{figure}

We have also examined the impact of a smaller weight, 8, which corresponds to having a similar number of data points for the
$t \bar t$ data as for the single strongest jet data set included in the CT14HERA2, that of the CMS 7 TeV inclusive jet cross section.
In Figs.~\ref{Fig:mj573w8g}-\ref{Fig:mj577w8g}, we show the results of ePump updated gluon PDFs when each $t\bar t$ data set is added to CT14HERA2mJ with weight 8. We find that the $t\bar t$ data have similar size effects with weight 8, as observed with weight 19. 
This is true especially for the ($m_{t\bar t} \,$,$\, y_{t\bar t}$) data set (ID 577), where we find almost the same impact on the  gluon PDF as for jets, except that jet data lead to a smaller error band in the $x$ range between 0.1 to 0.2, where the correlations of the $t\bar{t}$ data and gluon PDF were observed to be weaker. 

\begin{figure}[h] 
\includegraphics[width=0.45\textwidth]{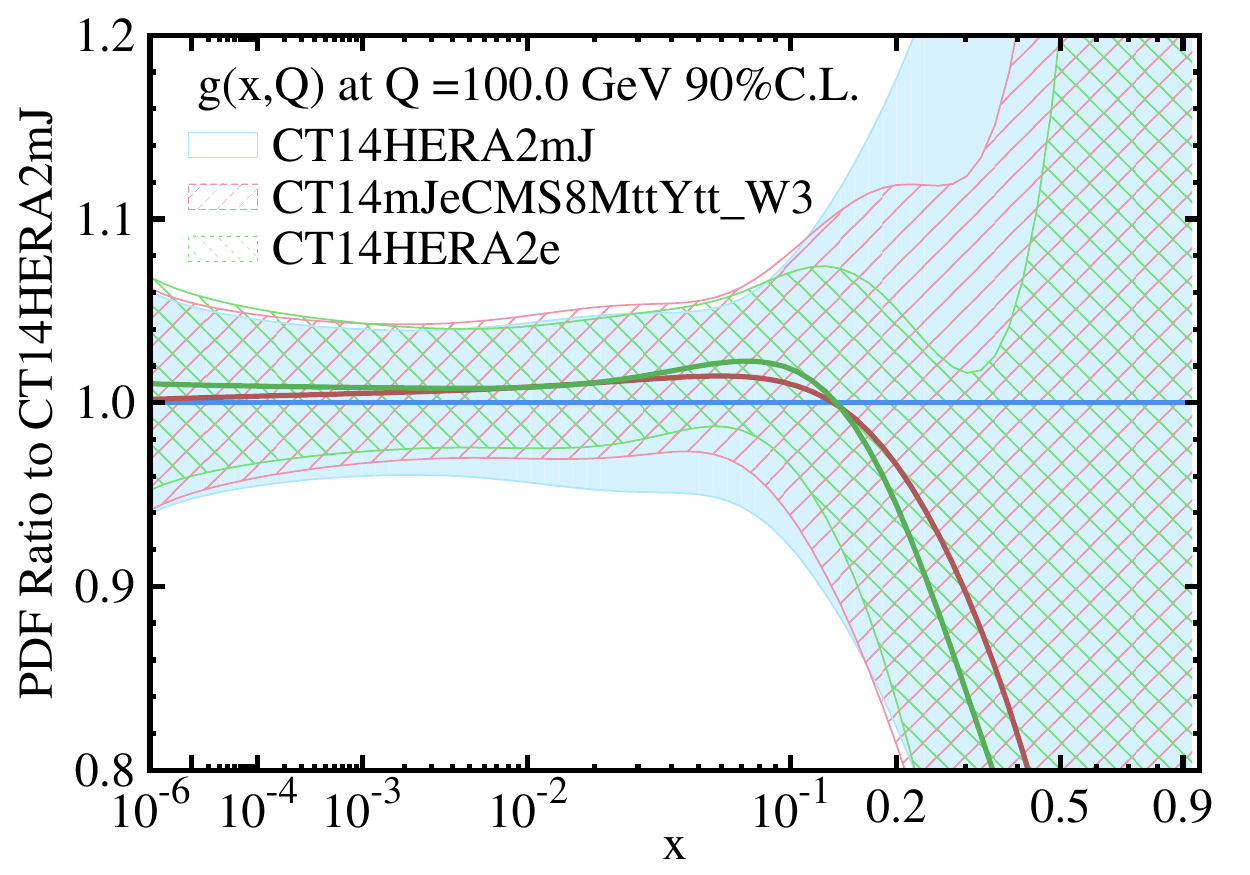}
\includegraphics[width=0.45\textwidth]{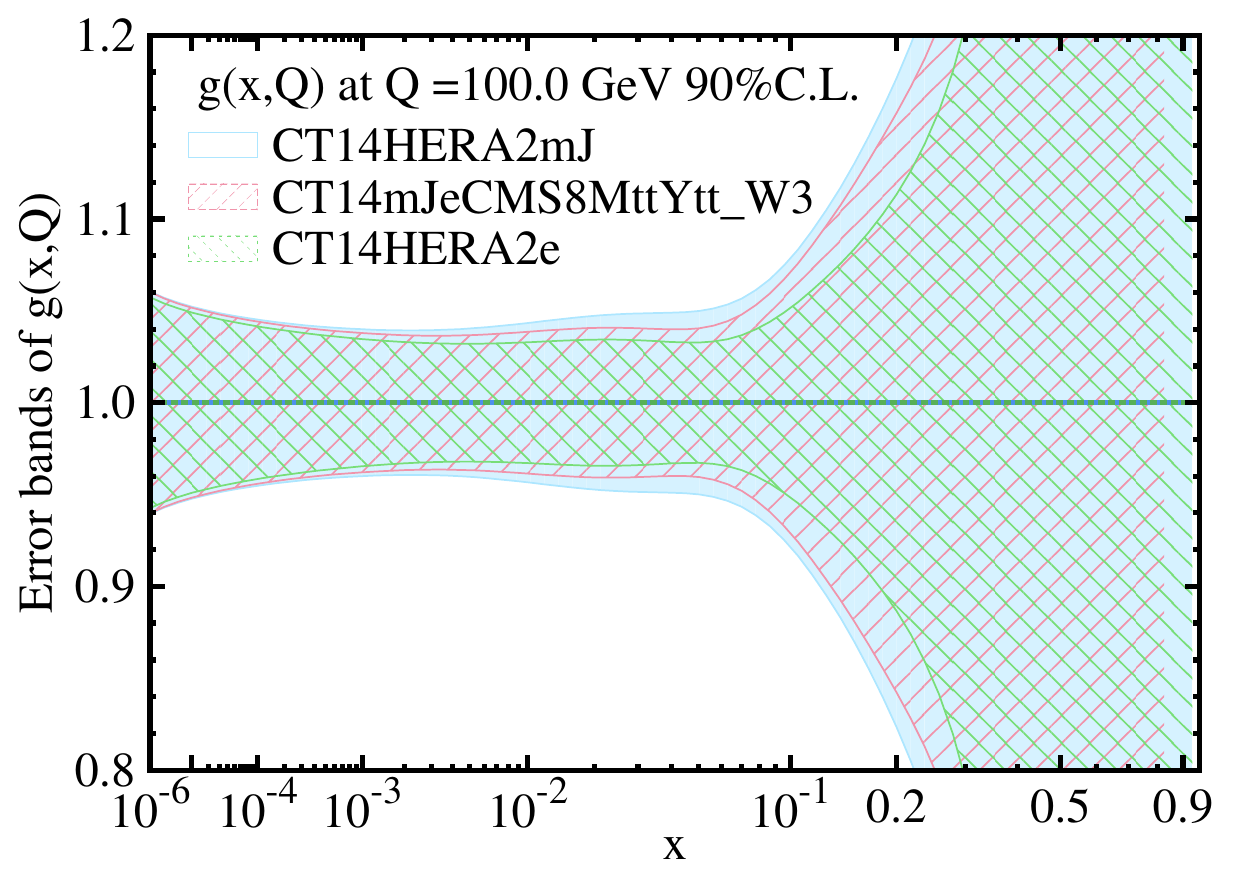}
\caption{The updated gluon PDF and its error band when adding data set 577 (which is differential in $m_{t\bar t}$ and $y_{t\bar t}$), using ePump with weight=3,  compared to CT14HERA2mJ and CT14HERA2e. Left: PDF central values. Right: Error bands.}
\label{Fig:mj577w3g}
\end{figure}

\begin{figure}[h] 
\includegraphics[width=0.45\textwidth]{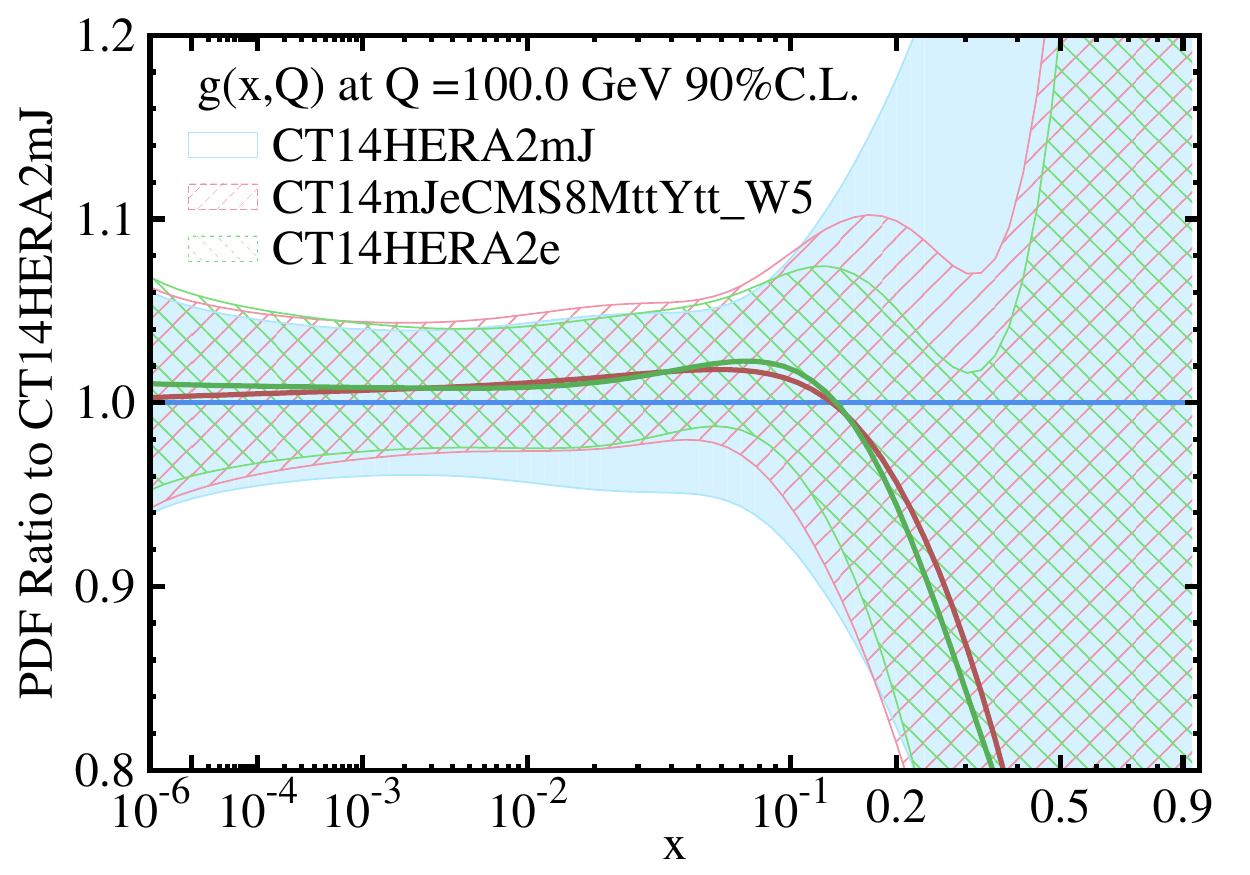}
\includegraphics[width=0.45\textwidth]{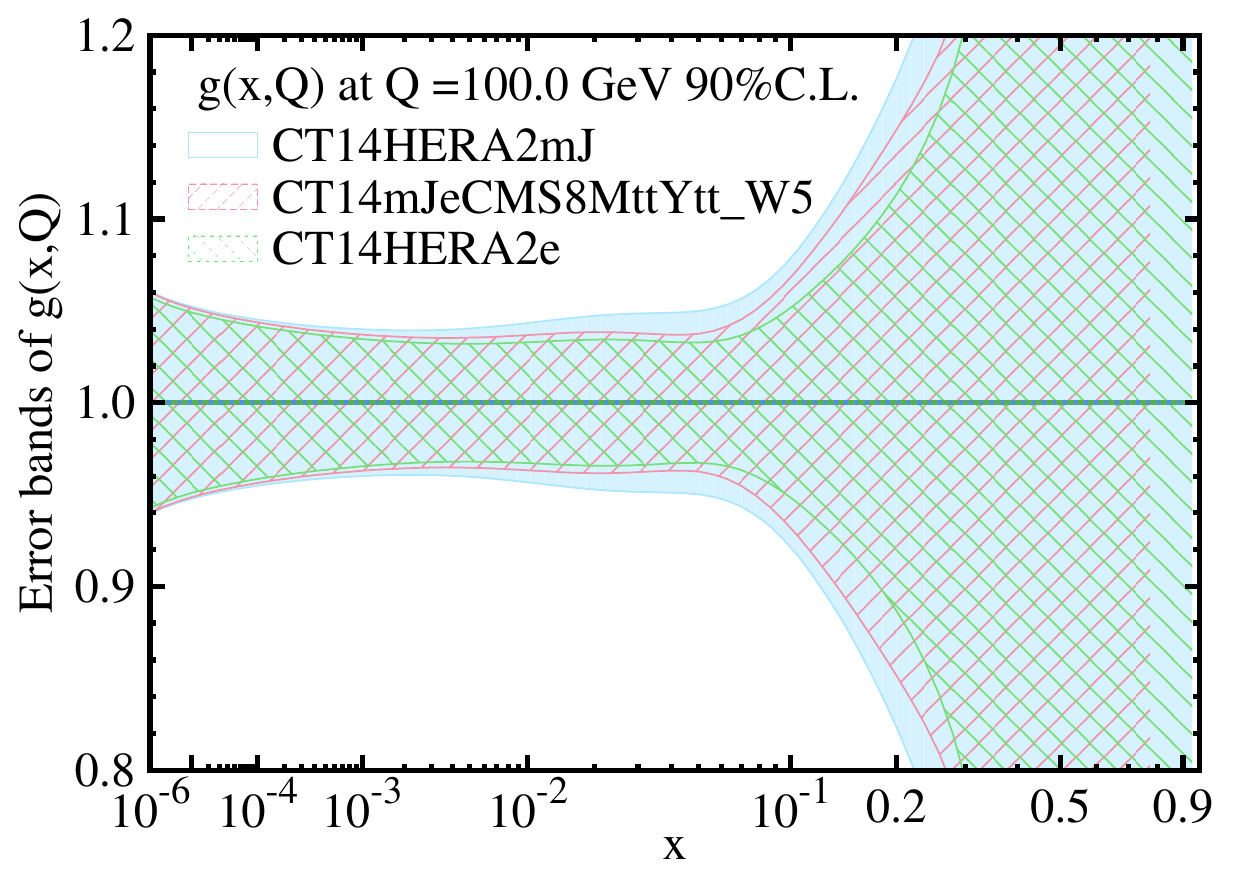}
\caption{The updated gluon PDF and its error band when adding data set 577 (which is differential in $m_{t\bar t}$ and $y_{t\bar t}$), using ePump with weight=5, compared to CT14HERA2mJ and CT14HERA2e. Left: PDF central values. Right: Error bands.}
\label{Fig:mj577w5g}
\end{figure}

It is also useful to examine the impact of even smaller weights, 3 and 5, in Figs.~\ref{Fig:mj577w3g} and~\ref{Fig:mj577w5g} for the ($m_{t\bar t} \,$,$\, y_{t\bar t}$) data set (ID 577). Here we see, as expected, intermediate results, which nonetheless indicate that even moderate increases in integrated luminosity samples could potentially lead to a noticeable impact by the double-differential top data. 

An alternative way of displaying the impact of a new data set on the resulting PDF distributions is to examine
 the length of the shift vector $d^0$~\cite{Schmidt:2018hvu},  of the best-fit position in PDF parameter space, from the original set of parameters for CT14HERA2 to those preferred by the fit with the inclusion of the new data set. The vector $d^0$ is 27-dimensional, corresponding to the number of free parameters in the CT14HERA2 global PDF fit. A value of $d^0$ of the order of 1 indicates that  the new best-fit vector touches the 90\% CL boundary, i.e. there is a very large impact (change) from this new data set, while  a value of $d^0$ smaller than 0.1 would imply no large change of the PDFs.  
In Table~\ref{table:ttb2DH} and \ref{table:ttb2Dmj}, the values of $d^0$ are shown for the results of including data sets 573-577 with various weights, using CT14HERA2 and CT14HERA2mJ respectively.
The values of $d^0$ increase with weight, as expected. The impact is greater with CT14HERA2mJ than with CT14HERA2. The largest values of $d^0$ results from the ($y_{t}, p^t_T$) and ($m_{t\bar t}, p_T^{t\bar t}$) data 
(IDs 573 and 574) for CT14HERA2, and the ($y_{t}, p^t_T$) 
and  ($m_{t\bar t}, y^{t\bar t}$) data (IDs 573 and 577) for CT14HERA2mJ. Note that a large value of $d^0$ results from the pull of that data set away from the gluon PDF obtained in CT14HERA2 or CT14HERA2mJ. The smallest values of $d^0$ are from those data sets that lead to the smallest apparent differences between either CT14HERA2 or CT14HERA2mJ.

\begin{table}[htbp]
	\begin{center}
		\caption{$t\bar t$ data list. Shift lengths of best-fit point when added to CT14HERA2 using ePump with various weights.}
		\begin{tabular}{| c | c | c | c | c | c | c |}
			\hline
			ID & data & \multicolumn{5}{c|}{$d^0$ CT14HERA2}\\
			&  & $w=1$ &  $w=3$ & $w=5$ & $w=8$ & $w=19$ \\
			\hline
			573 & $\sigma^{-1}\, d^2\sigma/dy_t dp^t_T$ & 0.06  & 0.16 & 0.25 & 0.37 & 0.73 \\
			\hline
			574 & $\sigma^{-1}\, d^2\sigma/dm_{t\bar t} dp^{t\bar t}_T$ & 0.10 & 0.28 & 0.43 & 0.61 & 1.06\\
			\hline
			575 & $\sigma^{-1}\, d^2\sigma/dm_{t\bar t} d\Delta \eta_{t\bar t}$ & 0.03  & 0.08 & 0.13 & 0.20 & 0.44 \\
			\hline
			576 & $\sigma^{-1}\, d^2\sigma/dm_{t\bar t} dy_t$ & 0.02 & 0.06 & 0.09 & 0.13 & 0.26\\
			\hline
			577 & $\sigma^{-1}\, d^2\sigma/dm_{t\bar t} dy_{t\bar t}$ & 0.04 & 0.10 & 0.15 & 0.21 & 0.35\\
			\hline
		\end{tabular}
		\label{table:ttb2DH}
	\end{center}
\end{table}

\begin{table}[htbp]
	\begin{center}
		\caption{$t\bar t$ data list. Shift lengths of best-fit point when added to CT14HERA2mJ using ePump with various weights.}
		\begin{tabular}{| c | c | c | c | c | c | c |}
			\hline
			ID & data &  \multicolumn{5}{c|}{$d^0$ CT14HERA2mJ}\\
			&  & $w=1$ &  $w=3$ & $w=5$ & $w=8$ & $w=19$ \\
			\hline
			573 & $\sigma^{-1}\, d^2\sigma/dy_t dp^t_T$ & 0.22 & 0.45 & 0.60 & 0.75 & 1.1 \\
			\hline
			574 & $\sigma^{-1}\, d^2\sigma/dm_{t\bar t} dp^{t\bar t}_T$ & 0.10 & 0.24 & 0.34 & 0.43 & 0.61\\
			\hline
			575 & $\sigma^{-1}\, d^2\sigma/dm_{t\bar t} d\Delta \eta_{t\bar t}$  & 0.05 & 0.15 & 0.24 & 0.36 & 0.73 \\
			\hline
			576 & $\sigma^{-1}\, d^2\sigma/dm_{t\bar t} dy_t$  & 0.10 & 0.25 & 0.34 & 0.43 & 0.59\\
			\hline
			577 & $\sigma^{-1}\, d^2\sigma/dm_{t\bar t} dy_{t\bar t}$ & 0.22 & 0.43 & 0.54 & 0.62 & 0.75\\
			\hline
		\end{tabular}
		\label{table:ttb2Dmj}
	\end{center}
\end{table}

\section{\label{sec:conclusion}Discussion and conclusions} 

The LHC can be correctly characterized as a top factory. Precise measurements of the $t\bar{t}$ final state allows for a better understanding of the production mechanisms and in particular, can allow for a determination of the gluon distribution, especially at high $x$, where it is currently relatively unconstrained. The determination of the gluon distribution, and indeed of all of the PDFs, needs to take place in the context of a global PDF fit, which includes a wide variety of data, including top production. Up to now, only singly differential top measurements have been included in global PDF fits. Double-differential measurements have the potential of providing more detailed information on the gluon distribution. With the double-differential measurements  taken by CMS, and the recent calculation of these observables to NNLO, it is now possible to use the double-differential data in a global PDF fit at NNLO, the order needed for precision determinations. 

Including the CMS double-differential top data with the nominal weight of one does not greatly impact the gluon distribution due to the greater influence of the inclusive jet data. A more sizeable impact is observed in the fit when the jet data is removed. 
We have seen that applying a  weight factor of 19 for the CMS double-differential $t\bar t$ data leads to a similar constraining power on the gluon distribution function as the jet data included in the CT14HERA2 global PDF fit. However, an almost equivalent constraining power can be reached using a lower  weight value of 8.
Such a sample is effectively present in the current 13 TeV data taken in Run 2 (especially allowing for the impact of the increased center-of-mass energy).  However, the LHC jet data will also increase proportionately. Even now, the 8 TeV CMS jet data set is more constraining than the 7 TeV data set, as will be shown in the CT18 paper~\cite{Hou:2019efy}. It is not clear in such an enlarged set of data what the relative influences of the top and inclusive jet data would be, but a greater integrated luminosity may have a larger impact on the top data as compared to the jet data, both in terms of the relative statistical and the relative systematic errors. Furthermore, due to the different composition of hard scattering processes contributing to the production of $t \bar t$ and jet productions at the LHC, precision $t \bar t$ data may constrain the gluon-PDF error band in somewhat different (large) $x$ regions as compared to jet data, cf. Figs.~\ref{Fig:mj573w19g} and~\ref{Fig:mj574w19g}. In addition, it may be possible to combine more than one double-differential set of observables, if the statistical correlations are taken into account, further strengthening the impact of the data. 
Finally, we note that similar conclusion about the impact of the CMS 8 TeV double differential data on gluon PDFs also holds for the recently released CT18 NNLO PDFs from the CTEQ-TEA group~\cite{Hou:2019efy}. Its detailed discussion is presented in the Appendix.

\begin{acknowledgments}
A.M. thanks the Department of Physics at Princeton University for hospitality during the completion of this work. 
The work of M.C. was supported by the Deutsche Forschungsgemeinschaft under grant 396021762 - TRR 257 and by a grant from Bundesministerium für Bildung und Forschung (BMBF). 
The work of A.M. and A.P. has received funding from the European Research Council (ERC) under the European Union's Horizon 2020 research and innovation programme (grant agreement No 683211); it was also supported by the UK STFC grants ST/L002760/1 and ST/K004883/1. A.P. is a cross-disciplinary post-doctoral fellow supported by funding from the University of Edinburgh and Medical Research Council (core grant to the MRC Institute of Genetics and Molecular Medicine).
The work of S.D. and I.S. was supported by the National Natural Science Foundation of China under the Grant No.~11965020 and No.~11847160.
We thank our CTEQ-TEA colleagues for support and discussions. This work was supported by the U.S. National Science Foundation under Grants No. PHY-1719914 and  PHY-1707812.
C.-P. Yuan is also grateful for the support from
the Wu-Ki Tung endowed chair in particle physics.
\end{acknowledgments}

\appendix
\section{Impacts of the CMS 8 TeV double-differential $t\bar{t}$ data on CT18 global-fit gluon PDF}
After this paper was written, the new global-fit PDF set of the CTEQ-TEA group, CT18, became available~\cite{Hou:2019efy}. In this global fit, the same conclusion holds that the CMS 8 TeV double-differential $t\bar{t}$ data do not significantly constrain the gluon PDF due to the inclusion of the other data sets (particularly, the HERA I+II combined data and the Tevatron and LHC jet data) in the CT18 global fit.  
It is the purpose of this appendix to show that 
the conclusions drawn in this paper about the  impact of various CMS 8 TeV double differential $t \bar t$ data on CT14HERA2 gluon PDF also hold when using the CT18 PDFs.

The CMS 8 TeV double differential ($y_t,p^t_T$) data set (ID 573) 
was chosen for CT18 analysis based on the compatibility with the global PDF fit.
Since we do not intend to perform a new global analysis after removing the data set 573 in the original CT18 fit, we shall instead compare to an updated CT14HERA2 fit with the inclusion of  
the data set 573  using the ePump updating package and refer to this 
set of new PDFs as CT14eCMS8YtPt in the following. 
In order to compare the potential impact of each double-differential $t\bar{t}$ data set, from ID 574 to ID 577, on these two fits (CT18 and CT14eCMS8YtPt), we shall add them one at a time with ePump and compare their impact on the central value and  error band of gluon PDFs. 
Here, we are not concerned that by doing so, the impact of the CMS double-differential $t\bar{t}$ data 
is double counted. The purpose of the exercise done in this Appendix is to demonstrate  
the same trend found about the impact of these $t \bar t$ data on gluon PDFs of 
CT18 and CT14eCMS8YtPt. 
We note that in the CT18 global analysis, we only include one of these double-differential $t\bar{t}$ data sets, i.e., the ($y_t,p^t_T$) data set (ID 573), to avoid double-counting the impact of the CMS 8 TeV data.

\begin{figure}[h] 
	\includegraphics[width=0.45\textwidth]{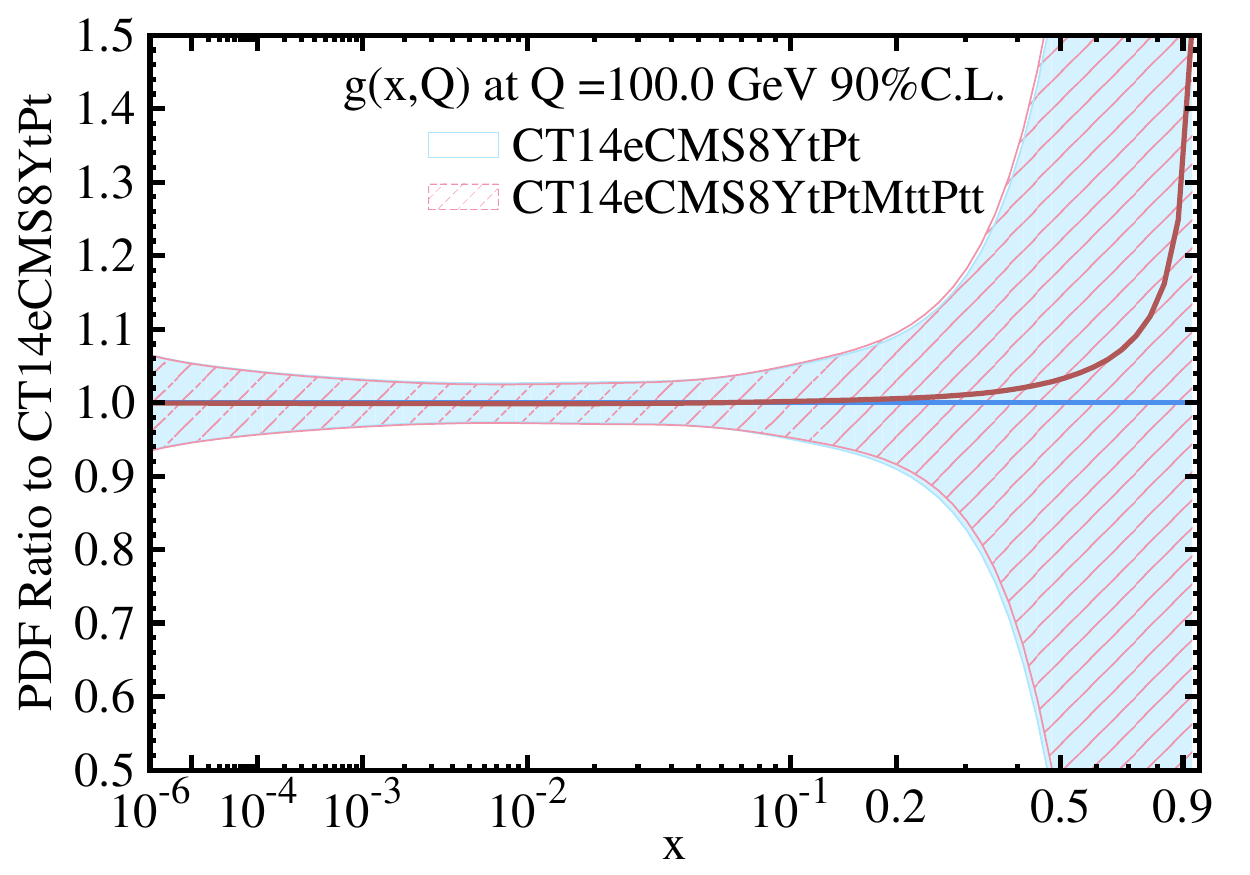}
	\includegraphics[width=0.45\textwidth]{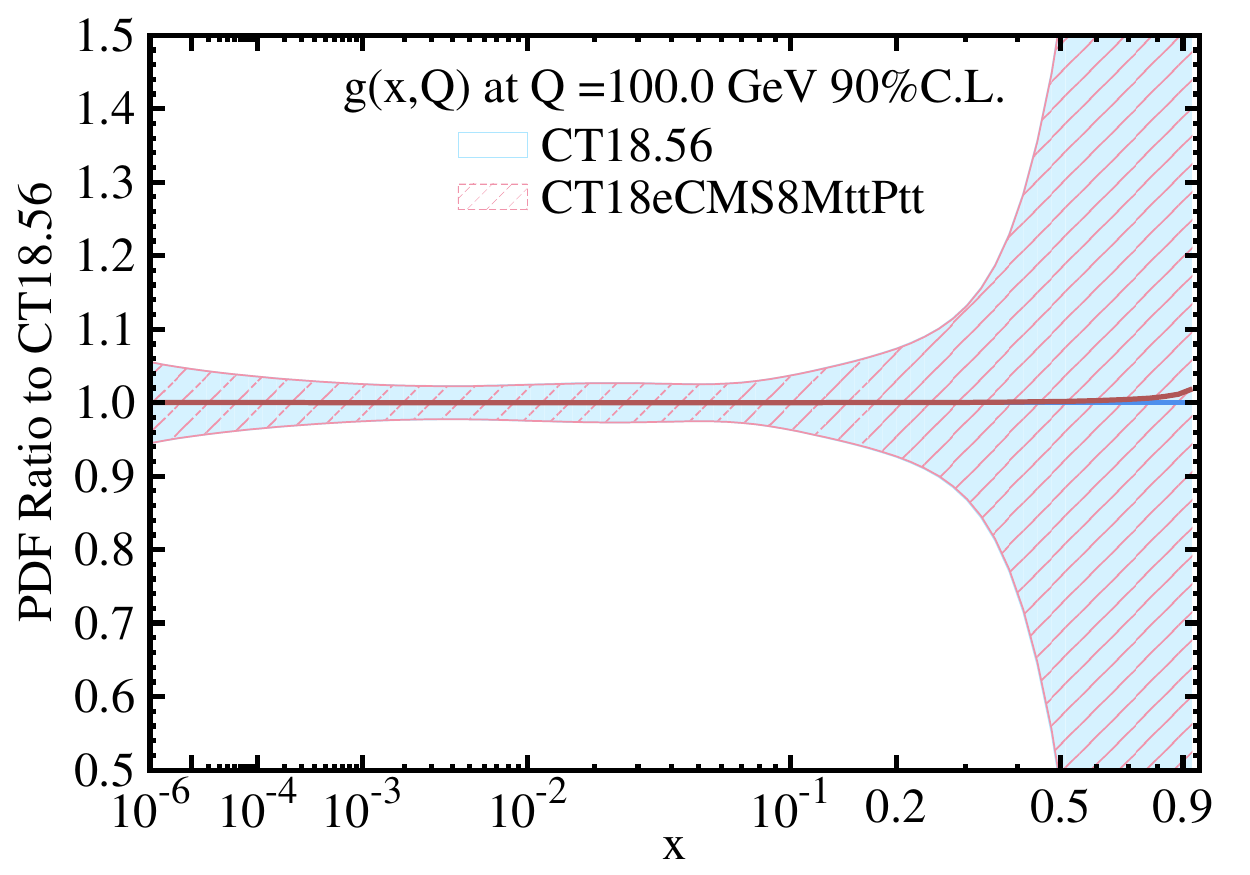}
	\caption {
The impact of 
the CMS 8 TeV double differential ($m_{t\bar t}, p_T^{t\bar t}$) data set 
(ID 574)
on gluon PDF of CT14eCMS8YtPt (left panel) and CT18 (right panel). 
CT14eCMS8YtPt is an updated CT14HERA2 fit with the inclusion of the ($y_t,p^t_T$) data set (ID 573), using the ePump updating package, and CT18  includes the exact same data set (ID 573) in its global analysis. 
}
\label{g-574}
\end{figure}

\begin{figure}[h] 
	\includegraphics[width=0.45\textwidth]{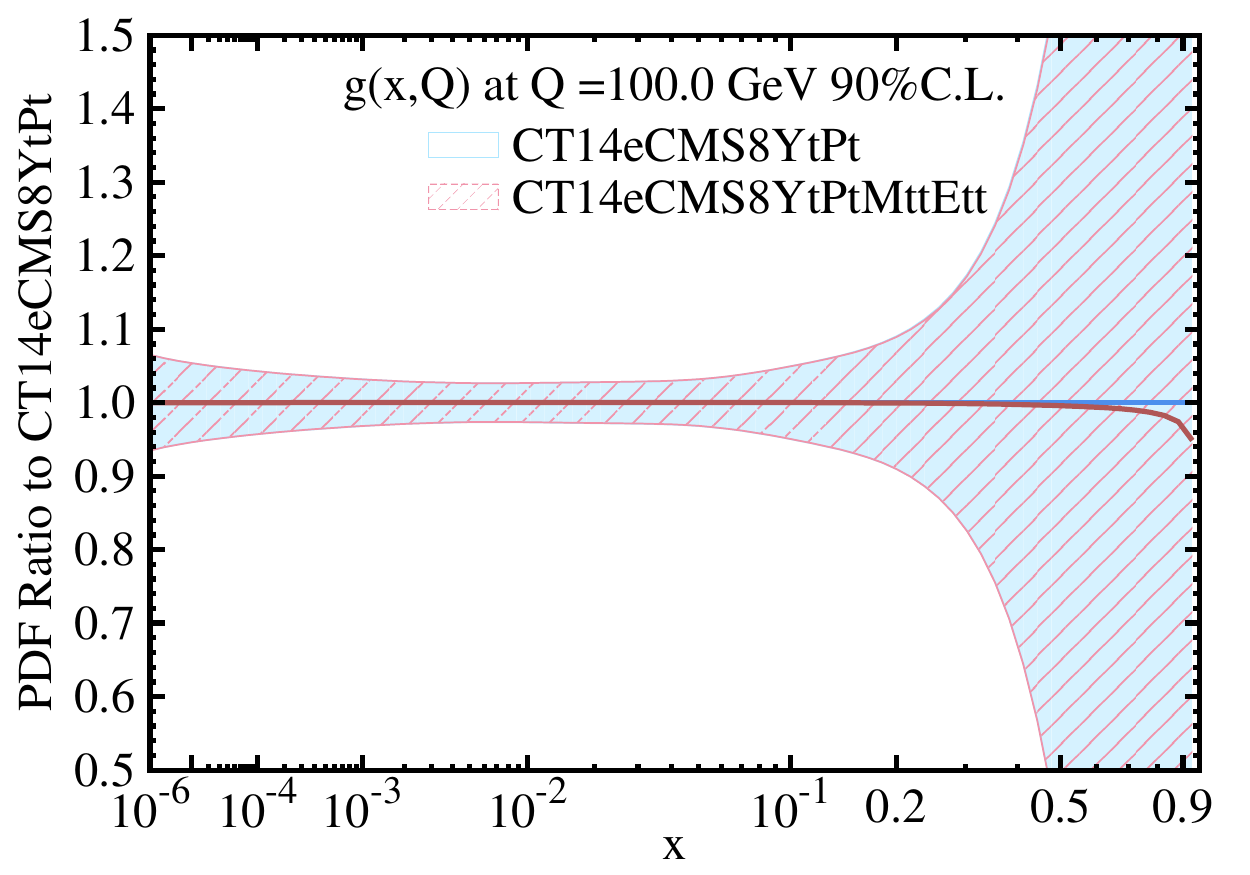}
	\includegraphics[width=0.45\textwidth]{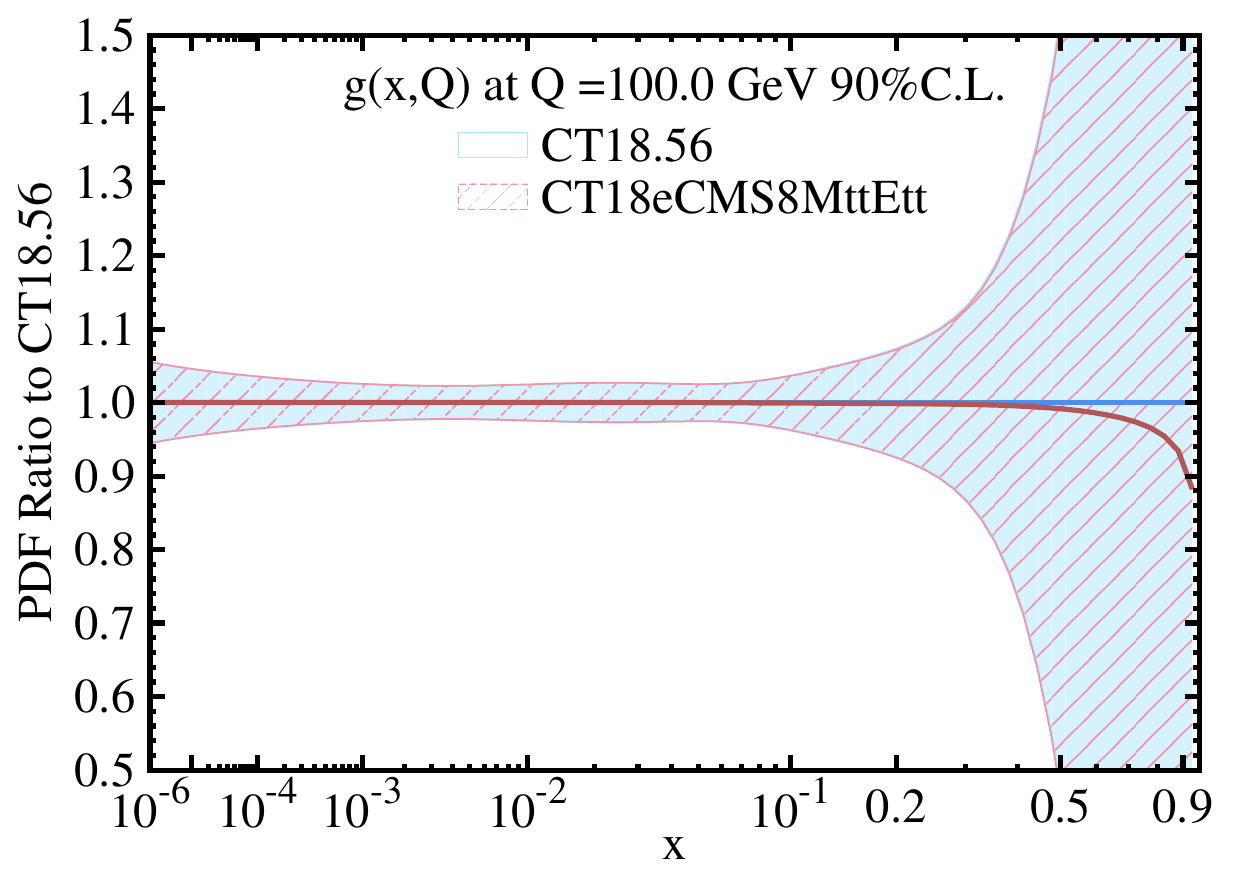}
	\caption {
 Similar to Fig.~\ref{g-574}, but the  ($m_{t\bar t}, \Delta \eta_{t\bar t}$)
 data set (ID 575)
 is added, using ePump, to update CT14eCMS8YtPt and CT18, respectively. 
}\label{g-575}
\end{figure}

\begin{figure}[h] 
	\includegraphics[width=0.45\textwidth]{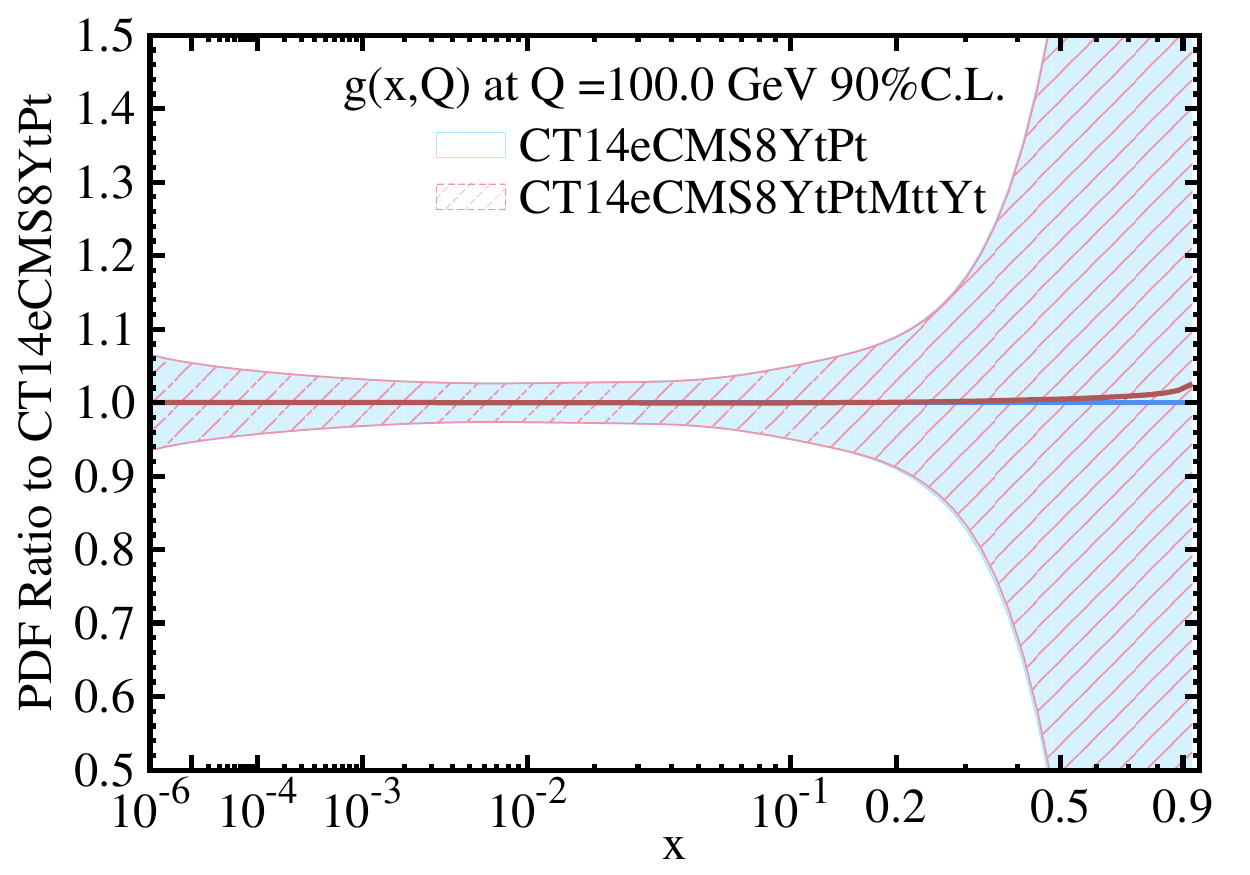}
	\includegraphics[width=0.45\textwidth]{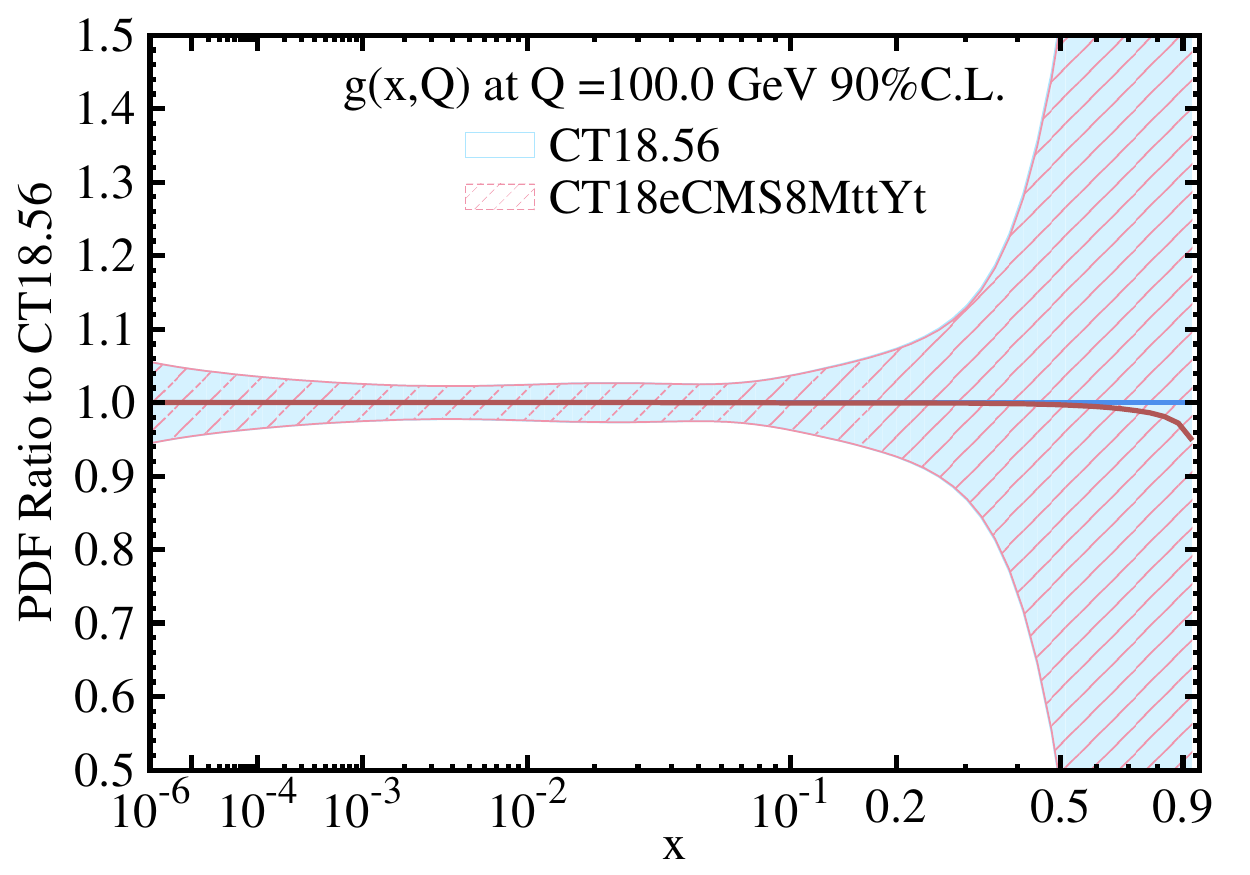}
	\caption { 
 Similar to Fig.~\ref{g-574}, but the ($m_{t\bar t},y_t$) data set  
 (ID 576)
is added, using ePump, to update CT14eCMS8YtPt and CT18, respectively. 
}\label{g-576}
\end{figure}

\begin{figure}[h] 
	\includegraphics[width=0.45\textwidth]{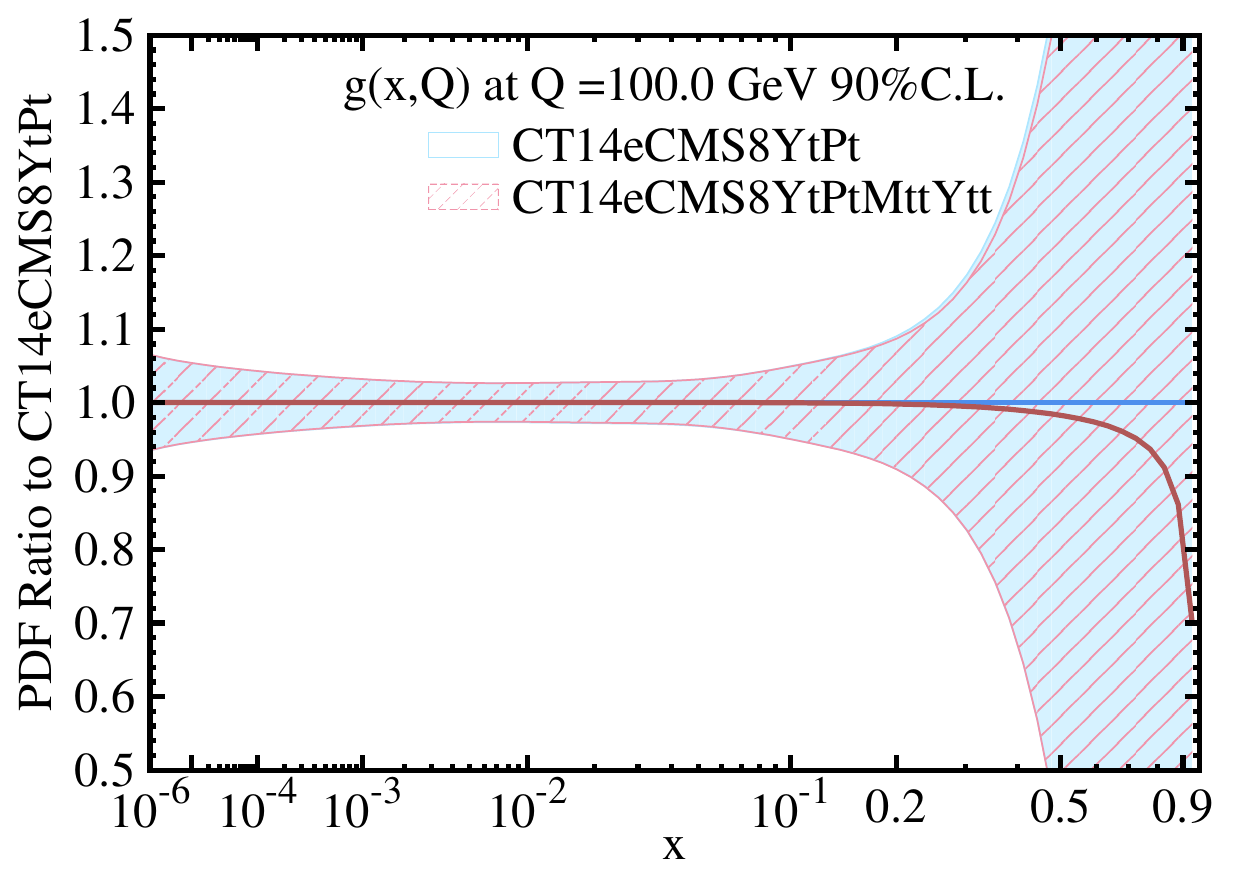}
	\includegraphics[width=0.45\textwidth]{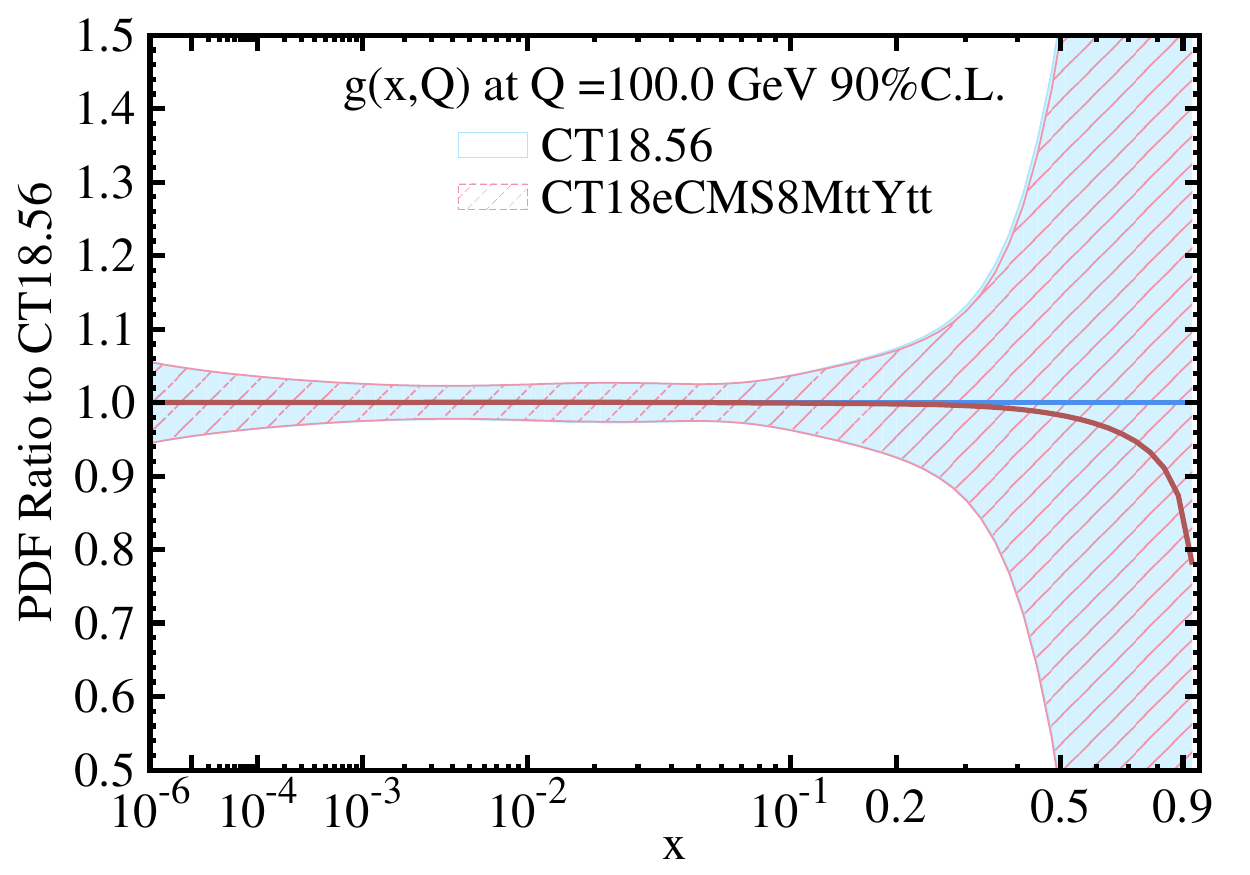}
	\caption { 
 Similar to Fig.~\ref{g-574}, but the  ($m_{t\bar t}, y^{t\bar t}$) data set (ID 577)
is added, using ePump, to update CT14eCMS8YtPt and CT18, respectively. 
}\label{g-577}
\end{figure}

In Fig.~\ref{g-574}, we compare $g$-PDFs obtained by adding the CMS 8 TeV double differential ($m_{t\bar t}, p_T^{t\bar t}$) data set 
(ID 574) to CT18 and CT14eCMS8YtPt, both with ePump. Here, we have excluded the effect from the last two error PDF eigen-sets, which were introduced in both the CT14HERA2 and CT18 fits to better describe the error band of gluon PDF in the small-$x$ region. 
Similar comparison for adding the other data set one at a time can be found in  Figs.~\ref{g-575} - \ref{g-577}, where we see that each of the double-differential $t\bar{t}$ data set is constraining the gluon PDF in the similar way for both CT18 and CT14eCMS8YtPt. 
\footnote{
It appears that in Fig.~\ref{g-576}, the inclusion of the $t\bar{t}$ ($m_{t\bar{t}} , y_t$) data set (ID 576) leads to a harder $g$ in the  
updated CT14eCMS8YtPt fit and a softer $g$ in the updated CT18 fit, for $x$ larger than about 0.5. 		
However, both changes in the $g$-PDFs are negligible as compared to $g$-PDF error band in the large $x$ region, where the nonperturbative parametrization forms of the PDFs play an important role. Hence, we do not deem the apparent difference to be significant.
}
We note that their detailed features are not identical because the total data sets included in the CT18 and CT14HERA2 fits are different, as described in Ref.~\cite{Hou:2019efy}.
Hence, we expect similar conclusion drawn in this work about the impact of the CMS 8 TeV double differential data on gluon PDFs (of CT14HERA2 NNLO PDFs) also holds for the recently released CT18 NNLO PDFs from the CTEQ-TEA group~\cite{Hou:2019efy}.

\end{document}